\documentclass[10pt,fleqn,journal,compsoc]{IEEEtran}
\usepackage{makeidx}
\usepackage{graphicx}
\usepackage{multirow}
\usepackage{cite}
\usepackage{subfigure}
\usepackage{pifont}
\usepackage{algorithm}
\usepackage{algorithmic}
\usepackage{balance}
\usepackage{mdwlist}
\usepackage{multirow,graphicx,url,epsfig,algorithmic,amsmath,amstext,amsthm}
\usepackage{amssymb}
\usepackage{amsmath}

\usepackage{makeidx}
\usepackage{graphicx}
\usepackage{subfigure}
\usepackage{mdwlist}
\usepackage{times,comment}
\usepackage{pstricks}
\usepackage{pst-node}
\usepackage{pst-tree}
\usepackage{verbatim}
\usepackage{subfigure}
\usepackage{balance}
\usepackage{pifont}
\usepackage{cite}
\usepackage{dsfont}
\usepackage{booktabs}

\begin{document}

\title{STor: Social Network based Anonymous Communication in Tor}
\author{
Peng Zhou, Xiapu Luo, Ang Chen, and Rocky K. C. Chang\\
Department of Computing, The Hong Kong Polytechnic University, Hunghom, Hong Kong\\
\texttt{\{cspzhouroc,csxluo,csachen,csrchang\}@comp.polyu.edu.hk}
}

\IEEEcompsoctitleabstractindextext{
\begin{abstract}
Anonymity networks hide user identities with the help of relayed anonymity routers. However, the state-of-the-art anonymity networks do not provide an effective trust model. As a result, users cannot circumvent malicious or vulnerable routers, thus making them susceptible to malicious router based attacks (e.g., correlation attacks). In this paper, we propose a novel social network based trust model to help anonymity networks circumvent malicious routers and obtain secure anonymity. In particular, we design an input independent fuzzy model to determine trust relationships between friends based on qualitative and quantitative social attributes, both of which can be readily obtained from existing social networks. Moreover, we design an algorithm for propagating trust over an anonymity network. We integrate these two elements in STor, a novel social network based Tor. We have implemented STor by modifying the Tor's source code and conducted experiments on PlanetLab to evaluate the effectiveness of STor. Both simulation and PlanetLab experiment results have demonstrated that STor can achieve secure anonymity by establishing trust-based circuits in a distributed way. Although the design of STor is based on Tor network, the social network based trust model can be adopted by other anonymity networks.
\end{abstract}

}
\maketitle
\IEEEdisplaynotcompsoctitleabstractindextext
\IEEEpeerreviewmaketitle

\section{Introduction}
\label{sec:intro}
Anonymity networks, which hide user identity by using relayed anonymity routers, play a very important role in protecting user privacy. However, without a trust model, the state-of-the-art anonymity networks, such as \cite{mix82,onion,Crowds,freedom00,webmix00,babel,mixminion,Tor,p2pTor1,p2pTor2,p2pTor3}, are vulnerable to various malicious router based attacks \cite{Johnson_CSF_09,Syverson_CCS_11}. Therefore, introducing an effective trust model to anonymity networks remains a critically important problem. In this paper, we propose a novel social network based trust model and apply it to the Tor network \cite{Tor}, which is one of the most dominant low-latency anonymity network today \cite{Syverson_CCS_11} and is used by around 250,000 users \cite{Goldberg06onthe,whoTor}. Although the design and implementation of the proposed trust model is based on Tor, this model can be easily applied to other anonymity networks.

A Tor user accesses Internet services through dynamically encrypted circuits. Each circuit usually consists of three Tor routers, preventing an attacker from tracking back a user and protecting the private content through nested encryptions. However, without a trust-based onion routing algorithm, a number of malicious router based attacks on Tor have successfully demonstrated that Tor's anonymity could be hampered if one or more Tor routers in a circuit become malicious. Examples of such attacks include correlation attacks \cite{low-cost,locating,low-resource,one-cell,cell-counter,ZFGBW09}, congestion attacks \cite{EDG09}, disclosure attacks \cite{AK03}, and latency based client location attacks \cite{HVC10}.

Tor \cite{path} uses guard routers at the entry point and selects exit routers according to the exit node policy to prevent malicious routers. Moreover, Tor relies on a group of directory servers to check each router's credibility according to its uptime \cite{path,uptime,tune-up}. These mechanisms, however, are insufficient for trust-based anonymous communication due to the following reasons. 
First, checking router identities based on uptime alone can be easily bypassed by an attacker. For example, an attacker can set up a malicious router and operate it normally for a period of time to gain the directory servers' trust. Second, without trust-based routing algorithm, Tor basically considers every candidate router with the same trust. Thus, Tor is unable to select routers based on their capability of providing secure anonymity when forming the circuits. Third, the central directory servers are potential targets of various attacks, notably targeted intrusion, IP address blacklisting, and DDoS attacks. For example, network wardens (e.g., The Great Firewall of China \cite{gfc}) can block Tor networks by blacklisting directory servers. The private bridges made available since version $0.2.1$ \cite{bridge}, which are designed to help users affected by the blocking, are often set up in an ad hoc manner. Fourth, the existing Tor architecture is not scalable, because a user is required to maintain up-to-date information about all the Tor routers \cite{scalable}.

To effectively evade the malicious router based attacks, we propose a novel trust-based anonymity network, called STor (Social network based Tor), which employs social networks to help users circumvent malicious routers. STor provides secure anonymity to its users by ``overlaying'' social networks on top of the current Tor infrastructure. More precisely, STor uses the existing trust relationships from real-world social networks to determine a trust score for each router. As the trust information is based on the router's owner, STor is able to identify not only malicious routers, but also the routers that are vulnerable to being compromised. A STor user thus selects a router based on the trust score computed from the relationship between the user and the router's owner. In particular, the user will exclude a router which is owned by someone who is not in his friendship circle (i.e., no relationship with the user). Another advantage of using social networks is to eliminate the need for the central directory servers. Therefore, STor can operate as a pure distributed system to provide trust-based anonymity services. Finally, STor is scalable, because each user is only required to maintain up-to-date information about their friends which is only a subset of the entire Tor network. However, the uncertainty and vagueness of the real-world social networks make the relationships between friends imprecise \cite{MJSJV06,CCR00}, thus designing an effective trust model based on social networks to facilitate anonymity networks is not trivial.

In summary, we have made three important contributions in this paper:
\begin{enumerate}
  \item We introduce a novel social network based trust model to further secure anonymity networks. In particular, we have designed STor, the Tor network enhanced by the social network based trust model. STor users thus select routers by taking into account their trust in those routers.
  \item We have designed an input independent fuzzy model to handle the uncertainty and vagueness of social relationships, thus determining the direct trust according to various social attributes in different social networks. We have also developed an algorithm to propagate indirect trust to a friend's friends which form a friendship circle. Both of them are the major elements in STor.
  \item We have implemented STor by modifying Tor's source code and performed a comprehensive evaluation of the effectiveness of STor in both simulation and experiments over PlanetLab platform. The experimental results show that STor can effectively establish trust-based circuits in a decentralized fashion and help users obtain secure anonymity.
\end{enumerate}

The remainder of this paper is organized as follows. Section \ref{sec:threatandSTor} presents the threat model considered in this paper. Section \ref{sec:design} explains how trust relationships in social networks are used in the design of STor, including the quantification and propagation of trust. Section \ref{sec:fuzzy} elaborates the qualification process of various social attributes by using a novel input independent fuzzy model. Section \ref{sec:experiment} presents both simulation and PlanetLab experiment results to evaluate STor. Section \ref{sec:relatedwork} reviews related work, followed by a conclusion in Section \ref{sec:conclusion}.


\section{Threat Model}
\label{sec:threatandSTor}

\begin{figure*}[ht!]
\centering
\includegraphics[width=0.70\textwidth]{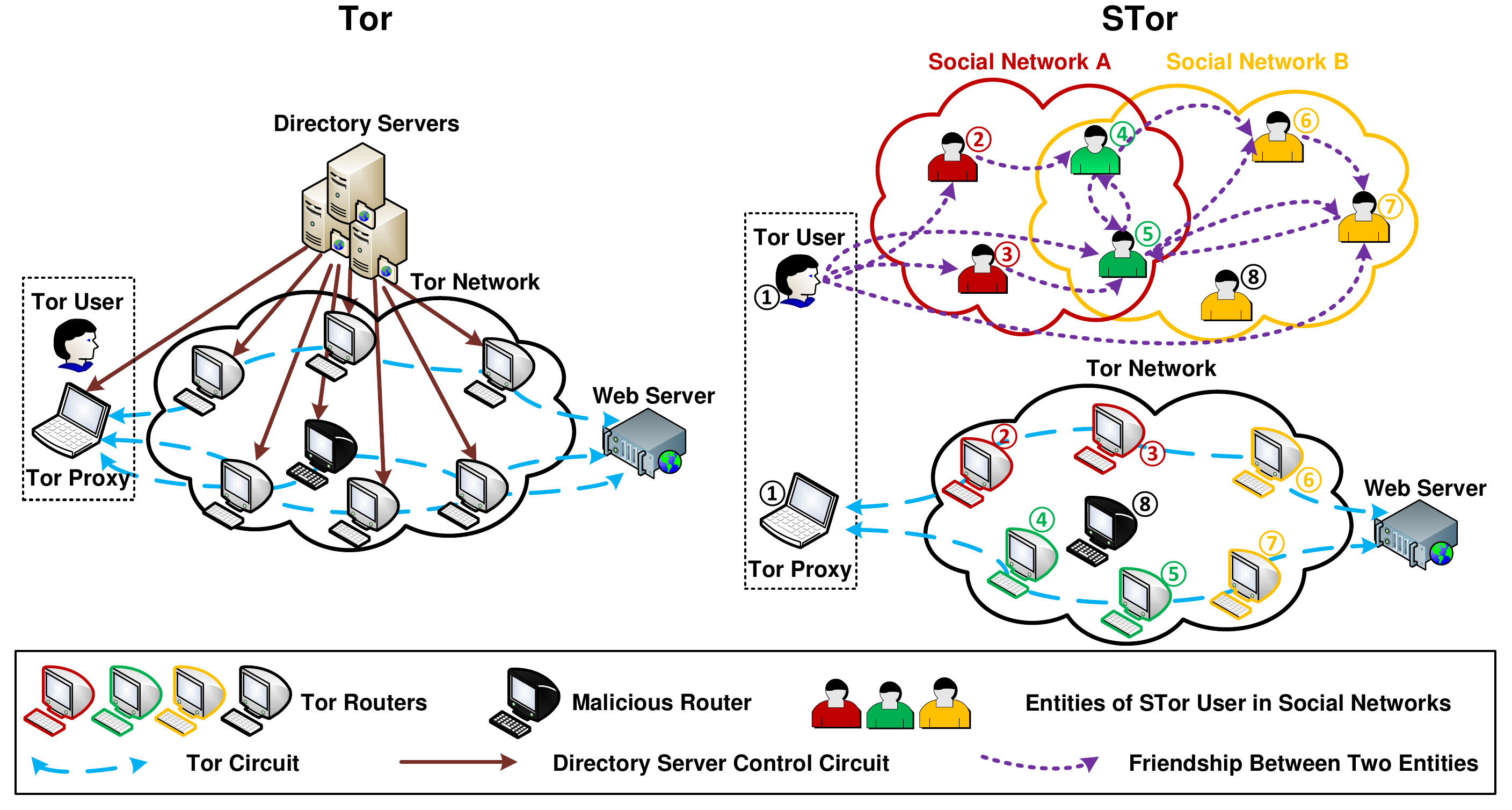}
\caption{Architecture of Tor and STor.}
\label{fig:stor}
\end{figure*}

The architecture of Tor is illustrated on the left side of Fig.~\ref{fig:stor}. When Tor users request anonymity service to visit a remote server, they ask the directory servers for a set of Tor routers that have already been determined as trustworthy due to their uptime to build encrypted circuits. After establishing the circuit, the Tor user sends data to the local Onion Proxy, which subsequently forwards the data to the remote server through the circuit. To provide sufficient anonymity, the circuit changes every ten minutes. In the Tor network, a larger number of candidate routers usually imply better anonymity service. We refer the anonymity service that is determined by the number of candidate routers to as \textit{baseline anonymity}, because it provides the necessary (but not sufficient) mechanisms for anonymity communication.

Directory servers were originally used for defeating Sybil attacks \cite{Sybil}, which generate a huge number of fake identities to attack distributed systems. However, attackers can easily obtain the Tor network's trust, as the directory servers verify router identities only based on their uptimes. Moreover, since the directory servers are publicly accessible, attackers could compromise them through DDoS, IP/DNS blocking, and targeted intrusion. To evade these attacks, a trust-based routing algorithm must be able to
\begin{itemize}
  \item Verify a Tor router's identity using a comprehensive set of parameters, including not only the router's uptime, but also its owner's reputation,
  \item Enable users to take trust into consideration when selecting routers to establish circuits, and
  \item Manage trust in a distributed way without being threatened by Sybil attacks.
\end{itemize}

STor uses the trust relationship from existing social networks to manage trust in the Tor network. As many recent studies have demonstrated that social networks can effectively help distributed systems evade Sybil attacks \cite{sybilanalysis,keep,wuhan,sumup,SybilInfer,SybilLimit,SybilGuard}, STor can therefore satisfy the three requirements and provide \textit{secure anonymity} by enhancing the baseline anonymity with the capability of circumventing malicious routers based on a trust model.

As STor determines the trust relationship between users based on existing social networks, attacks targeting social networks, like \cite{BBR10,BPHKBK10}, may compromise STor. In the paper, it is therefore essential to assume that the social network provides sufficient security for establishing trust among friends. In particular, we make two assumptions for the design of STor:
\begin{itemize}
  \item All entities in social networks can authenticate information about their friends.
  \item Existing social networks have comprehensive mechanisms to avoid leaking of private information, such as an entity's friendship circles and the IP addresses of their routers.
\end{itemize}
The first assumption is to guarantee that the trust relationship obtained from social networks is reliable for computing the trust of each router. The second assumption is to prevent the trust relationships from being leaked to the public domain. If an attacker were aware of an entity's trust relationships, he could compromise the entity's anonymity by tracking the routers of the entity's friends.

The right side of Fig.~\ref{fig:stor} illustrates the architecture of STor. Any router included in a circuit must be owned by the user's friends or friends of their friends from social networks $A$ and $B$. Furthermore, friends with a higher trust is more likely to be selected for the circuit formation. For instance, router \ding{179}, a malicious router, is excluded when circuits are built because its owner, the human entity \ding{179}, is neither a friend of user \ding{172} nor his friends' friend. Using this approach, STor can allocate trust-based circuits in a decentralized manner without the need of directory servers and exclude (potentially) malicious routers with the help of trust relationships from real-world social networks.

For user \ding{172}, some of their friends or some friends of friends (e.g., \ding{175}, \ding{176}) belong to both social networks $A$ and $B$, which are considered as merged entities by STor. That is, STor can manage trust relationships across multiple social networks. Section \ref{subsec:basicmethod} will further elaborate on this. Furthermore, each entity can generally possess more than one router in STor and access STor through multiple proxies. In the interest of clarity, we consider in the rest of this paper only the case that each user has a single router and uses a single proxy.

\section{Design of STor}
\label{sec:design}

STor, as a social network based Tor, harnesses the trust relationship from existing real-world social networks to help users obtain secure anonymity by using trust-based circuits. This section addresses the challenges in the design of STor, including the calculation of the direct and indirect trust, using the trust relationship to facilitate router selection. We also discuss the impacts on the performance and baseline anonymity provided by Tor.

\subsection{Modeling the trust relationship}
\label{subsec:basicmethod}
We model the social structure of STor as a weighted directed cyclic graph, $\mathbb{G}=\langle\mathbb{N}, \mathbb{L}\rangle$, whose nodes represent human entities and directed links indicate their uni-directional friendships. Each link has a trust value indicating the level of trust a person places in their friend in STor. Let $\mathbb{N}=\{1,2, \ldots  ,n\}$ denote the set of human entities (i.e., nodes on the graph) and $\mathbb{L}=\{(i\rightarrow j), i,j\in \mathbb{N}\}$ the set of uni-directional friendships between human entities (i.e., links on the graph), where $i\rightarrow j$ indicates that entity $i$ trusts entity $j$.
%

As discussed in Section \ref{sec:threatandSTor}, STor admits trust relationships from multiple social networks. Therefore, we have $\mathbb{G}=\mathop{\cup}\limits_{\mathcal{S}_s\in \mathbb{S}}\mathbb{G}^{s}=\langle\mathop{\cup}\limits_{\mathcal{S}_s\in \mathbb{S}}\mathbb{N}^{s}, \mathop{\cup}\limits_{\mathcal{S}_s\in \mathbb{S}}\mathbb{L}^{s}\rangle$, where, $\mathbb{S}=\{\mathcal{S}_s,~s=1,2,\ldots\}$ is the set of social networks used in STor. For a given social network $\mathcal{S}_s$, the subgraph $\mathbb{G}^{s}=\langle\mathbb{N}^{s}, \mathbb{L}^{s}\rangle$ represents the structure of $\mathcal{S}_s$. $\mathbb{N}^{s}$ and $\mathbb{L}^{s}$ are the respective sets in $\mathcal{S}_s$.

If entity $i$ can reach entity $j$ through an acyclic path comprising $r$ directed links in graph $\mathbb{G}$, entity $j$ belongs to the \textit{$r$-hop friendship circle} of entity $i$, denoted as $F_{i,r}=\{q|p\in F_{i,r-1}, (p\rightarrow q)\in \mathbb{L}\}$. $F_{i,r}$ has a recursive definition with $F_{i,1}=\{q| (i\rightarrow q)\in \mathbb{L}\}$ as the base case. An entity could be a member of multiple friendship circles of another entity. The \textit{friendship circle} of entity $i$ is therefore given by $F_{i}=F_{i,1}\cup F_{i,2}\cup \ldots  \cup F_{i,g}$, where $g$ is the maximum number of directed links between entity $i$ and their friends. We use $||F_{i}||$ to denote the number of entities in $F_{i}$. Obviously, $F_{i,1}$ includes entity $i$'s friends and $F_{i,r} (r>1)$ contains the friends of $i$'s friends. Additionally, we define $F_{i}^s$ as the friendship circle of entity $i$ in social network $\mathcal{S}_s\in \mathbb{S}$ and subsequently $F_{i}=\mathop{\cup}\limits_{\mathcal{S}_s\in \mathbb{S}}F_{i}^s$.

\begin{figure}[ht!]
\centering
\includegraphics[width=0.3\textwidth]{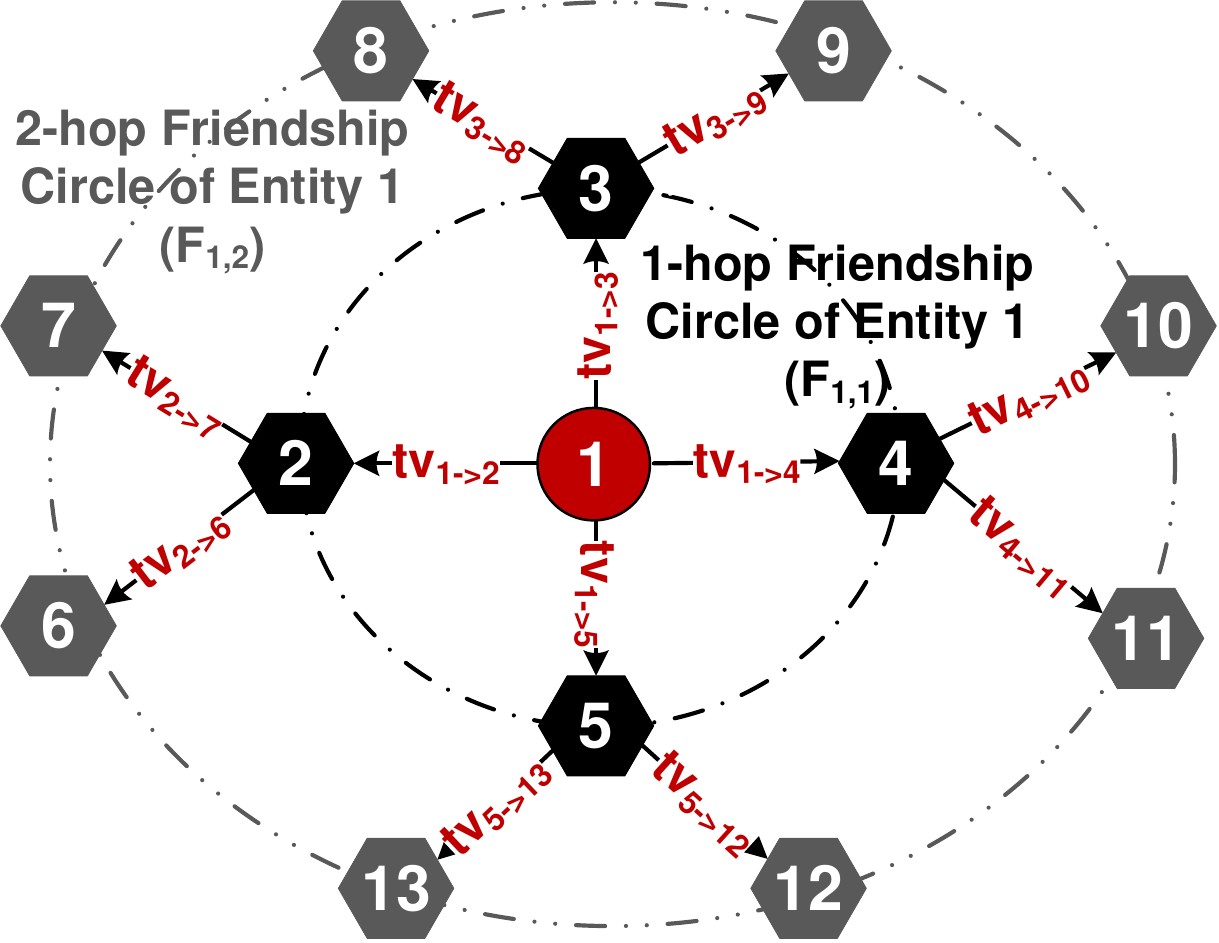}
\caption{Entity 1's friendship circle.}
\label{fig:friend_circle}
\end{figure}

Fig.~\ref{fig:friend_circle} shows an example of the friendship circles for entity $1$ which has a total of $12$ friends and friends of friends. Four of them are in their $1$-hop friendship circle (i.e., friends), while the rest are in their $2$-hop friendship circle (i.e., friends of friends). Each directed link is associated with a trust value, which will be defined next.

\subsection{Computing the direct trust}
\label{subsec:cttv}

We define \textit{trust value} as the degree of trust one entity assigns to his friend directly, denoted as $tv\in [0,1]$. Let $tv_{i\rightarrow j}$ represent the trust value entity $i$ gives to $j$, $j\in F_{i,1}$. As each entity can make friends in more than one social network, let $tv_{i\rightarrow j}^{s}$ denote the trust value that entity $i$ places in $j$ in social network $\mathcal{S}_s$. If entities $i$ and $j$ are friends in more than one social network, the highest trust value will be used, that is, $tv_{i\rightarrow j}=\mathop{max}\limits_{(i\rightarrow j)\in\mathbb{L}^{s}, \mathcal{S}_s\in\mathbb{S}}tv_{i\rightarrow j}^{s}$.

The trust value $tv_{i\rightarrow j}^{s}$ can be computed based on quantitative and qualitative social attributes. \textit{Quantitative social attributes} are quantifiable, such as the communication frequency between two friends and the duration of their friendship in the social network. Higher communication frequency and longer friendship generally result in a higher trust value. A Tor router's uptime is also considered as a quantitative social attribute. \textit{Qualitative social attributes}, on the other hand, represent qualitative features, such as the relationship between two people, and their majors and careers. For example, a family member should receive more trust than a stranger, and a friend with a major and a career in computer security should receive additional trust.

In the real social networks, the relationships between friends are imprecise because the environments are uncertainty and vagueness \cite{MJSJV06}, thus converting social attributes into a trust value can be effectively handled by the fuzzy model \cite{fuzzy}, which is especially useful for imprecise categories \cite{CCR00}. The fuzzy model can help calibrate the measure of social trust by using set membership relevant to substantive knowledge from social attributes \cite{CCR00}. However, as there is a lack of quantifiable units, the qualitative attributes, which are important elements in STor, cannot be converted by using a fuzzy model directly. To overcome this problem, we propose a novel \textit{input independent fuzzy model} to determine the trust value based on both quantitative and qualitative social attributes. Section \ref{sec:fuzzy} details this fuzzy process.

\subsection{Computing the indirect trust}
\label{subsec:tpm}

In STor, the degree of indirect trust is computed through trust propagation over the underlying \textit{friendship paths}. Let $\mathbb{P}_{i\rightarrow j}=\{\Gamma_{i\rightarrow j}^c,\ c=1,2,\ldots,C\}$ be the set containing all the friendship paths starting from entity $i$ to entity $j$, where the friendship path $\Gamma_{i\rightarrow j}^c=i\rightarrow h_{1}\rightarrow\ldots \rightarrow h_{r-1}\rightarrow j$ indicates trust propagation from entity $i$ to their $r$-hop friend $j$ through intermediate friends $h_{1}, \ldots , h_{r-1}$ and $||\mathbb{P}_{i\rightarrow j}||=C$ is the number of paths in the set. For each $\Gamma_{i\rightarrow j}^c$, a \textit{trust distance}, denoted as $td(\Gamma_{i\rightarrow j}^c)$, represents the degree of trust entity $i$ places in $j$ over this path.

Since $td(\Gamma_{i\rightarrow j}^c)$ should be a non-increasing function of $r$, we compute it by multiplying the trust value of each link:
\begin{eqnarray}
\begin{array}{c}
\label{equ:td}
td(\Gamma_{i\rightarrow j}^c) = tv_{i\rightarrow h_{1}}\times \ldots  \times tv_{h_{r-1}\rightarrow j}.
\end{array}
\end{eqnarray}
Moreover, we define a \textit{trust score} as the highest degree of trust entity $i$ can give to entity $j$, denoted as $ts_{i\Rightarrow j}$, by considering all possible friendship paths. Thus,
\begin{eqnarray}
\begin{array}{c}
\label{equ:ts}
ts_{i\Rightarrow j} = \mathop{max}\limits_{\forall \Gamma_{i\rightarrow j}^c\in \mathbb{P}_{i\rightarrow j}}td(\Gamma_{i\rightarrow j}^c).
\end{array}
\end{eqnarray}



\begin{figure}[ht!]
\centering
\includegraphics[width=0.3\textwidth]{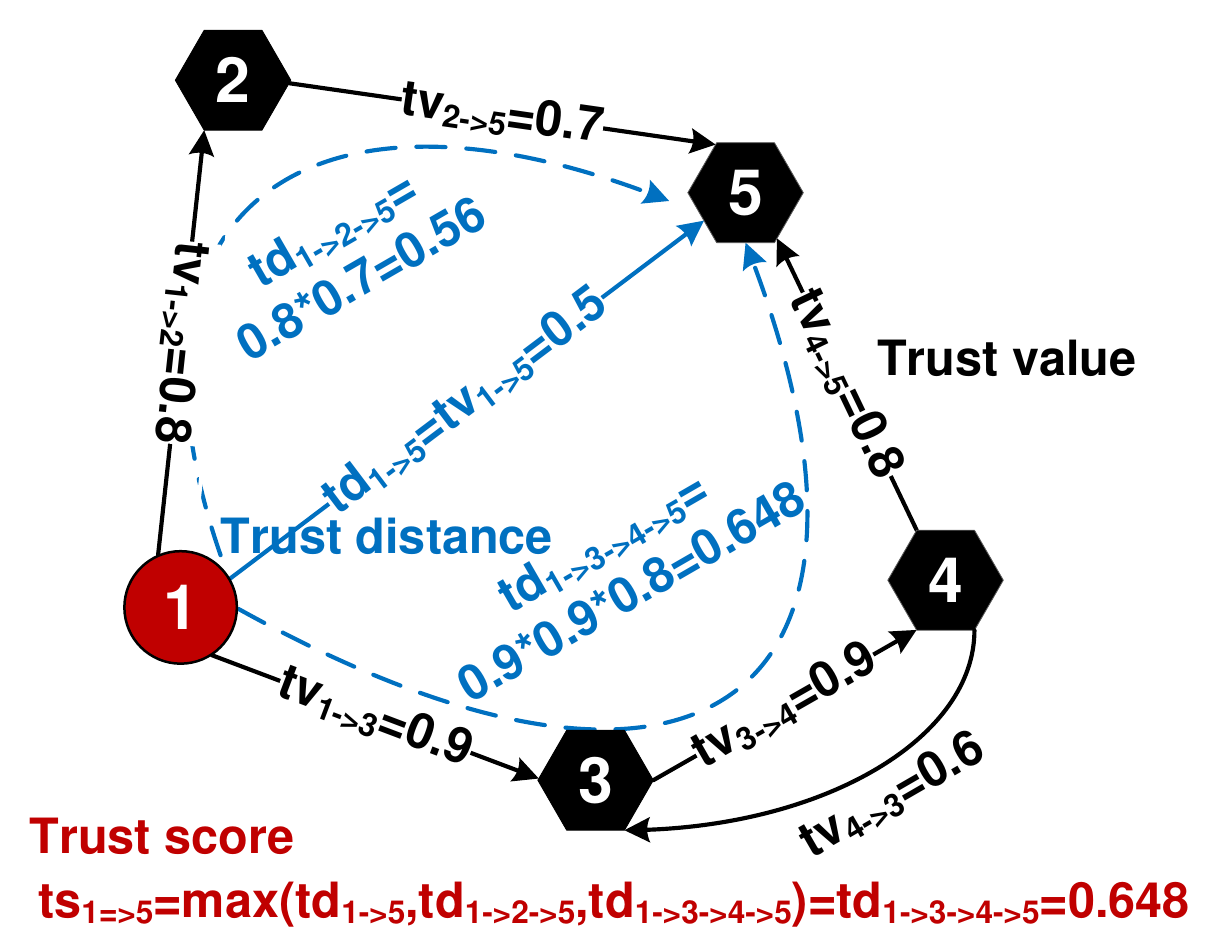}
\caption{An example for the trust value, trust distance, and trust score.}
\label{fig:longest_path}
\end{figure}

Fig.~\ref{fig:longest_path} illustrates an example of computing the trust value, trust distance, and trust score. In this example, entity $5$ belongs to $F_{1,1}$, $F_{1,2}$, and $F_{1,3}$ simultaneously and three different paths connect entity $1$ with entity $5$: $1\rightarrow 5$, $1\rightarrow 2\rightarrow 5$, and $1\rightarrow 3\rightarrow 4\rightarrow 5$. The trust score $ts_{1\Rightarrow 5}$ equals $0.648$, which is the maximum trust distance given by the last path.

The problem of calculating the trust score from each entity to his friends can be formulated as a longest path problem with a non-increasing distance function (i.e., Eq.~(\ref{equ:td})). It is well known that the shortest path problem can be solved in polynomial time when the distance function is non-decreasing in a weighted directed cyclic graph \cite{lpp}. Therefore, by converting the distance function from non-decreasing to non-increasing, the longest path problem in STor can be solved efficiently by converted it into a shortest path problem. A dynamic programming algorithm can subsequently be used to solve this problem in polynomial time.

According to Eq.~(\ref{equ:td}) and $tv\in[0,1]$, we have $td\in [0,1]$. Therefore,
\begin{equation*}
\mathop{max}\limits_{\forall \Gamma_{i\rightarrow j}^c\in \mathbb{P}_{i\rightarrow j}}td(\Gamma_{i\rightarrow j}^c)\equiv 1-\mathop{min}\limits_{\forall \Gamma_{i\rightarrow j}^c\in \mathbb{P}_{i\rightarrow j}}(1-td(\Gamma_{i\rightarrow j}^c)).
\end{equation*}
Based on this, the definition of trust scores can be transformed to $ts_{i\Rightarrow j}=1-\mathop{min}\limits_{\forall \Gamma_{i\rightarrow j}^c\in \mathbb{P}_{i\rightarrow j}}(1-td(\Gamma_{i\rightarrow j}^c))$. By defining \textit{distrust distance} as $utd(\Gamma_{i\rightarrow j}^c)=1-td(\Gamma_{i\rightarrow j}^c)$, which is a non-decreasing distance function, the longest trust distance problem is converted to a shortest distrust distance problem. The Fibonacci heap can be used to implement the typical Dijkstra algorithm \cite{dijkstra} to determine the $||F_{i}||$ shortest paths from each entity $i$ to his friends and friends of friends (i.e., $\forall j\in F_i$) with a time complexity of $O(||F_{i}||\times log(||F_{i}||))$.

\subsection{Trust-based Onion Routing Algorithm}
\label{subsec:newalgorithm}

In Tor, directory servers select routers from the available ones to construct circuits. Let $b_{j}$ denote the available bandwidth of entity $j$. The probability of selecting router $j$ is ${b_{j}}/{\sum_{k=1}^{n}b_{k}}$ \cite{path} for $n$ available routers. In contrast, entity $i$ in STor can only select Onion routers from candidates that are confined to $F_{i}$ and the probability of selecting router $j$ (i.e., $Pr_{ij}$) is
\begin{equation}
\label{equ:pj}
Pr_{ij}=\frac{(1-\omega) \times ts_{i\Rightarrow j}+\omega \times BW_{j}}{\sum_{k\in F_{i}}((1-\omega) \times ts_{i\Rightarrow k}+\omega \times BW_{k})},
\end{equation}
where $BW_{j}={b_{j}}/{max_{x\in F_{i}}(b_{x})}$ is a normalized bandwidth and $\omega\in [0,1]$ is a parameter to balance the trust score with bandwidth. A small $\omega$ gives more weight to the trust score, whereas a large $\omega$ gives more weight to bandwidth. With this new algorithm, the STor users can only select routers from their friendship circle and have a high probability of choosing routers with a high trust score.

\subsection{Effects on Tor networks}
\label{subsec:impact}
\subsubsection{Effect on performance}
\label{subsubsec:perimpact}
Although parameter $\omega$ in Eq. (\ref{equ:pj}) can be increased to give a high probability of connecting to routers with large bandwidth, the candidate routers are still restricted to a user's friendship circle in STor, which is only a subset of the Tor network. As a result, the circuit established in STor may not enjoy the best performance, because other routers outside the friendship circle cannot be used. This problem could be resolved by encouraging users to invite more friends to join STor, especially those with high bandwidth routers. Particularly, STor can build up a similar recruiting platform, like BRAIDS \cite{JHK10}, to benefit users who successfully request their friends to participate in STor.

\subsubsection{Effect on baseline anonymity}
\label{subsubsec:anoimpact}
In anonymity networks, the baseline anonymity service can be achieved by forming circuits from a large set of candidate routers. STor obtains secure anonymity by using the trust-based routing algorithm to circumvent malicious routers. However, the STor users can only select routers from their friendship circle. As a result, the baseline anonymity could be degraded. To abate this possible degradation, users are encouraged to make more friends with existing users in STor or introduce more outside friends to join STor. Enlarging the friendship circle unfortunately may run into the risk of including malicious routers, because it is more likely to include friends with low trust. To address this problem, we use a threshold, $ts_{h}$, to filter out the friends with low trust scores. If an entity's trust score is lower than $ts_{h}$, his router will not be considered for circuit establishment. The friendship circle is therefore refined to a \textit{trustworthy friendship circle} $TF_{i}=\{j\in F_{i}, ts_{i\Rightarrow j}\geq ts_{h}\}$, and $||TF_i||$ is the size of $TF_i$.

%

\section{Determination of Trust Value according to Social Attributes}
\label{sec:fuzzy}
In Section \ref{subsec:cttv} we derive an algorithm to merge trust values from multiple social networks and mention the use of an input independent fuzzy model to convert both quantitative and qualitative attributes into a trust value in each social network. This section details this process. Section \ref{subsec:tfm} elucidates the challenges of the conversion by using the traditional fuzzy model. Section \ref{subsec:safm} highlights the advantages of the input independent fuzzy model and demonstrates its theoretical design. This model is subsequently applied to STor in Section \ref{subsec:astor}. As an example, a real case for the conversion by using the input independent fuzzy model is illustrated in Section \ref{subsec:exam}.

\subsection{Challenges in the use of traditional fuzzy model}
\label{subsec:tfm}
In the social realm, as social environments are always uncertainty and vagueness, the social relationships between friends are imprecise \cite{MJSJV06}. The fuzzy model \cite{fuzzy} is therefore an appropriate technique to handle the conversion from social attributes to the trust values \cite{MJSJV06}.

Traditionally, the fuzzy model \cite{fuzzy} includes various quantitative input variables (i.e., called crisp inputs in \cite{fuzzy}), each of which is associated with a relevant input fuzzy set. Each input fuzzy set consists of a group of qualitative values (i.e., called linguistic values in \cite{fuzzy}) and these values can be converted into membership functions whose elements have degrees of membership. An quantitative input variable can obtain their membership grade from each membership function in their relevant input fuzzy set. Each grade can be mapped to a corresponding membership function in a corresponding output fuzzy set specified by the fuzzy rules. Finally, the fuzzy model uses defuzzification methods to determine quantitative outputs. Note that the quantitative input variable and their relevant input fuzzy set have to be related to the same property. For example, if the quantitative input variable is a value of temperature, its relevant input fuzzy set must be the qualitative description of temperature.

To integrate the quantitative social attributes into the trust value by the traditional fuzzy model, we can use their quantitative values as the quantitative input variables and design an input fuzzy set with a set of reasonable membership functions to represent imprecise categories for each quantitative attribute. Unlike the quantitative attributes, the lack of quantitative values make the qualitative social attributes hard to be converted by using the traditional fuzzy model directly.

To overcome this challenge, we propose a novel fuzzy model, called the input independent fuzzy model, to determine social relationships. In this model, the quantitative input variables can obtain their membership grade over unrelated input fuzzy sets (i.e., input independent). As a result, by allowing quantitative social attributes to get their membership grade over different qualitative social attributes, the trust value can be determined based on both quantitative and qualitative social attributes by using the input independent fuzzy model.

\subsection{The input independent fuzzy model}
\label{subsec:safm}
The input independent fuzzy model considers each qualitative social attribute as an input fuzzy set, and converts their qualitative values to membership functions. By allowing the quantitative social attributes to obtain their membership grade over unrelated qualitative social attributes, a quantitative trust value is yielded according to both qualitative and quantitative social attributes.

\begin{figure*}[ht!]
     \centering
     \subfigure[The input independent fuzzy model.]{
           \label{fig:socialfuzzy}
          \includegraphics[width=.45\textwidth]{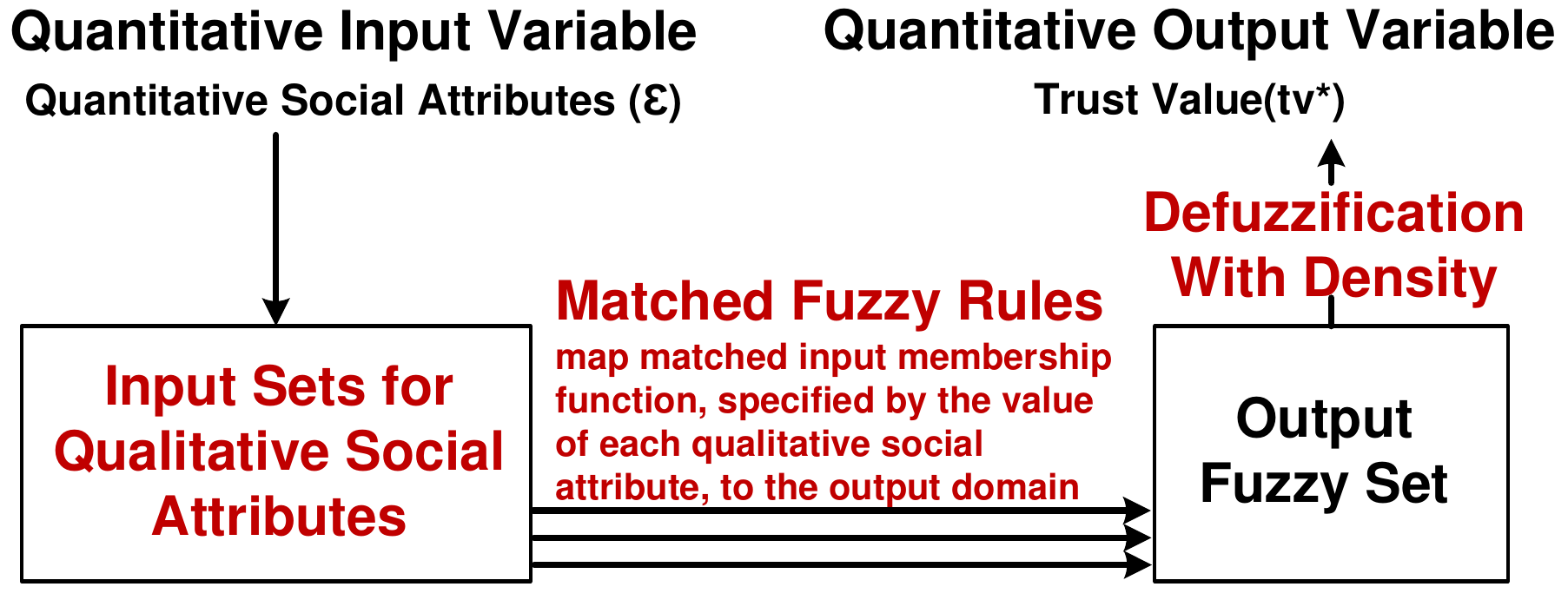}}
 \hspace{2ex}
 \subfigure[The traditional fuzzy model.]{
           \label{fig:trafuzzy}
          \includegraphics[width=.45\textwidth]{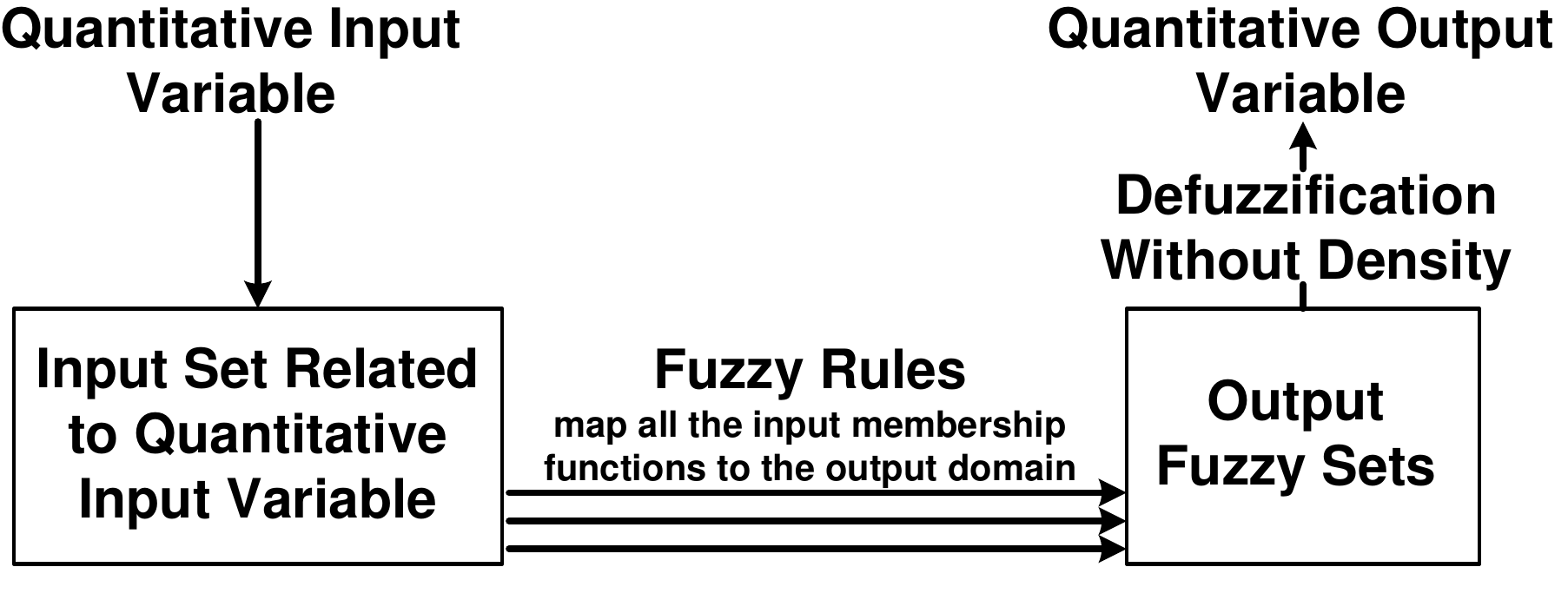}}
     \caption{Difference between the input independent fuzzy model and the traditional fuzzy model.}
     \label{fig:difftands}
\end{figure*}

Comparing with the traditional fuzzy model, there are three major improvements in this new model, as shown in Fig.~\ref{fig:difftands}:
\begin{itemize}
  \item The quantitative input variable can calculate their membership grade over unrelated input fuzzy sets that are defined by qualitative social attributes (i.e., input independent).
  \item The quantitative input variable can only obtain their membership grade from an exclusively selected membership function in each input fuzzy set. This selected membership function is specified by the qualitative value of each qualitative social attribute.
  \item A density constant is introduced to balance the defuzzification when different output membership functions surround different sizes of area.
\end{itemize}

As a result, the input independent fuzzy model shows its advantages for determining the trust value as follows.
\begin{itemize}
  \item This model is able to handle the uncertainty and vagueness of the social relationships, which are usual features in the social realm \cite{MJSJV06}, thus helping convert social attributes into the trust value.
  \item With the help of quantitative social attributes, qualitative social attributes can be converted by using this model. Both quantitative and qualitative social attributes can thus be used to determine the trust value simultaneously.
  \item This model is flexible as unlimited number of social attributes can be taken into account. Moreover, the qualitative social attributes can define different number of membership functions to implement different levels of conversion accuracy from social attributes to the trust value.
\end{itemize}

\subsubsection{The quantitative input variable}
\label{subsec:qiv}
We use $\mathcal{E}$ to denote the quantitative input variable in this model and compute it based on all the quantitative social attributes. In each social network, $\mathcal{E}$ is calculated as:
\begin{eqnarray}
\label{eqn:quantitative}
\begin{array}{c}
\mathcal{E}=\mathop{\sum}\limits_{\forall E_{e} \in \mathbb{E}}(\pi_{e}\times \overline{\mathcal{E}}(E_e)),\\
where,\ \mathop{\sum}\limits_{\forall E_{e} \in \mathbb{E}}(\pi_{e})=1,\\
\overline{\mathcal{E}}(E_e)\in [0,1], \ \pi_{e}\in [0,1], \ e\in [1,2,\ldots].
\end{array}
\end{eqnarray}
$\mathbb{E}=\{E_{e}, e\in [1,2,3,\ldots]\}$ represents the set of all the quantitative social attributes in each social network and $E_e$ is the $e$th quantitative social attribute in $\mathbb{E}$. $\overline{\mathcal{E}}(E_e)$ is the normalized quantified value of quantitative attribute $E_e$ while $\pi_{e}$ is the weight of $E_e$.

For example, in social network $S_s$, $\mathcal{E}_{i\rightarrow j}^{s}(E_e)$ is used to represent the quantifiable value of quantitative attribute $E_e$ between entity $i$ and $j\in F_{i}^s$. Thus their normalized quantitative value $\overline{\mathcal{E}}_{i\rightarrow j}^{s}(E_e)$ can be calculated as $\overline{\mathcal{E}}_{i\rightarrow j}^{s}(E_e)={\mathcal{E}_{i\rightarrow j}^{s}(E_e)}/{\mathop{max}\limits_{p\in F_{i}^s}(\mathcal{E}_{i\rightarrow p}^{s}(E_e))}$, where, $F_{i}^s$ is the friendship circle of entity $i$ in social network $\mathcal{S}_s$, as defined in Section \ref{subsec:basicmethod}.

\subsubsection{The fuzzy sets}
\label{subsec:tfs}


The input independent fuzzy model enables the ability to calculate the membership of quantitative input variable over unrelated input fuzzy sets, and use membership functions to represent the qualitative values in each qualitative social attribute. $\mathcal{E}$ can thus obtain their membership over any qualitative social attributes in each social network.
More precisely, each qualitative social attribute is defined as an input fuzzy set and their qualitative values are converted into membership functions, as:
\begin{eqnarray}
\begin{array}{c}
\mathcal{I}=\mathop{\bigcup}\limits_{\forall \mathcal{A}_{k}\in \mathbb{A}}\mathcal{I}^{\mathcal{A}_{k}}, \
\mathcal{I}^{\mathcal{A}_{k}}=\{(\ \mathcal{I}_{p}^{\mathcal{A}_{k}},\ \mu_{\mathcal{I}_{p}^{\mathcal{A}_k}}(\mathcal{E})\ )\},\\
\mu_{\mathcal{I}_{p}^{\mathcal{A}_k}}(\mathcal{E})=\{\mathcal{E}:\mathcal{I}_{p}^{\mathcal{A}_k}\rightarrow [0,1]\}, \\
where,\ \mathcal{A}_k\in \mathbb{A},\ p\in [1,2,\ldots],\ \mathcal{E}\in D_{\mathcal{I}_{p}^{\mathcal{A}_{k}}}\subseteq [0,1],
\end{array}
\end{eqnarray}
where $\mathbb{A}=\{\mathcal{A}_{k}, k\in [1,2,3,\ldots]\}$ represents all the qualitative social attributes in the considered social network. $\mathcal{A}_{k}$ is the $k$th social attribute in $\mathbb{A}$ and $\mathcal{I}^{\mathcal{A}_{k}}$ represents the input fuzzy set of $\mathcal{A}_{k}$. $\mathcal{I}_{p}^{\mathcal{A}_{k}}$ is the $p$th qualitative value of qualitative social attribute $\mathcal{A}_{k}$, while $\mu_{\mathcal{I}_{p}^{\mathcal{A}_k}}(\mathcal{E})$ is their membership function that can be used to map $\mathcal{E}$ to the degree of membership for this qualitative social attribute. $D_{\mathcal{I}_{p}^{\mathcal{A}_{k}}}$ is the definition domain of the membership function $\mu_{\mathcal{I}_{p}^{\mathcal{A}_k}}(\mathcal{E})$.

Unlike the input fuzzy sets, we have only one output fuzzy set in the input independent fuzzy model. This output fuzzy set is relevant to the quantitative output (i.e., trust value) but includes various qualitative descriptions of it.
We define the output fuzzy set as:
\begin{eqnarray}
\begin{array}{c}
\mathcal{O}=\{(\ \mathcal{O}_{q},\ \mu_{\mathcal{O}_{q}}(tv)\ )\},\
\mu_{\mathcal{O}_{q}}(tv)=\{tv:\mathcal{O}_{q}\rightarrow [0,1]\}, \\
where,\ q\in [1,2,\ldots],\ tv\in D_{\mathcal{O}_{q}}\subseteq [0,1],
\end{array}
\end{eqnarray}
where $\mathcal{O}_{q}$ is the $q$th qualitative description of output fuzzy set $\mathcal{O}$ and can be represented as a membership function,  $\mu_{\mathcal{O}_{q}}(tv)$. $D_{\mathcal{O}_{q}}$ is the definition domain of $\mu_{\mathcal{O}_{q}}(tv)$.

\subsubsection{The fuzzy rules}
\label{subsec:tfr}
The fuzzy logic rule for correlating the $p$th membership function in the $k$th input set with a corresponding $q$th output membership function is defined as:
\begin{center}
\begin{description}
  \item[Rule $k.p$]
    \begin{tabular}{l}
        IF The Qualitative Value of $\mathcal{A}_k$ belongs to $\mathcal{I}_{p}^{\mathcal{A}_k}$,\\
        THEN Trust Value is $\mathcal{O}_{q}$.
    \end{tabular}
\end{description}
\end{center}

For each qualitative social attribute $\mathcal{A}_k\in \mathbb{A}$, if its qualitative value belongs to $\mathcal{I}_{p=p_{\mathbf{m}}}^{\mathcal{A}_k}$, the corresponding fuzzy rule $k.p_{\mathbf{m}}$ is used to map $\mathcal{I}_{p_{\mathbf{m}}}^{\mathcal{A}_k}$ to $\mathcal{O}_{q_{\mathbf{m}}}$. We call the fuzzy rule, $k.p_{\mathbf{m}}$ which is selected by the qualitative value of $\mathcal{A}_k$, as the matched rule. For each matched rule, the input membership grade $\mu_{\mathcal{I}_{p_{\mathbf{m}}}^{\mathcal{A}_k}}(\mathcal{E})$, calculated according to the quantitative input variable $\mathcal{E}$, can be used to truncate the corresponding membership function in the output fuzzy set. Thus, there is a truncated membership function in the output domain for each qualitative social attribute $\mathcal{A}_k$, in which the input membership function is selected by the matched rule $k.p_{\mathbf{m}}$:
\begin{eqnarray}
\label{eqn:single}
\begin{array}{l}
\mu_{k.p_{\mathbf{m}}}(\mathcal{E},tv)=\left\{
\begin{array}{l}
\mu_{\mathcal{I}_{p_{\mathbf{m}}}^{\mathcal{A}_k}}(\mathcal{E}), \mu_{\mathcal{O}_{q_{\mathbf{m}}}}(tv)\geq\mu_{\mathcal{I}_{p_{\mathbf{m}}}^{\mathcal{A}_k}}(\mathcal{E})\\
\mu_{\mathcal{O}_{q_{\mathbf{m}}}}(tv), \mu_{\mathcal{O}_{q_{\mathbf{m}}}}(tv)<\mu_{\mathcal{I}_{p_{\mathbf{m}}}^{\mathcal{A}_k}}(\mathcal{E})
\end{array}
\right.\\
where,\ \mathcal{E}\in D_{\mathcal{I}_{p_{\mathbf{m}}}^{\mathcal{A}_{k}}}\subseteq [0,1],\ tv\in D_{\mathcal{O}_{q_{\mathbf{m}}}}\subseteq [0,1].
\end{array}
\end{eqnarray}

\subsubsection{The defuzzification process}
\label{subsec:tdp}
In the defuzzification process, the output trust value $tv^*(\mathcal{E})$ is calculated as the center of mass of a shape. The shape is surrounded by the union of the truncated membership functions $\mu_{k.p_{\mathbf{m}}}(\mathcal{E},tv)$ for all the considered qualitative social attributes $\mathcal{A}_k\in \mathbb{A}$ and the $tv$ axis in the output domain.

For each qualitative social attribute $\mathcal{A}_{k=K}$, if its qualitative value matches the rule $K.p_{\mathbf{m}}$, the positions in the $tv$ axis weighted by the masses of the truncated membership function $\mu_{K.p_{\mathbf{m}}}(\mathcal{E},tv)$ can be computed as:
\begin{eqnarray}
\label{eqn:mptv}
\begin{array}{l}
\mathcal{MP}_{K.p_{\mathbf{m}}}(\mathcal{E})=\int_{tv=0}^{1}tv\times\rho_{q}\times\mu_{K.p_{\mathbf{m}}}(\mathcal{E},tv)d(tv).
\end{array}
\end{eqnarray}
and the total mass of this truncated membership function as:
\begin{eqnarray}
\label{eqn:mtv}
\begin{array}{l}
\mathcal{M}_{K.p_{\mathbf{m}}}(\mathcal{E})=\int_{tv=0}^{1}\rho_{q}\times\mu_{K.p_{\mathbf{m}}}(\mathcal{E},tv)d(tv).
\end{array}
\end{eqnarray}

Therefore, $tv^*(\mathcal{E})$ can be computed as the center of the mass according to Greiner's algorithm \cite{gravity}:
\begin{eqnarray}
\label{eqn:tvstar}
\begin{array}{l}
tv^*(\mathcal{E})=\frac{\mathop{\sum}\limits_{\mathcal{A}_{k}\in \mathbb{A}}\mathcal{MP}_{k.p_{\mathbf{m}}}(\mathcal{E})}{\mathop{\sum}\limits_{\mathcal{A}_{k}\in \mathbb{A}}\mathcal{M}_{k.p_{\mathbf{m}}}(\mathcal{E})}.
\end{array}
\end{eqnarray}
Particularly, when only one social attribute $\mathcal{A}_K$ is used, its output trust value can be computed as:
\begin{eqnarray}
\label{eqn:singletvstar}
\begin{array}{l}
tv_{k=K,p=p_{\mathbf{m}}}^*(\mathcal{E})=\frac{\mathcal{MP}_{K.p_{\mathbf{m}}}(\mathcal{E})}{\mathcal{M}_{K.p_{\mathbf{m}}}(\mathcal{E})}.
\end{array}
\end{eqnarray}

Unlike the the center of gravity defuzzification algorithm in the traditional fuzzy model \cite{defuzzy}, here a density constant $\rho_{q}$, which is associated with each of the corresponding output membership function $\mu_{\mathcal{O}_{q}}(tv)$, is introduced to the center of mass algorithm in Eq.~(\ref{eqn:mptv}) and Eq.~(\ref{eqn:mtv}). $\rho_{q}$ must satisfy the following identical equation:
\begin{eqnarray}
\label{eqn:iequ}
\begin{array}{c}
\int_{tv=0}^{1}\rho_{q_x}\times\mu_{\mathcal{O}_{q_x}}d(tv)\equiv\int_{tv=0}^{1}\rho_{q_y}\times\mu_{\mathcal{O}_{q_y}}d(tv),\\
where,\ \forall \mathcal{O}_{q_x},\mathcal{O}_{q_y}\subseteq \mathcal{O}.
\end{array}
\end{eqnarray}
This equation guarantees that all the output membership functions provide the same weight to the final trust value $tv^{*}(\mathcal{E})$.

As a larger $\mathcal{E}$ leads to a higher trust value, the fuzzy sets and rules should be designed to let function $tv^*(\mathcal{E})$ be non-decreasing, thus the inequality Eq.~(\ref{eqn:ineq}) should be satisfied.
\begin{eqnarray}
\label{eqn:ineq}
\begin{array}{l}
(tv^*(\mathcal{E}))^{'}=\frac{d(tv^*(\mathcal{E}))}{d(\mathcal{E})}\geq 0.
\end{array}
\end{eqnarray}

\subsection{Application in STor}
\label{subsec:astor}
%
In STor, each qualitative social attribute forms an input fuzzy set, while only one output fuzzy set is used. Note that more membership functions declarations in a fuzzy set can lead to finer-grained quantitative conversion. In our model, STor can be designed with different numbers of membership functions in each input fuzzy set (i.e., the qualitative social attribute) and output fuzzy set to achieve different levels of conversion accuracy for the trust value when necessary. As a case study, STor defines three membership functions for each input fuzzy set and use five membership functions in the output set.

For each qualitative social attribute $\mathcal{A}_k$, their qualitative values, $\mathcal{I}_{p}^{\mathcal{A}_{k}}$, are defined as:
\begin{eqnarray}
\begin{array}{c}
\mathcal{I}_{p}^{\mathcal{A}_{k}}=\left\{
\begin{array}{ll}
(POSITIVE)^{k}, &p=1\\
(NEUTRAL)^{k}, &p=2\\
(NEGATIVE)^{k}, &p=3
\end{array}
\right.\\
where,\ \mathcal{A}_k\in \mathbb{A},\ p\in [1,2,3].
\end{array}
\end{eqnarray}
And the qualitative values of the output set, $\mathcal{O}_{q}$, are:
\begin{eqnarray}
\begin{array}{c}
\mathcal{O}_{q}=\left\{
\begin{array}{ll}
LARGEST, &q=1\\
LARGE, &q=2\\
NORMAL, &q=3\\
SMALL, &q=4\\
SMALLEST, &q=5
\end{array}
\right.\\
where,\ q\in [1,2,3,4,5].
\end{array}
\end{eqnarray}

If the qualitative value of qualitative social attribute $\mathcal{A}_k$ belongs to $(POSITIVE)^{k}$, it indicates that this value will cause a large trust value. If the value belongs to $(NEGATIVE)^{k}$, a small trust value will be yielded. $(NEUTRAL)^{k}$, unlike the other two, will result in the intermediate trust value. In the output set, both $\mathcal{O}_{1}=LARGEST$ and $\mathcal{O}_{2}=LARGE$ indicate larger trust values but with different magnitudes. In contrast, $\mathcal{O}_{4}=SMALL$ and $\mathcal{O}_{5}=SMALLEST$ reflect different levels of small trust values.

Based on these definitions, following fuzzy rules for each qualitative social attribute $\mathcal{A}_k$ are designed:
\begin{center}
\begin{small}
\begin{description}
  \item[Rule $k.1$]
    \begin{tabular}{l}
        IF The Qualitative Value of $\mathcal{A}_k$ belongs to $\mathcal{I}_{1}^{\mathcal{A}_{k}}=POSITIVE^k$,\\
        THEN Trust Value is
        $\{
        \begin{array}{ll}
        i.\mathcal{O}_{1}=LARGEST.\\
        ii.\mathcal{O}_{2}=LARGE.
        \end{array}
        $
    \end{tabular}
  \item[Rule $k.2$]
    \begin{tabular}{l}
        IF The Qualitative Value of $\mathcal{A}_k$ belongs to $\mathcal{I}_{2}^{\mathcal{A}_{k}}=NEUTRAL^k$,\\
        THEN Trust Value is $\mathcal{O}_{3}=NORMAL$.
    \end{tabular}
  \item[Rule $k.3$]
    \begin{tabular}{l}
        IF The Qualitative Value of $\mathcal{A}_k$  belongs to $\mathcal{I}_{3}^{\mathcal{A}_{k}}=NEGATIVE^k$,\\
        THEN Trust Value is
        $\{
        \begin{array}{ll}
        i.\mathcal{O}_{4}=SMALL.\\
        ii.\mathcal{O}_{5}=SMALLEST.
        \end{array}
        $
    \end{tabular}
\end{description}
\end{small}
\end{center}
%

As there is a very strong link between triangular plots and most forms of social attributes \cite{CCR00}, STor employs the triangular membership function definition~\cite{whytriangular} to declare membership functions both in input and output fuzzy sets. Since fuzzy rules map $(POSITIVE)^{k}$ to larger trust values while $(NEGATIVE)^{k}$ to smaller ones, and considering larger $\mathcal{E}$ yielding larger trust values, membership function of $(POSITIVE)^{k}$ should be increasing when $\mathcal{E}$ increases and that of $(NEGATIVE)^{k}$ needs to be decreasing. For $(NEUTRAL)^{k}$, their membership function is designed to increases at first then goes down. By considering inequality Eq.~(\ref{eqn:ineq}), we thus define the membership functions in input fuzzy sets as follows:
%
\begin{eqnarray}
\label{eqn:inputm}
\begin{array}{ll}
\mu_{\mathcal{I}_{1}^{\mathcal{A}_{k}}}(\mathcal{E})&=\mathcal{E},\ \ \ \ \ \ \ \ \ \ \ \mathcal{E}\in[0, 1],\\
\mu_{\mathcal{I}_{2}^{\mathcal{A}_{k}}}(\mathcal{E})&=\left\{
\begin{array}{ll}
\mathcal{E},& \mathcal{E}\in[0, 0.5],\\
1-\mathcal{E},& \mathcal{E}\in[0.5, 1],
\end{array}
\right.\\
\mu_{\mathcal{I}_{3}^{\mathcal{A}_{k}}}(\mathcal{E})&=1-\mathcal{E},\ \ \ \ \ \ \mathcal{E}\in[0, 1].
\end{array}
\end{eqnarray}
Similarly, the membership functions in the output fuzzy sets $\mathcal{O}_{q},\ q\in [1,2,3,4,5]$ are defined as:
\begin{eqnarray}
\label{eqn:outputm}
\begin{array}{ll}
\mu_{\mathcal{O}_{1}}(tv)&=\left\{
\begin{array}{ll}
4\times tv-3, &tv\in[0.75, 1],\\
0, &tv\in others,
\end{array}
\right.\\
\mu_{\mathcal{O}_{2}}(tv)&=\left\{
\begin{array}{ll}
4\times tv-2, &tv\in[0.5, 0.75],\\
4-4\times tv, &tv\in[0.75, 1],\\
0, &tv\in others,
\end{array}
\right.\\
\mu_{\mathcal{O}_{3}}(tv)&=\left\{
\begin{array}{ll}
4\times tv-1, &tv\in[0.25, 0.5],\\
3-4\times tv, &tv\in[0.5, 0.75],\\
0, &tv\in others,
\end{array}
\right.\\
\mu_{\mathcal{O}_{4}}(tv)&=\left\{
\begin{array}{ll}
4\times tv, &tv\in[0, 0.25],\\
2-4\times tv, &tv\in[0.25, 0.5],\\
0, &tv\in others,
\end{array}
\right.\\
\mu_{\mathcal{O}_{5}}(tv)&=\left\{
\begin{array}{ll}
1-4\times tv, &tv\in[0, 0.25],\\
0, &tv\in others.
\end{array}
\right.
\end{array}
\end{eqnarray}

\begin{figure*}[ht!]
     \centering
     \subfigure[The Output Fuzzy Set.]{
           \label{fig:output}
          \includegraphics[width=.238\textwidth]{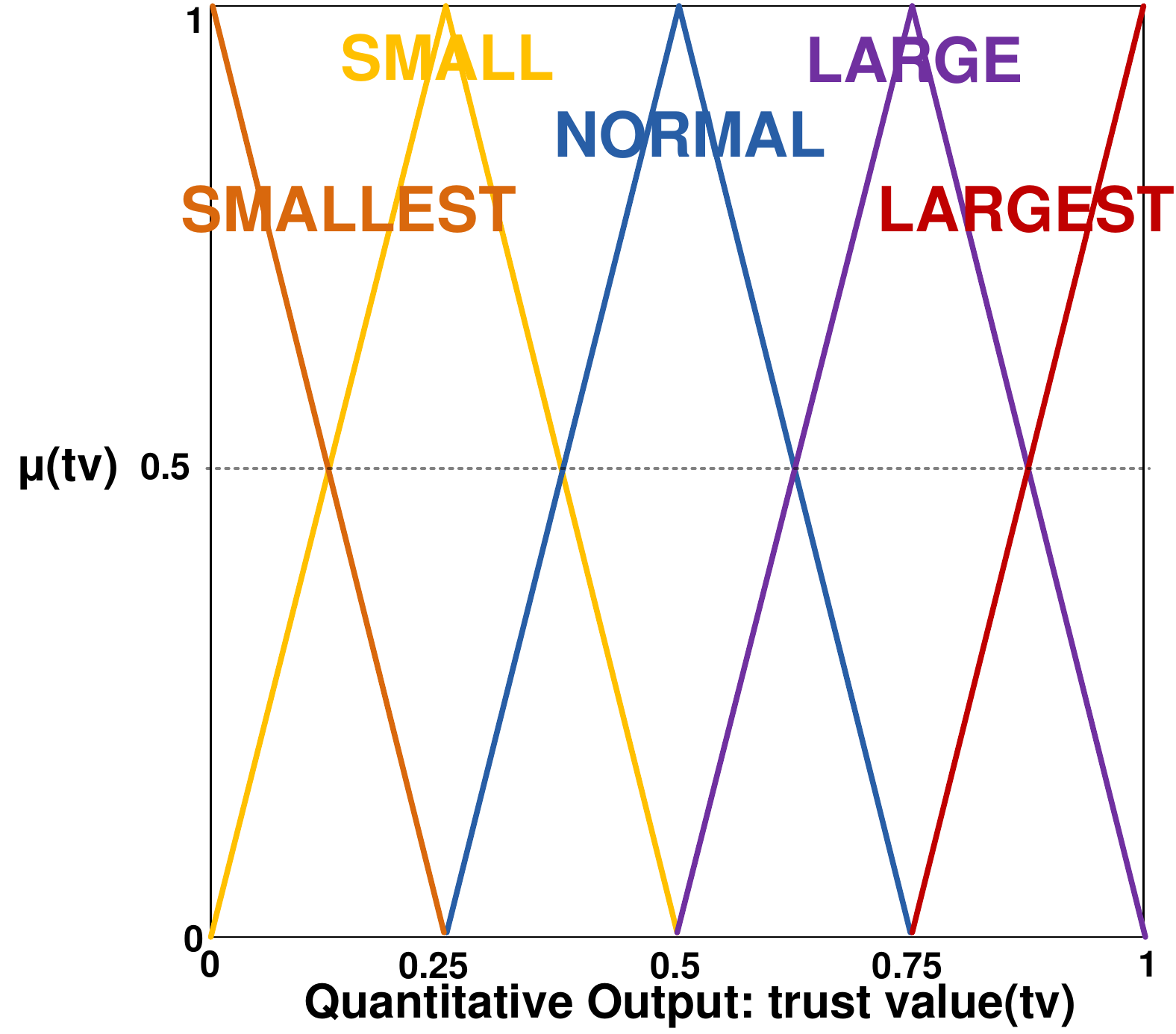}}
     \subfigure[The Input Independent Input Fuzzy Set.]{
           \label{fig:sainput}
          \includegraphics[width=.238\textwidth]{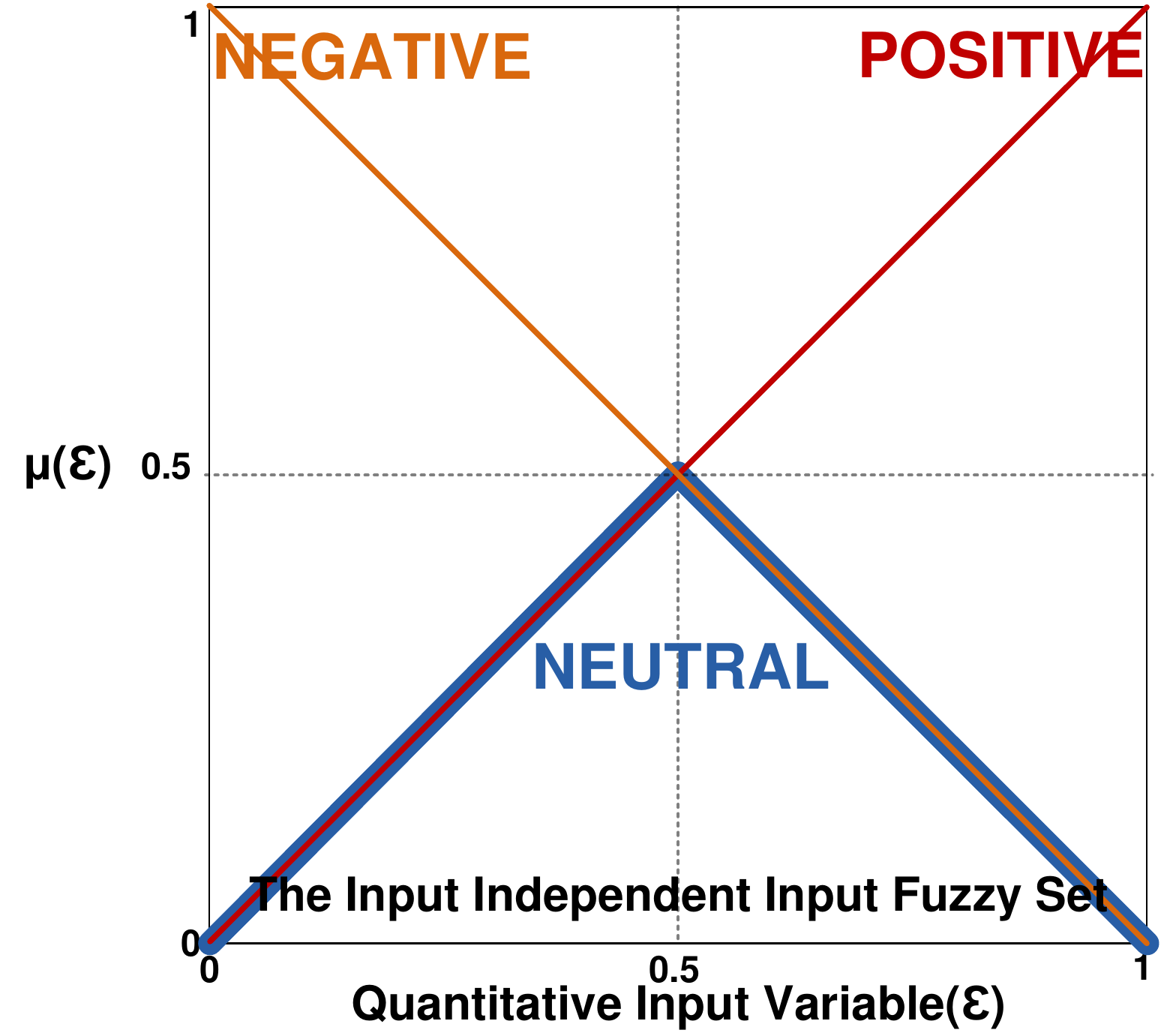}}
     \subfigure[The Input Fuzzy Set where the attribute is $\mathcal{A}_1=Major$.]{
           \label{fig:expertise}
          \includegraphics[width=.238\textwidth]{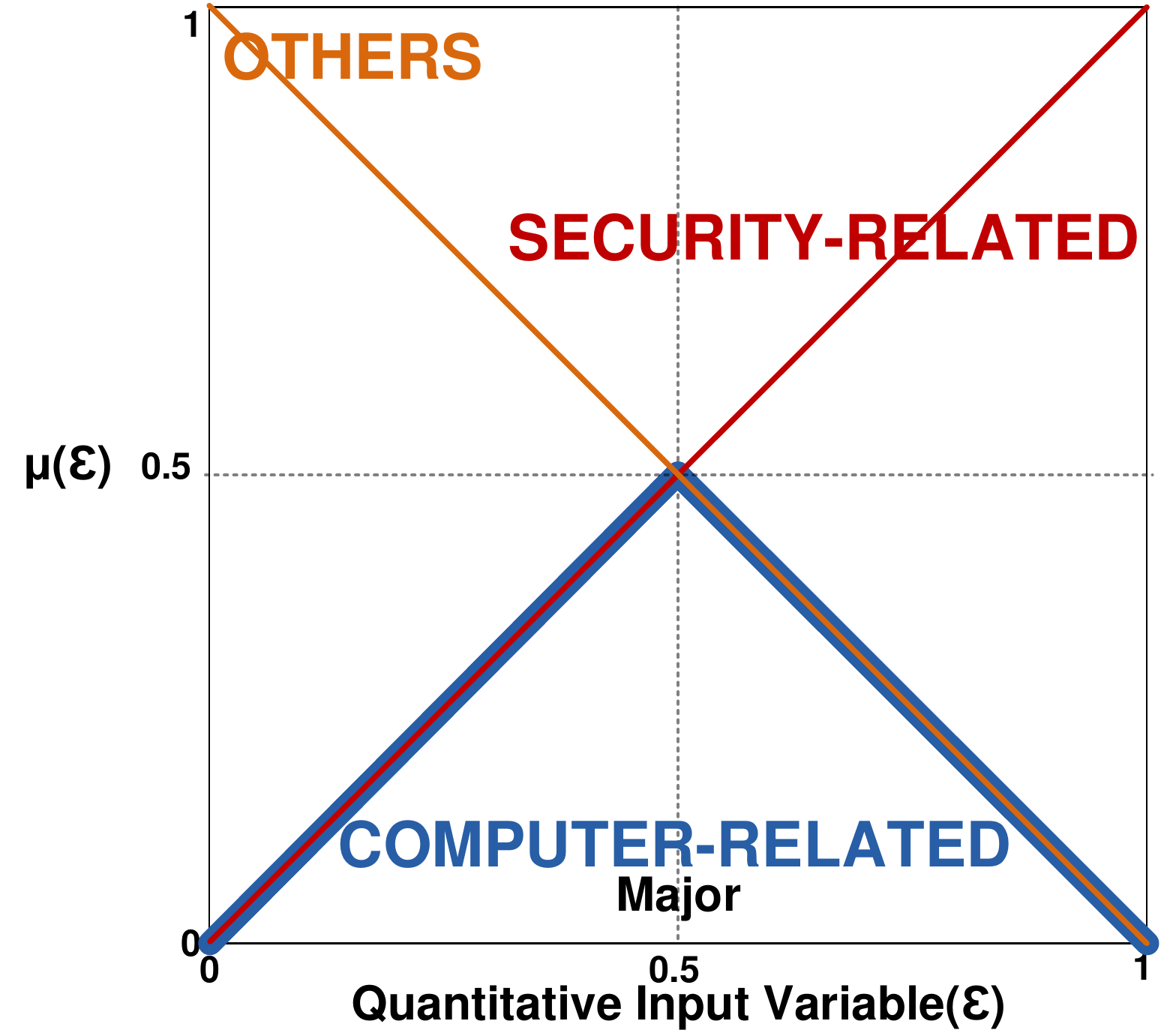}}
     \subfigure[The Input Fuzzy Set where the attribute is $\mathcal{A}_2=Relationship$.]{
           \label{fig:relation}
          \includegraphics[width=.238\textwidth]{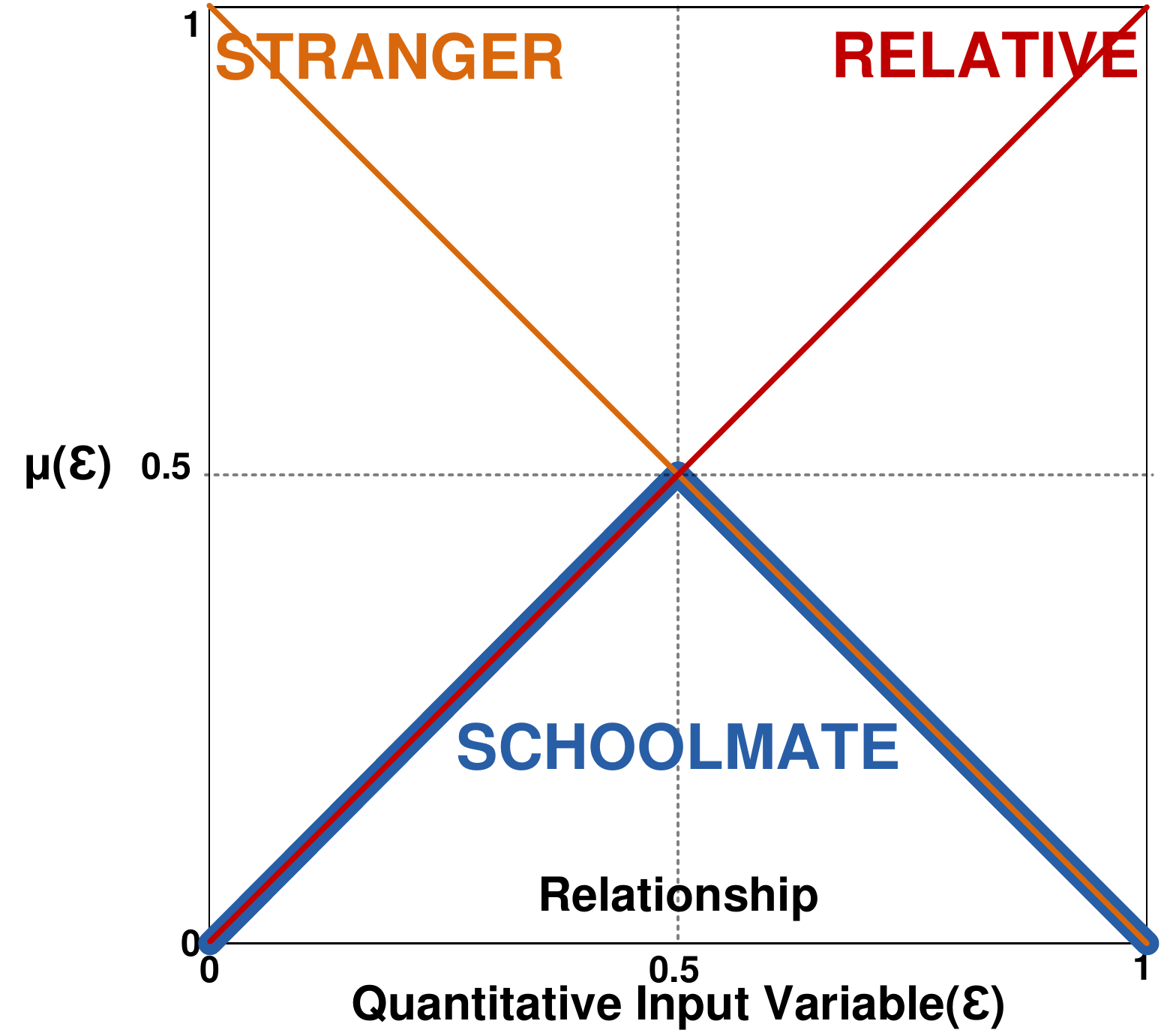}}
     \caption{Membership Functions Definition in Fuzzy Sets.}
     \label{fig:sa_input}
\end{figure*}

Fig.~\ref{fig:output} and \ref{fig:sainput} illustrate the membership functions of $\mathcal{O}$ and $\mathcal{I}$ in STor, respectively.

To meet the requirement of Eq.~(\ref{eqn:iequ}), $\rho_{q}$ can be set as:
\begin{eqnarray}
\label{eqn:rho}
\rho_{q}&=\left\{
\begin{array}{ll}
2, &q=1,\ q=5\\
1, &q=2,\ q=3,\ q=4.\\
\end{array}
\right.
\end{eqnarray}

Considering Eqs.~(\ref{eqn:single}-\ref{eqn:mtv}), Eq.~(\ref{eqn:singletvstar}), and Eqs.~(\ref{eqn:inputm}-\ref{eqn:rho}), for each qualitative social attribute $\mathcal{A}_{k=K}$ with each rule, $K.1.i$, $K.1.ii$, $K.2$, $K.3.i$ and $K.3.ii$, $\mathcal{MP}_{K.p_{\mathbf{m}}}(\mathcal{E})$, $\mathcal{M}_{K.p_{\mathbf{m}}}(\mathcal{E})$ and $tv_{k=K,p=p_{\mathbf{m}}}^*(\mathcal{E})$ can be calculated as follows. Appendix A details the calculation process.
\begin{eqnarray}
\begin{array}{l}
\mathcal{MP}_{K.1.i}(\mathcal{E})=-\frac{1}{48}(\mathcal{E}^3+9\mathcal{E}^2-21\mathcal{E}),\ \mathcal{E}\in[0,1],\\
\mathcal{MP}_{K.1.ii}(\mathcal{E})=-\frac{3}{16}(\mathcal{E}^2-2\mathcal{E}),\ \ \ \ \ \ \ \ \ \ \mathcal{E}\in[0,1],\\
\left\{
\begin{array}{l}
\mathcal{MP}_{K.2}(\mathcal{E})=-\frac{1}{8}(\mathcal{E}^2-2\mathcal{E}),\ \ \ \ \ \ \ \ \ \ \mathcal{E}\in[0,0.5],\\
\mathcal{MP}_{K.2}(\mathcal{E})=-\frac{1}{8}(\mathcal{E}^2-1),\ \ \ \ \ \ \ \ \ \ \ \ \mathcal{E}\in[0.5,1],
\end{array}
\right.\\
\mathcal{MP}_{K.3.i}(\mathcal{E})=-\frac{1}{16}(\mathcal{E}^2-1),\ \ \ \ \ \ \ \ \ \ \ \ \ \mathcal{E}\in[0,1],\\
\mathcal{MP}_{K.3.ii}(\mathcal{E})=-\frac{1}{48}(\mathcal{E}^3-1),\ \ \ \ \ \ \ \ \ \ \ \ \mathcal{E}\in[0,1],\\
\end{array}
\end{eqnarray}

\begin{eqnarray}
\begin{array}{l}
\mathcal{M}_{K.1.i}(\mathcal{E})=-\frac{1}{4}(\mathcal{E}^2-2\mathcal{E}),\ \ \ \mathcal{E}\in[0,1],\\
\mathcal{M}_{K.1.ii}(\mathcal{E})=-\frac{1}{4}(\mathcal{E}^2-2\mathcal{E}),\ \ \mathcal{E}\in[0,1],\\
\left\{
\begin{array}{l}
\mathcal{M}_{K.2}(\mathcal{E})=-\frac{1}{4}(\mathcal{E}^2-2\mathcal{E}),\ \mathcal{E}\in[0,0.5],\\
\mathcal{M}_{K.2}(\mathcal{E})=-\frac{1}{4}(\mathcal{E}^2-1),\ \ \ \mathcal{E}\in[0.5,1],
\end{array}
\right.\\
\mathcal{M}_{K.3.i}(\mathcal{E})=-\frac{1}{4}(\mathcal{E}^2-1),\ \ \ \ \ \mathcal{E}\in[0,1],\\
\mathcal{M}_{K.3.ii}(\mathcal{E})=-\frac{1}{4}(\mathcal{E}^2-1),\ \ \ \ \mathcal{E}\in[0,1].\\
\end{array}
\end{eqnarray}

\begin{eqnarray}
\begin{array}{l}
tv^*_{K.1.i}(\mathcal{E})=\frac{\mathcal{E}^2+9\mathcal{E}-21}{12(\mathcal{E}-2)},\ \mathcal{E}\in[0,1],\\
tv^*_{K.1.ii}(\mathcal{E})=\frac{3}{4},\ \ \ \ \ \ \ \ \ \ \mathcal{E}\in[0,1],\\
tv^*_{K.2}(\mathcal{E})=\frac{1}{2},\ \ \ \ \ \ \ \ \ \ \ \ \mathcal{E}\in[0,1],\\
tv^*_{K.3.i}(\mathcal{E})=\frac{1}{4},\ \ \ \ \ \ \ \ \ \ \ \mathcal{E}\in[0,1],\\
tv^*_{K.3.ii}(\mathcal{E})=\frac{\mathcal{E}^2+\mathcal{E}+1}{12(\mathcal{E}+1)},\ \ \ \mathcal{E}\in[0,1].\\
\end{array}
\end{eqnarray}

\begin{figure*}[ht!]
     \centering
     \subfigure[$tv^*(\mathcal{E})$ for rules $K.1.i$ and $K.1.ii$.]{
           \label{fig:k1tv}
          \includegraphics[width=.238\textwidth]{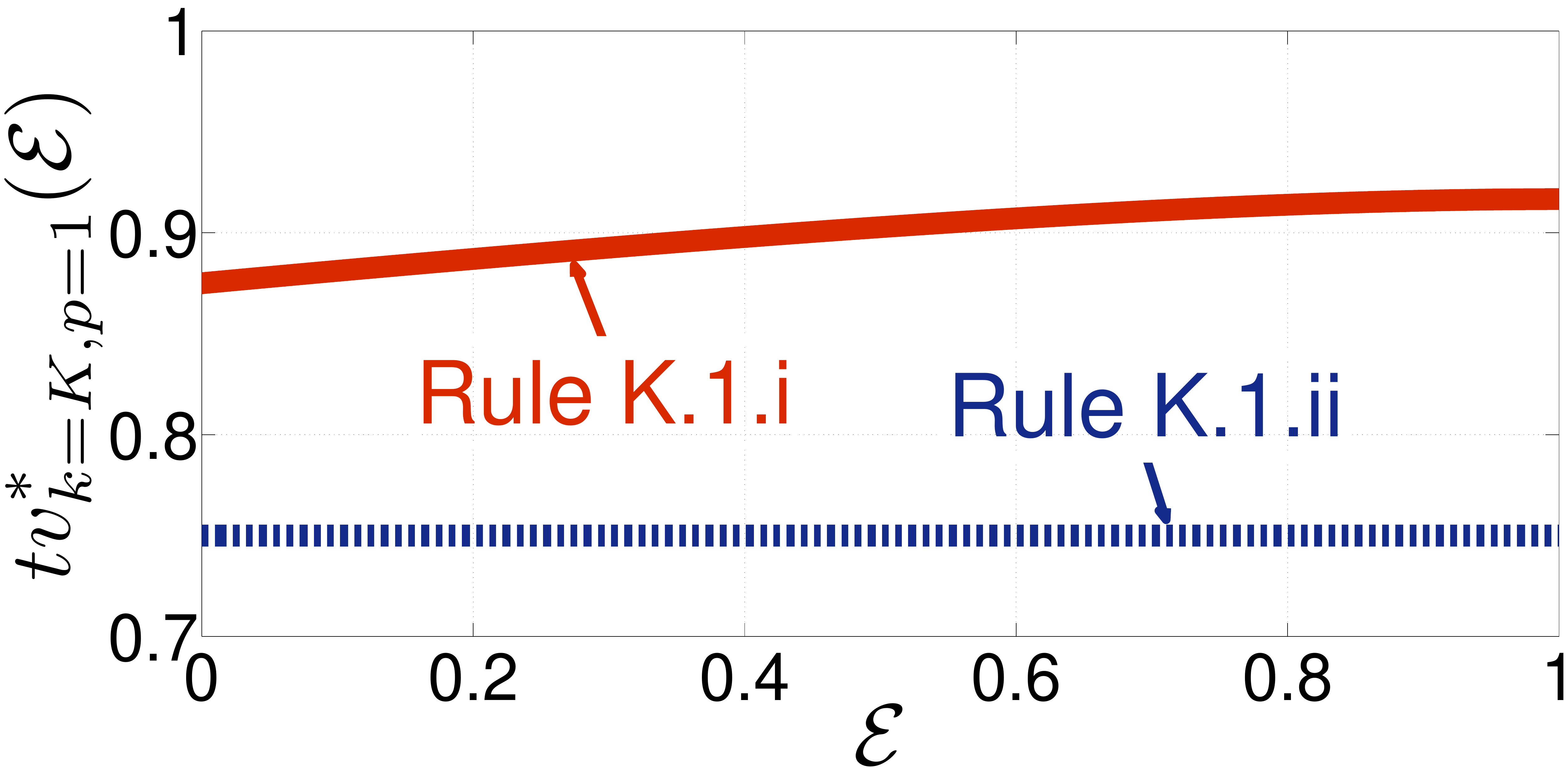}}
     \subfigure[$\mathcal{MP}(\mathcal{E})$ for rules $K.1.i$ and $K.1.ii$.]{
           \label{fig:k1mp}
          \includegraphics[width=.238\textwidth]{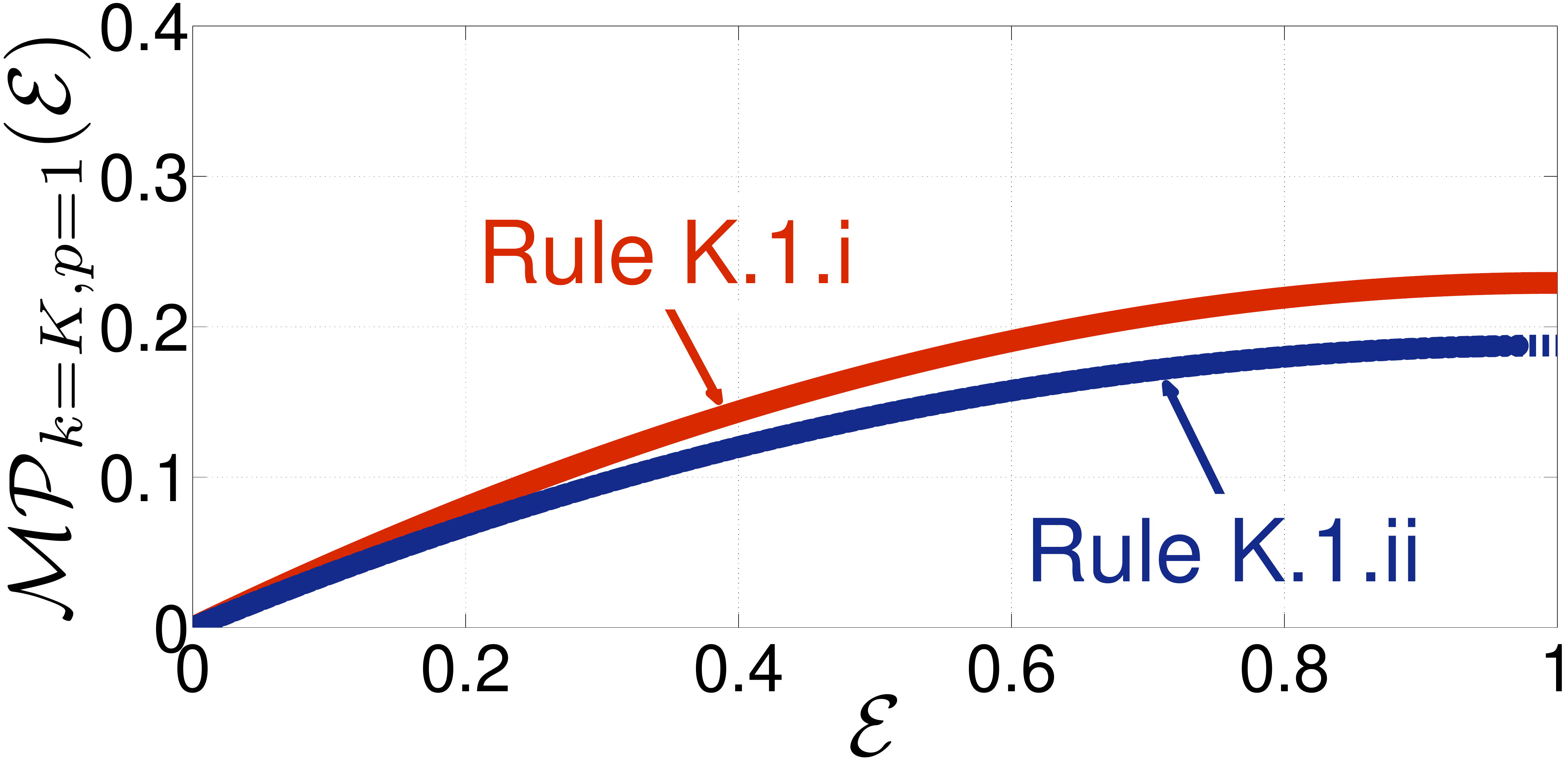}}
     \subfigure[$tv^*(\mathcal{E})$ for rules $K.3.i$ and $K.3.ii$.]{
           \label{fig:k3tv}
          \includegraphics[width=.238\textwidth]{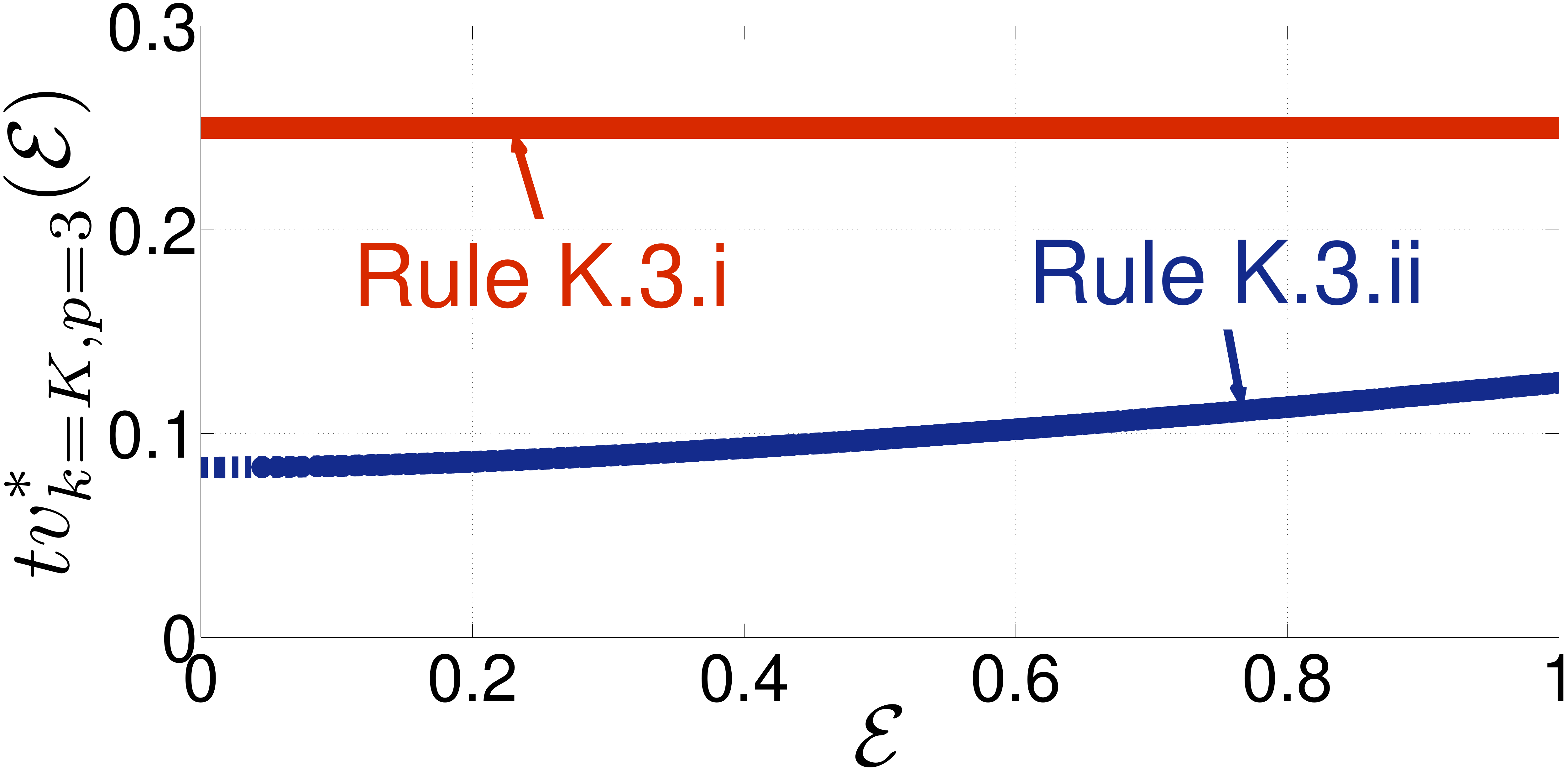}}
     \subfigure[$\mathcal{MP}(\mathcal{E})$ for rules $K.3.i$ and $K.3.ii$.]{
           \label{fig:k3mp}
          \includegraphics[width=.238\textwidth]{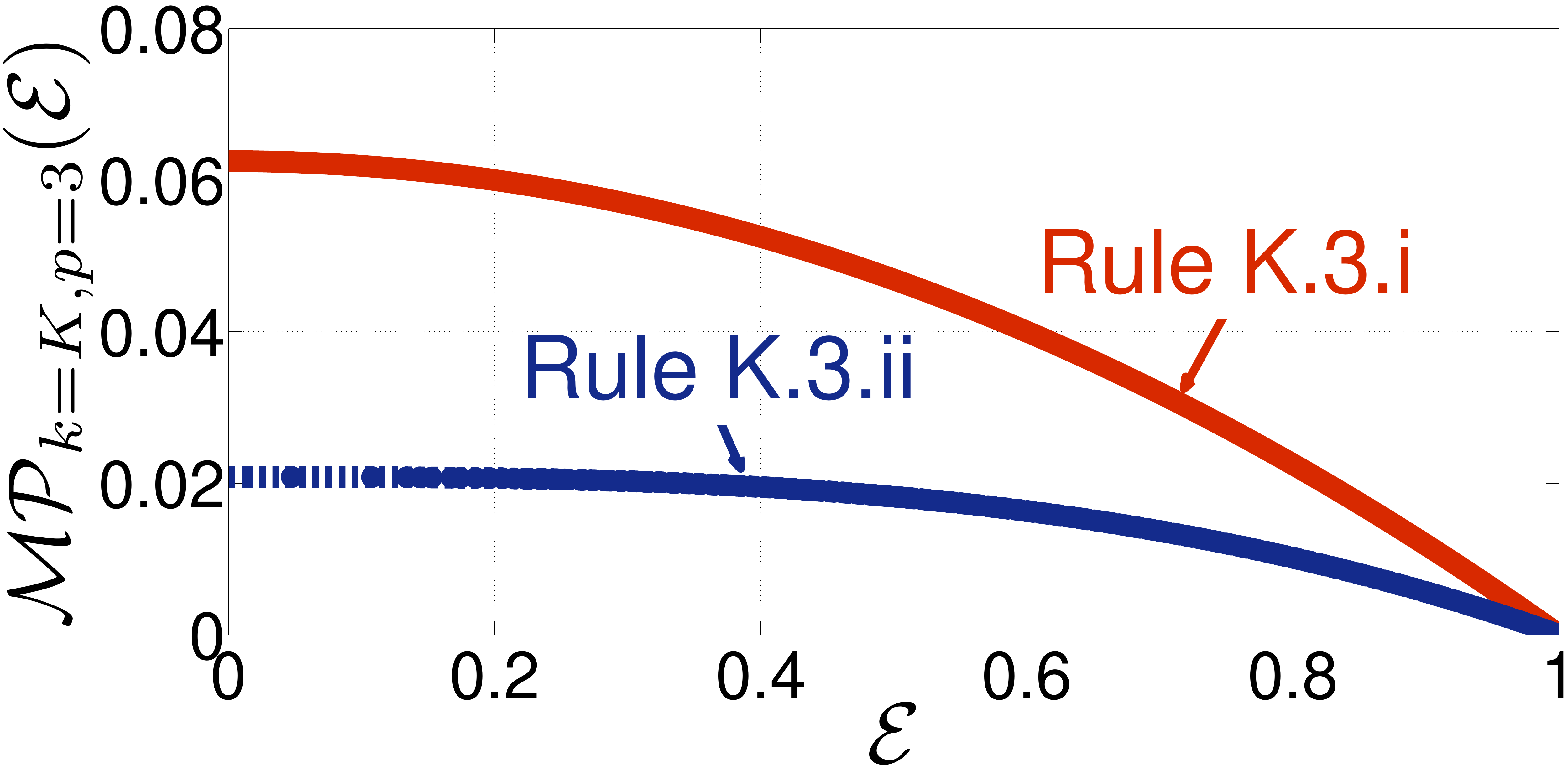}}
     \caption{$\mathcal{MP}(\mathcal{E})$ and $tv^*(\mathcal{E})$ for a single social attribute $\mathcal{A}_{k=K}$ with rules $K.1$ and $K.3$.}
     \label{fig:mptv}
\end{figure*}

Rule $K.1$ has two candidate rules, $K.1.i$ and $K.1.ii$, to map each $\mathcal{I}_{1}^{\mathcal{A}_{k}}=POSITIVE^k$ to $\mathcal{O}_{1}=LARGEST$ and $\mathcal{O}_{2}=LARGE$, respectively. These two rules reflect different levels of a large trust that $(POSITIVE)^k$ can result in. As shown in Figs.~\ref{fig:k1tv}-\ref{fig:k1mp}, rule $K.1.i$ has the ability to convert any of the $\mathcal{E}$ to a higher $tv^*(\mathcal{E})$ and $\mathcal{MP}(\mathcal{E})$ than rule $K.1.ii$. That is, candidate rule $K.1.i$ can be used as rule $K.1$ when the value $POSITIVE^K$ of qualitative social attribute $\mathcal{A}_{k=K}$ has a stronger possibility of leading to a larger trust value.

Similarly, rule $K.3$ uses $K.3.i$ and $K.3.ii$, which indicate $(NEGATIVE)^k$ can lead to different degrees of small trust. Fig.~\ref{fig:k3tv} and \ref{fig:k3mp} demonstrate that the rule $K.3.ii$ can map all the $\mathcal{E}$ to smaller $tv^*(\mathcal{E})$ and $\mathcal{MP}(\mathcal{E})$ values than the rule $K.3.i$. Therefore, there is a tendency to set rule $K.3$ as $K.3.ii$ if the value $NEGATIVE^K$ of qualitative social attribute $\mathcal{A}_{k=K}$ has a stronger possibility of leading to a smaller trust value.

STor is capable of including an unlimited number of qualitative and quantitative social attributes in the conversion of the trust value.
Table \ref{table:sas} lists $10$ possible qualitative social attributes that come from popular social networks on the Internet, including Facebook, LinkedIn, MSN, and QQ. It also details which qualitative values taken from them can be classified into particular $\mathcal{I}_{p}^{\mathcal{A}_{k}}$. More qualitative social attributes can be included in STor even if they are not listed in the table.

\begin{table*}[ht!]
  \centering
  \caption{\footnotesize Qualitative social attributes from existing social networks.}
  \renewcommand{\arraystretch}{1.2}
  \label{table:sas}
  \scriptsize
\begin{tabular}{cc|ccc|c}
  \toprule
  \textbf{$k$} &\textbf{Qualitative Attribute} ($\mathcal{A}_k$) &$\mathcal{I}_{1}^{\mathcal{A}_{k}}=(POSITIVE)^{k}$ &$\mathcal{I}_{2}^{\mathcal{A}_{k}}=(NEUTRAL)^{k}$ &$\mathcal{I}_{3}^{\mathcal{A}_{k}}=(NEGATIVE)^{k}$ &\textbf{Source Social Network}\\
  \midrule
  1 &Major &SECURITY-RELATED &COMPUTER-RELATED &OTHERS &Facebook, LinkedIn\\
  2 &Relationship &RELATIVE &SCHOOLMATE &STRANGER &Facebook, LinkedIn, MSN\\
  3 &Career &SECURITY-RELATED &COMPUTER-RELATED &OTHERS &Facebook, LinkedIn\\
  4 &Recommendation &POSITIVE &NONE &NEGATIVE &LinkedIn\\
  5 &Citizenship &FELLOW-CITIZEN &NEUTRAL-CITIZEN &ENEMY-CITIZEN &Facebook, MSN, QQ\\
  6 &Geolocation &SAME-CITY &SAME-COUNTRY &DIFFERENT-COUNTRY &Facebook, QQ\\
  7 &Religion &SAME RELIGION &NO RELIGION &DIFFERENT RELIGION &Facebook\\
  8 &Political View &SAME-VIEW &NEUTRAL-VIEW &DIFFERENT-VIEW &Facebook\\
  9 &Hometown &SAME-CITY &SAME-COUNTRY &DIFFERENT-COUNTRY &Facebook\\
  10 &Position &SAME-COMPANY &NEUTRAL-COMPANY &RIVAL-COMPANY &Facebook,LinkedIn\\
  ... &... &... &... &... &...\\
  \bottomrule
\end{tabular}
\end{table*}

\subsection{Example of trust value calculation}
\label{subsec:exam}
In this example, only $E_{1}=freq$ (i.e., the communication frequency between two friends) and $E_{2}=time$ (i.e., the duration of their friendship) are considered as quantitative social attributes. Assuming that the same weight, $\pi_{1}=\pi_{2}=0.5$, is assigned to each of them, then $\mathcal{E}$ between entities $i$ and $j$ in social network $\mathcal{S}_s$ can be calculated as $\mathcal{E}_{i\rightarrow j}^s=({\overline{\mathcal{E}}_{i\rightarrow j}^s(freq)+\overline{\mathcal{E}}_{i\rightarrow j}^s(time)})/{2}$ according to Eqn.~(\ref{eqn:quantitative}). Here, $\overline{\mathcal{E}}_{i\rightarrow j}^s(freq)={\mathcal{E}_{i\rightarrow j}^s(freq)}/{\mathop{max}\limits_{p\in F_{i}^s}(\mathcal{E}_{i\rightarrow p}^s(freq))}$, $\overline{\mathcal{E}}_{i\rightarrow j}^s(time)={\mathcal{E}_{i\rightarrow j}^s(time)}/{\mathop{max}\limits_{p\in F_{i}^s}(\mathcal{E}_{i\rightarrow p}^s(time))}$.

In social network $\mathcal{S}_s$, only qualitative social attributes $\mathcal{A}_1=Major$ and $\mathcal{A}_2=Relationship$ are considered. Figs.~\ref{fig:expertise}-\ref{fig:relation} illustrate the input fuzzy sets with their membership functions for these two attributes. Their fuzzy rules are defined below:
\begin{center}
\begin{description}
  \item[Rule $1.1$]
    \begin{tabular}{l}
        IF Major belongs to SECURITY-RELATED,\\
        THEN Trust Value is LARGE.
    \end{tabular}
  \item[Rule $1.2$]
    \begin{tabular}{l}
        IF Major belongs to COMPUTER-RELATED,\\
        THEN Trust Value is NORMAL.
    \end{tabular}
  \item[Rule $1.3$]
    \begin{tabular}{l}
        IF Major belongs to OTHERS,\\
        THEN Trust Value is SMALL.
    \end{tabular}
\end{description}
\end{center}

\begin{center}
\begin{description}
  \item[Rule $2.1$]
    \begin{tabular}{l}
        IF Relationship belongs to RELATIVE,\\
        THEN Trust Value is LARGEST.
    \end{tabular}
  \item[Rule $2.2$]
    \begin{tabular}{l}
        IF Relationship belongs to SCHOOLMATE,\\
        THEN Trust Value is NORMAL.
    \end{tabular}
  \item[Rule $2.3$]
    \begin{tabular}{l}
        IF Relationship belongs to STRANGER,\\
        THEN Trust Value is SMALLEST.
    \end{tabular}
\end{description}
\end{center}
In this definition, rules $1.1$ and $1.3$ adopt respective candidate rules $k.1.ii$ and $k.3.i$, while rule $2.1$ and $2.3$ implement respective candidates $k.1.i$ and $k.3.ii$. This is due to social attribute $\mathcal{A}_2=Relationship$ possessing a stronger ability to affect the trust value.

\begin{figure*}[ht!]
     \centering
     \subfigure[\footnotesize $\mathcal{A}_1$=Major, $\mathcal{I}_{1}^{\mathcal{A}_{1}}$=SECURITY-RELATED, $\mathcal{A}_2$=Relationship, $\mathcal{I}_{1}^{\mathcal{A}_{2}}$=RELATIVE.]{
           \label{fig:df_social_1}
          \includegraphics[width=.4\textwidth]{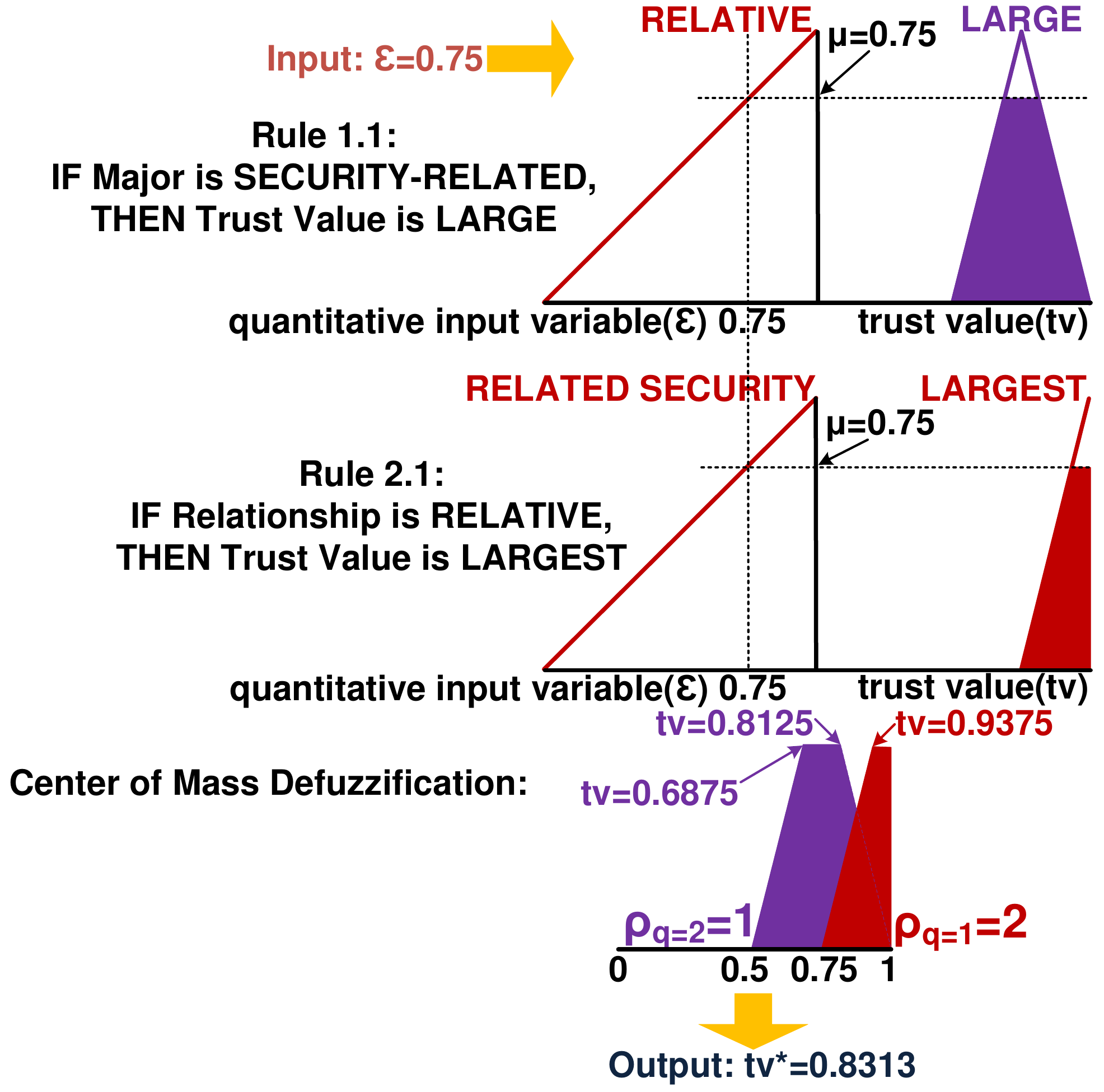}}
  \hspace{4ex}
     \subfigure[\footnotesize $\mathcal{A}_1$=Major, $\mathcal{I}_{2}^{\mathcal{A}_{1}}$=COMPUTER-RELATED, $\mathcal{A}_2$=Relationship, $\mathcal{I}_{3}^{\mathcal{A}_{2}}$=STRANGER.]{
           \label{fig:df_social_2}
          \includegraphics[width=.4\textwidth]{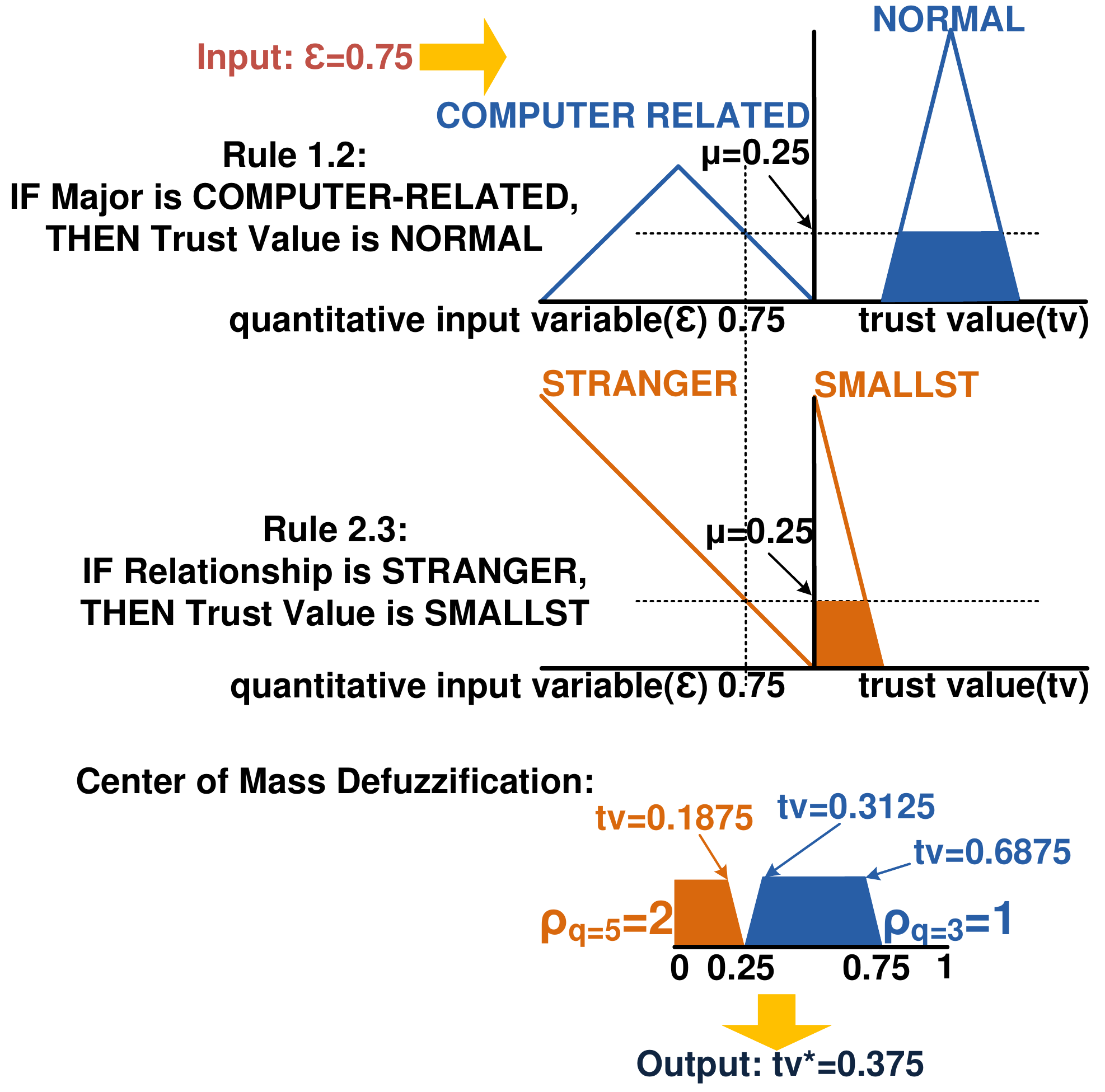}}
     \caption{Fuzzification and Defuzzification process in the input independent fuzzy model.}
     \label{fig:sa_input}
\end{figure*}

Figs.~\ref{fig:df_social_1}-\ref{fig:df_social_2} show the process of calculating the trust value through the input independent fuzzy model based on both qualitative and quantitative social attributes. Consider entity $i$'s two friends, entities $j_1$ and $j_2$, entity $j_1$ is a relative and majors in network security while entity $j_2$ is a stranger and majors in computer science. If both of them have the same $\mathcal{E}_{i\rightarrow j_1}^s=\mathcal{E}_{i\rightarrow j_2}^s=0.75$ with respect to entity $i$, the relative (i.e., $j_1$) will achieve a trust value of $tv_{i\rightarrow j_1}^{s}(0.75)=0.8313$ while the stranger (i.e., $j_2$) will only receive a trust value of $tv_{i\rightarrow j_2}^{s}(0.75)=0.375$.

\section{Evaluation}
\label{sec:experiment}
In this section, we first adopt the Markov random graph to model the architecture of STor with $2,500$ entities and deploy a testing network of Tor using PlanetLab nodes \cite{planetlab} to implement STor network. As a consequence, we randomize the social attributes of each entity in STor and calculate the trust scores among them with the help of the input independent fuzzy model and the trust propagation algorithm. By conducting the evaluations in both simulation and experiments over PlanetLab nodes, We show STor's ability to obtain secure anonymity and the impact on performance and baseline anonymity. Moreover, a better scalability has been demonstrated in STor compared to Tor.

\subsection{Experiment Setup}
\label{sec:exsetup}

\subsubsection{Simulation}
\label{sec:sisetup}
As reported at the 5th Oct. 2011 in \cite{TorStat}, there are about $2,389$ public routers in the Tor network. As a result, we simulate STor with $2,500$ user entities who run up STor routers. Each entity can also act as a STor user to acquire anonymity service from its friends. Following the models and methods presented in \cite{socialnetwork}, we use the Markov random graph model (i.e., P* model with the Markov dependency) to represent the underlying friendship graph $\mathbb{G}$ in STor. In accordance with \cite{socialnetwork}, the friendship between two entities is established with an exponential family of distributions yielded by the Hammersley-Clifford theorem. Particularly, we follow the trilogy papers \cite{SP96,SP99,SPL99} to select parameters in these exponential family of distributions. In this simulation setup, we finally establish the graph $\mathbb{G}$ of STor with $2,500$ friendship circles $F_i$, where, $1974\leq||F_i||\leq 1981$. Beside that, Each entity is associated with a value ranging in ($0$,$10$MB] as its bandwidth of the Tor or STor router.

\subsubsection{PlanetLab Platform}
\label{sec:plsetup}

\begin{figure*}[htbp!]
     \centering
     \subfigure[Major is SECURITY-RELATED, Relationship is RELATIVE.]{
           \label{fig:11_21_tv}
          \includegraphics[width=.238\textwidth]{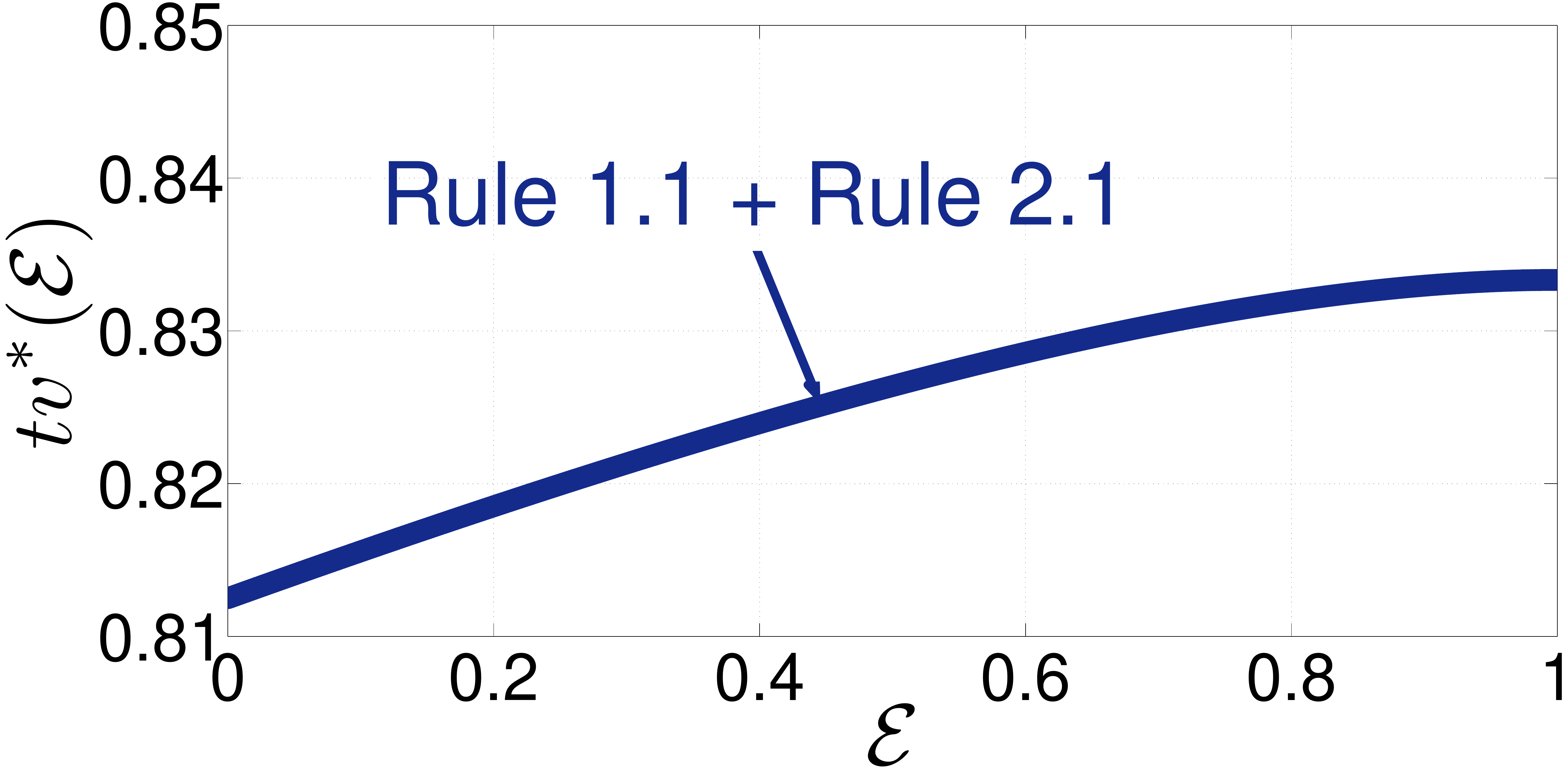}}
 \hspace{2ex}
 \subfigure[Major is SECURITY-RELATED, Relationship is SCHOOLMATE.]{
           \label{fig:11_22_tv}
          \includegraphics[width=.238\textwidth]{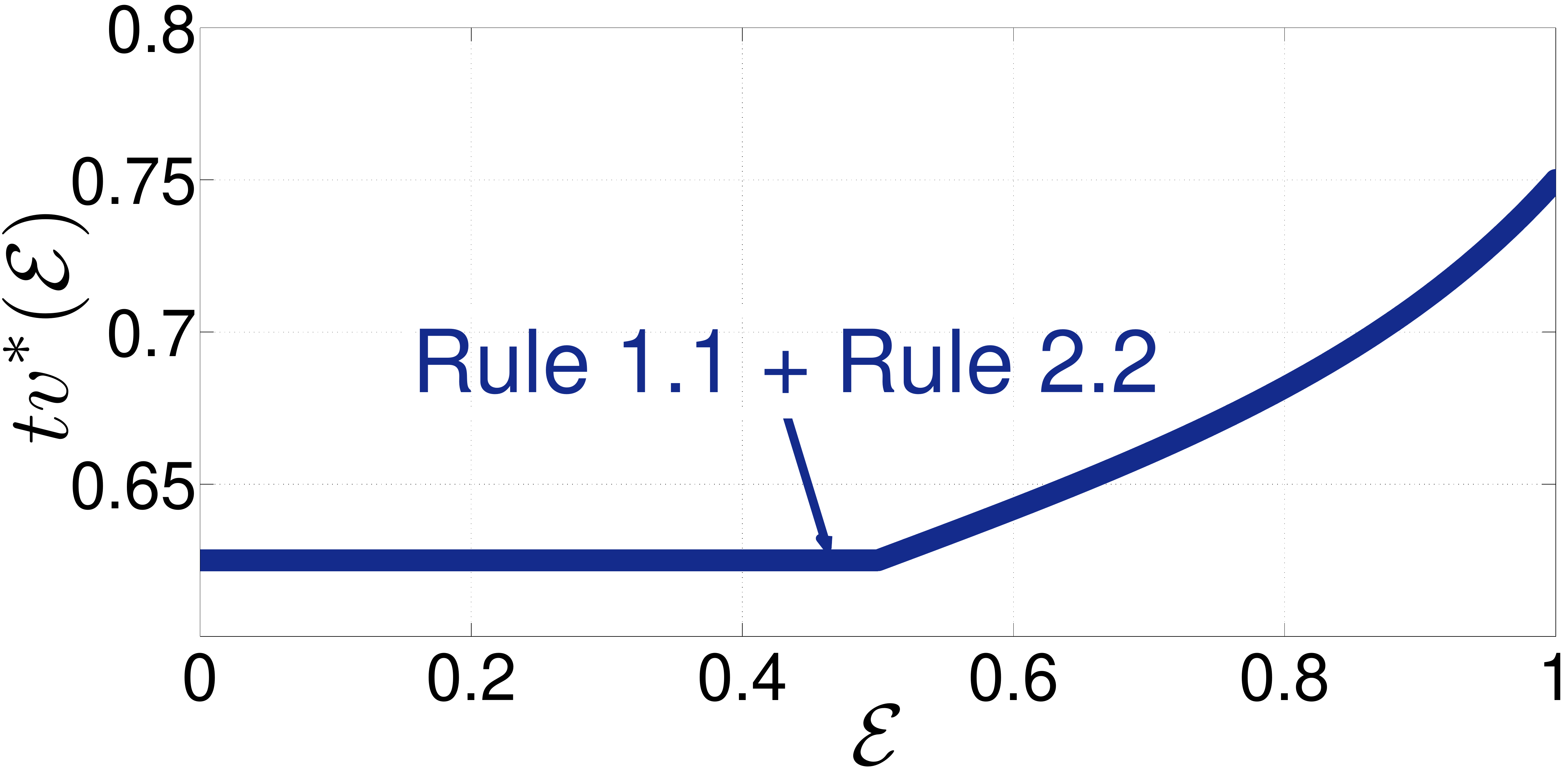}}
 \hspace{2ex}
 \subfigure[Major is SECURITY-RELATED, Relationship is STRANGER.]{
           \label{fig:11_23_tv}
          \includegraphics[width=.238\textwidth]{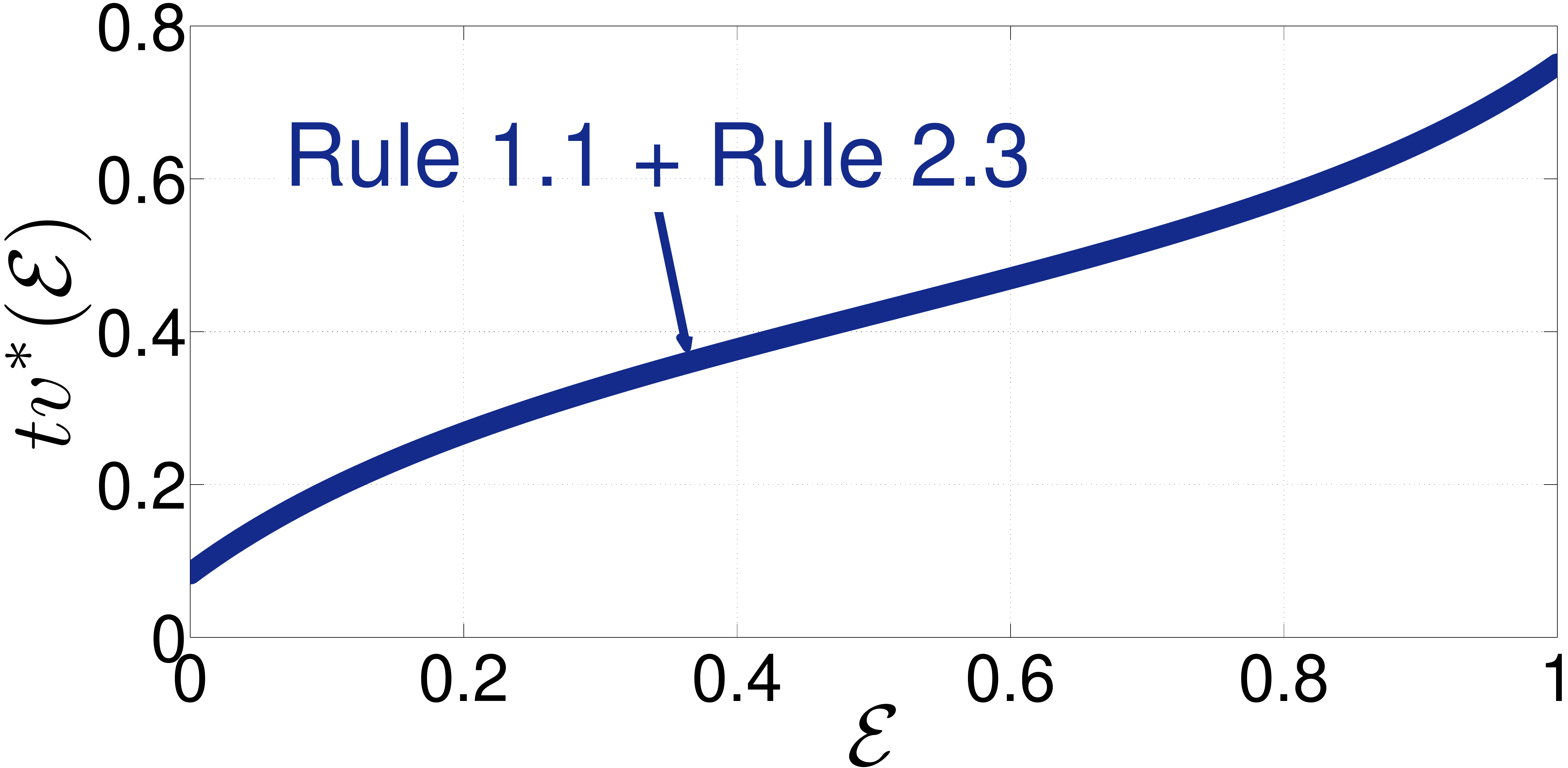}}
 \\
     \subfigure[Major is COMPUTER-RELATED, Relationship is RELATIVE.]{
           \label{fig:12_21_tv}
          \includegraphics[width=.238\textwidth]{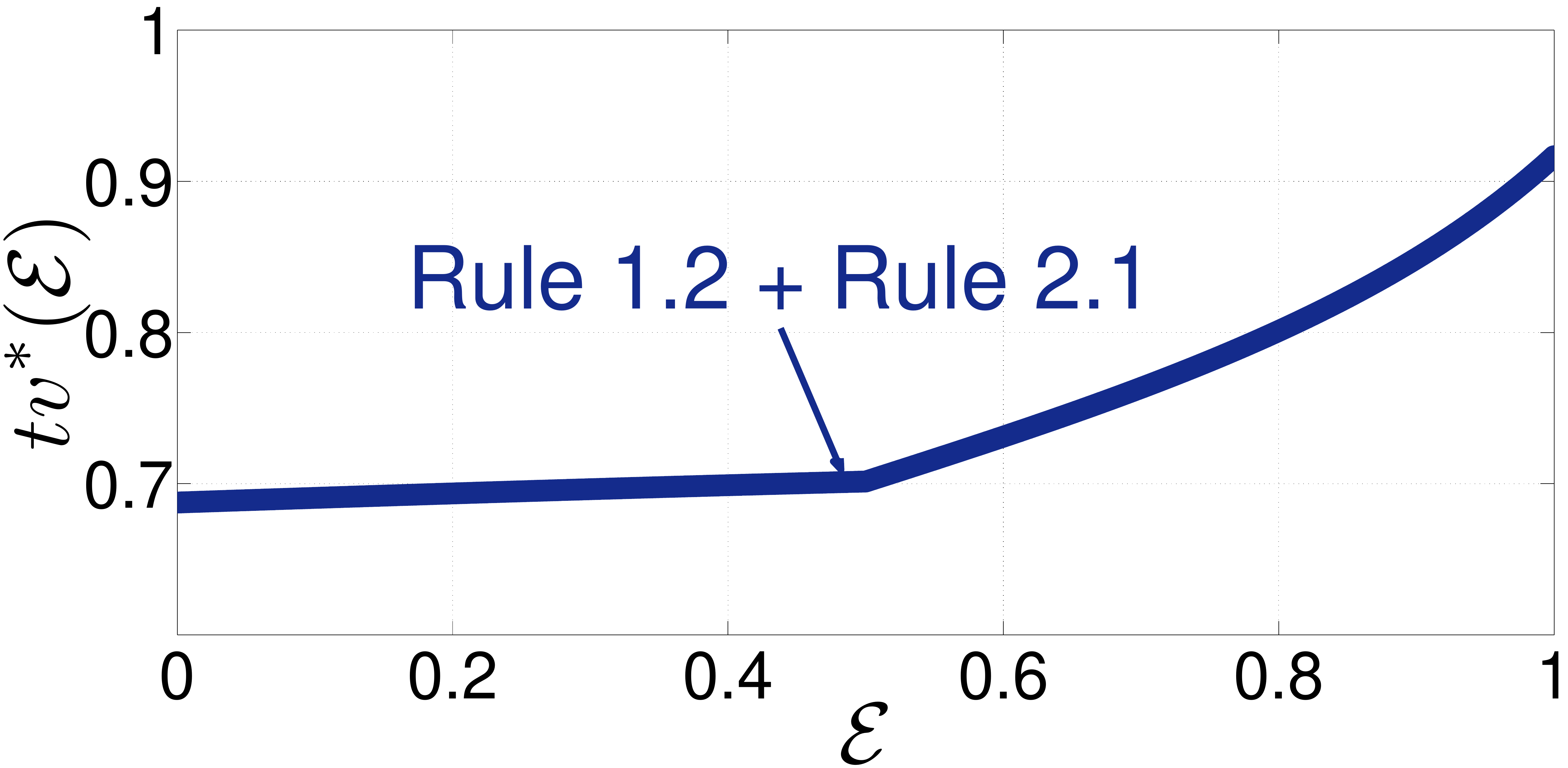}}
 \hspace{2ex}
     \subfigure[Major is COMPUTER-RELATED, Relationship is SCHOOLMATE.]{
           \label{fig:12_22_tv}
          \includegraphics[width=.238\textwidth]{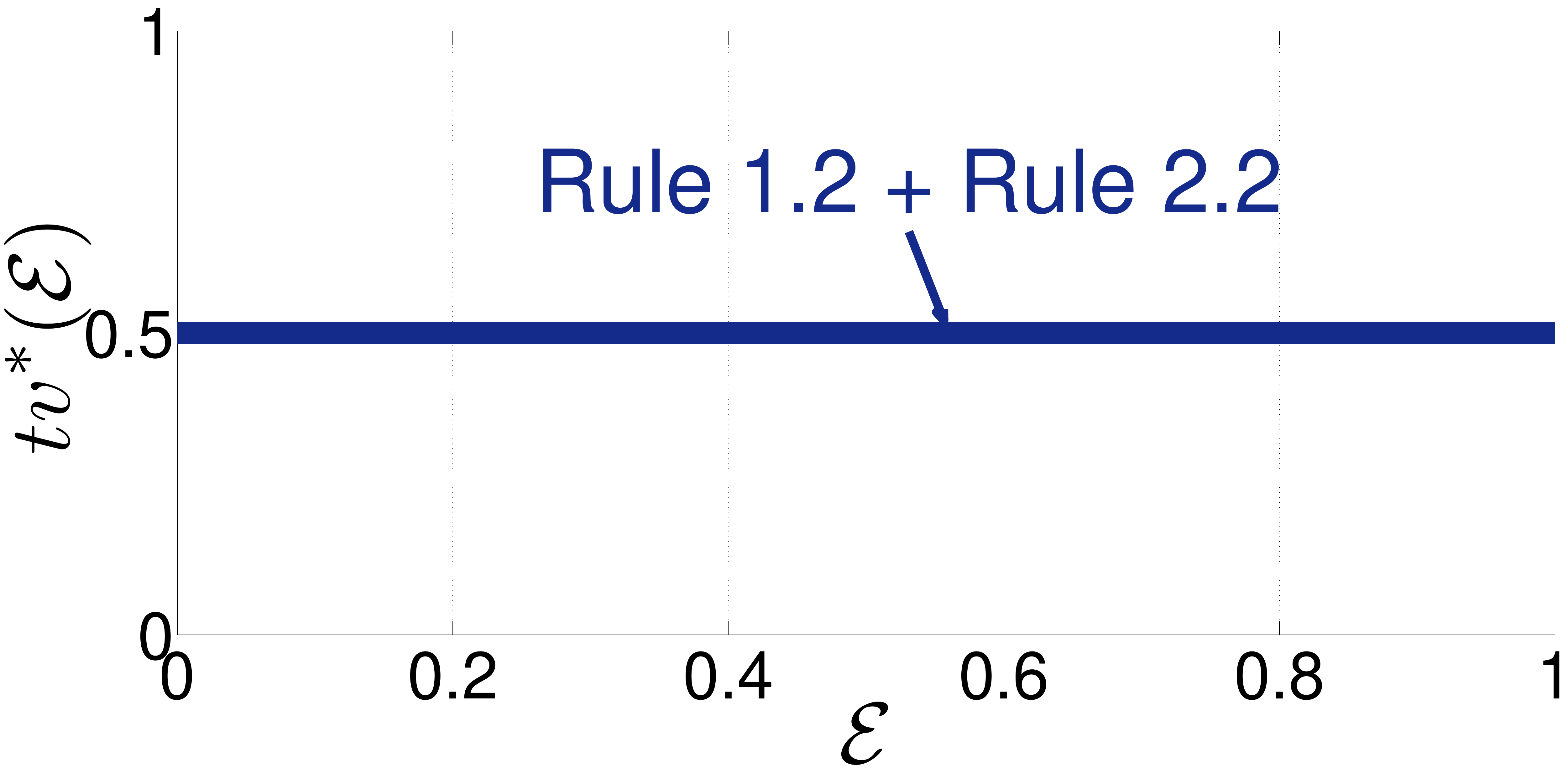}}
 \hspace{2ex}
     \subfigure[Major is COMPUTER-RELATED, Relationship is STRANGER.]{
           \label{fig:12_23_tv}
          \includegraphics[width=.238\textwidth]{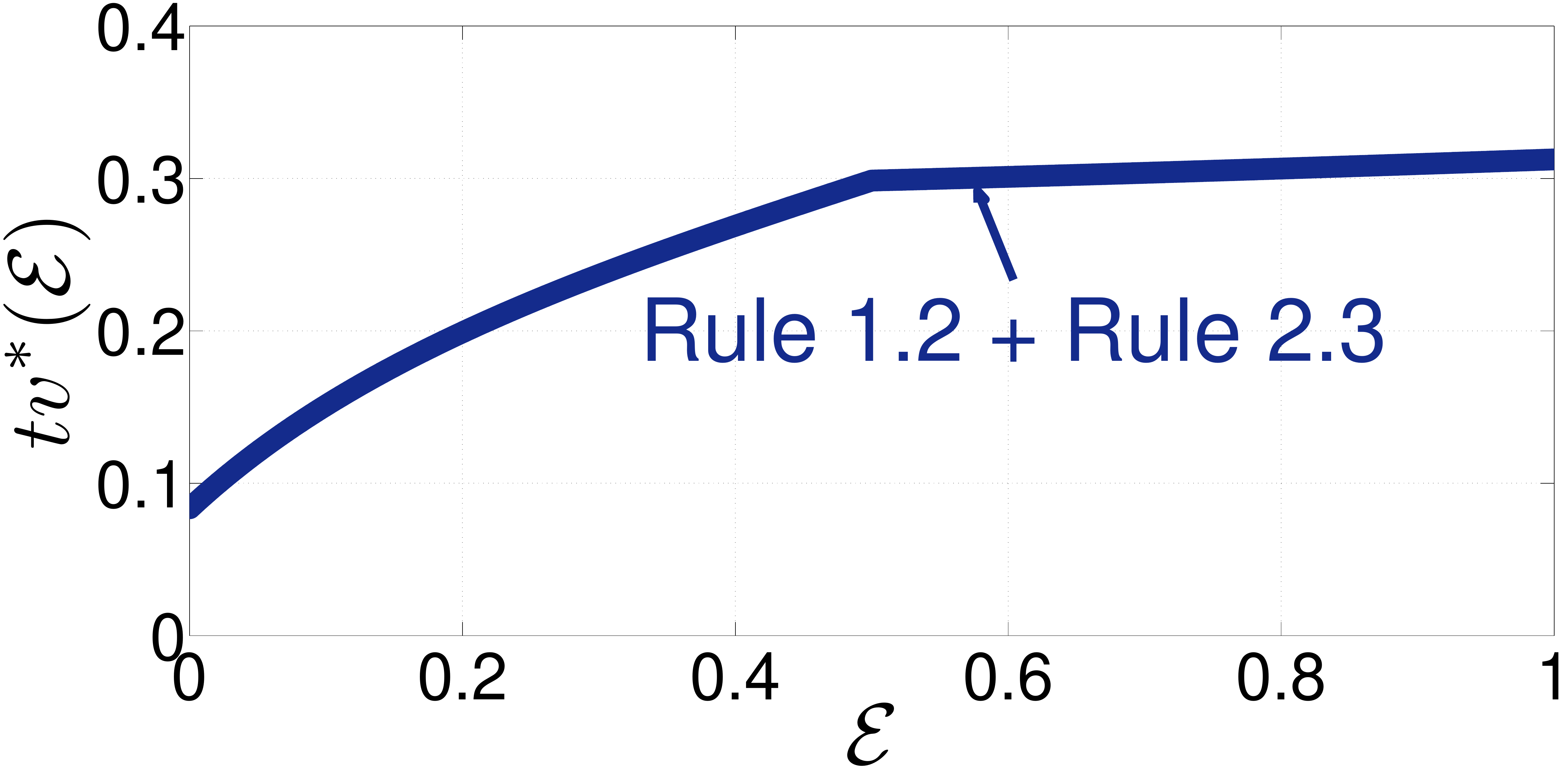}}
 \\
     \subfigure[Major is OTHERS, Relationship is RELATIVE.]{
           \label{fig:13_21_tv}
          \includegraphics[width=.238\textwidth]{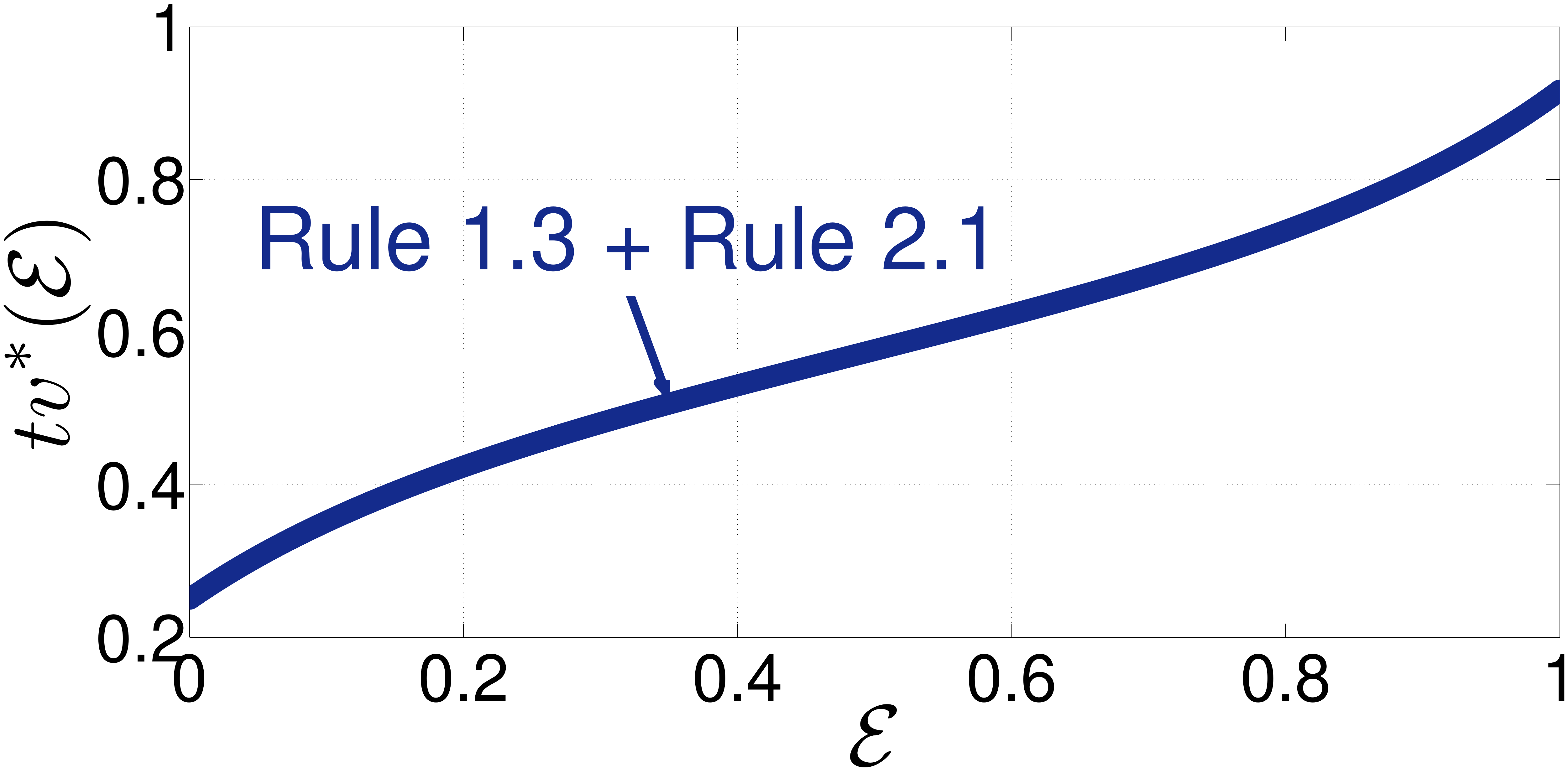}}
 \hspace{2ex}
     \subfigure[Major is OTHERS, Relationship is SCHOOLMATE.]{
           \label{fig:13_22_tv}
          \includegraphics[width=.238\textwidth]{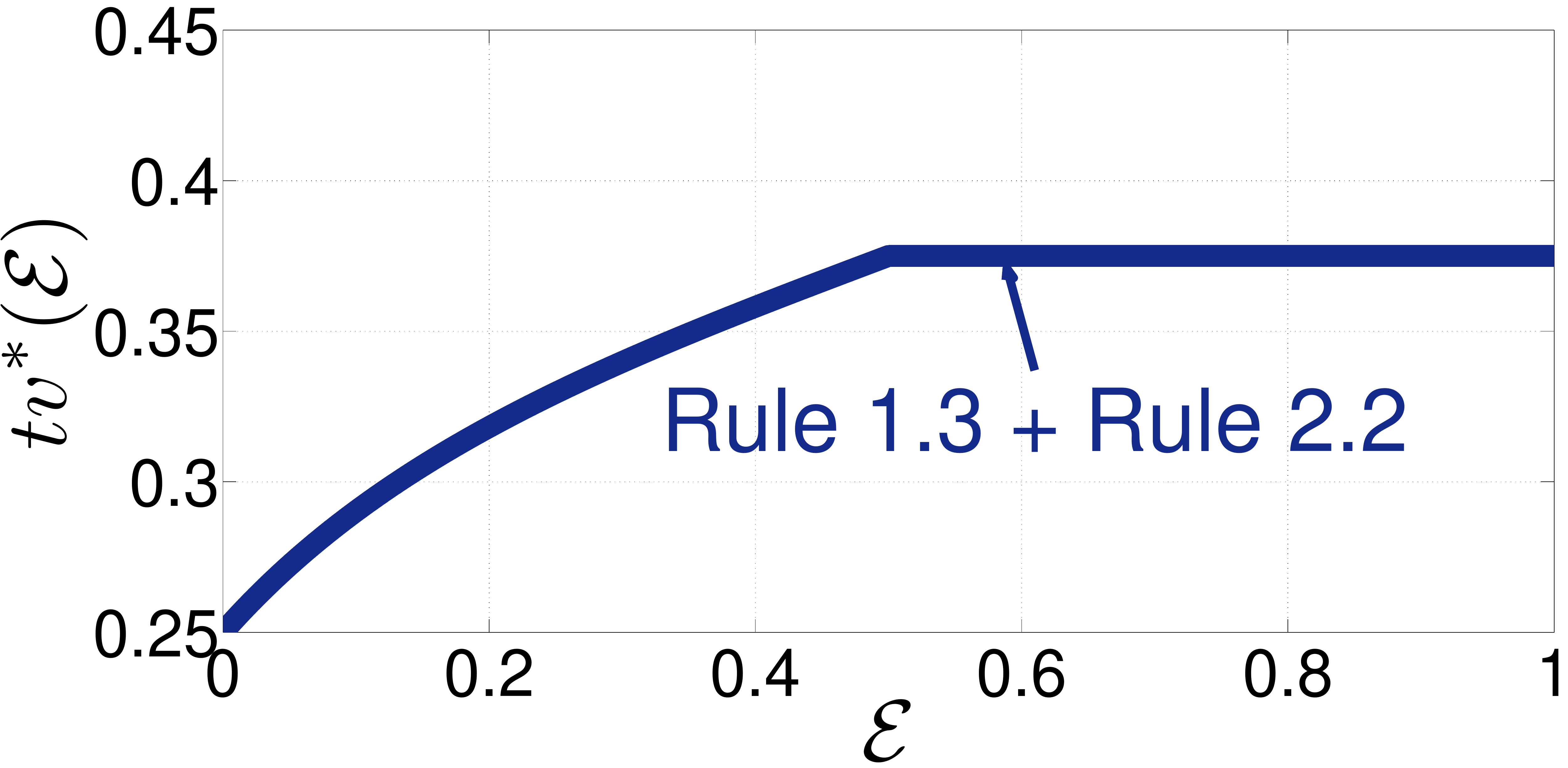}}
 \hspace{2ex}
     \subfigure[Major is OTHERS, Relationship is STRANGER.]{
           \label{fig:13_23_tv}
          \includegraphics[width=.238\textwidth]{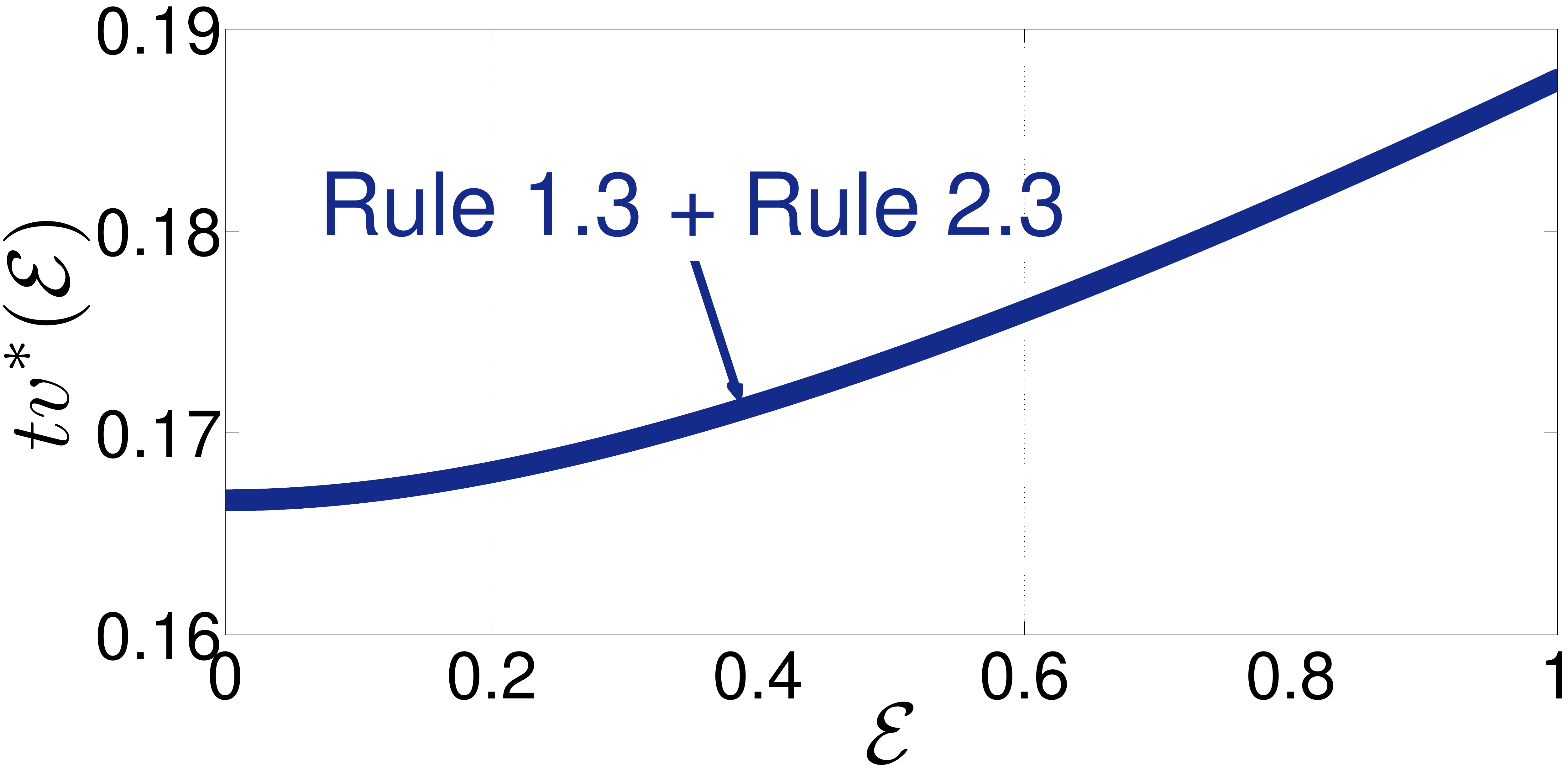}}
     \caption{Trust value for different Relationship and Major.}
     \label{fig:ls_tv}
\end{figure*}

To evaluate STor using the Internet, a private Tor network was set up in PlanetLab platform \cite{planetlab}. $101$ PlanetLab nodes were randomly chosen from around the globe and installed with the Tor (V0.2.1.26). A node in Japan (133.15.59.1) was chosen as the directory server and a node in Taiwan (140.109.17.181) was chosen as the an onion proxy. The remaining $99$ nodes were set up as onion routers, which reported bandwidths from $21.9$KB to $336.3$KB to the directory server. The configuration option ``TestingTorNetwork" was set to $1$ to allow the nodes to work as an independent private Tor network. In addition, the function $smartlist\_choose\_by\_bandwidth()$ in Tor's source code was modified to bypass the existing circuit selection algorithm and insert a new selection method in line with Eq.~(\ref{equ:pj}). We have disabled the uptime based policies for exit and entry nodes in the Tor as attackers are easy to bypass them by setting up malicious routers for a period of time to receive trust. As a consequence, we choose $100$ entities from the friendship graph $\mathbb{G}$, which is generated in Section \ref{sec:sisetup}. These entities include at least one friendship circle $F_i$ with $||F_i||=99$. Finally, we apply these entities to STor network in PlanetLab nodes.

%
\subsection{Turst Score Calculation}
\label{sec:fgm}
The input independent fuzzy model, defined in section \ref{subsec:safm}, is used to initialize the trust value in each directed link on the graph $\mathbb{G}$ that is generated in Section \ref{sec:sisetup}. Only qualitative social attributes $\mathcal{A}_1=Major$ and $\mathcal{A}_2=Relationship$ are considered in this case and the fuzzy rules defined in section \ref{subsec:exam} are adopted directly. Therefore, $tv^*(\mathcal{E})$ can be calculated according to Eq.(\ref{eqn:tvstar}) as $tv^*(\mathcal{E})=\frac{\mathcal{MP}_{1.p1}(\mathcal{E})+\mathcal{MP}_{2.p2}(\mathcal{E})}{\mathcal{M}_{1.p1}(\mathcal{E})+\mathcal{M}_{2.p2}(\mathcal{E})}$, for all rules $(1.p1+2.p2),\ p1,p2\in [1,2,3]$. Fig.~\ref{fig:ls_tv} demonstrates the trust values calculated by using the input independent fuzzy model for different social attributes. A relative who majors in network security obtains trust values ranging in around $0.81$ to more than $0.83$, but a stranger with an security major only receives trust values from less than $0.2$ to about $0.75$. The relative always receives higher trust values than the stranger. When we consider another stranger who majors in literature arts (i.e., belongs to OTHERS), the trust values drop to the range of $[0.16, 0.19]$. Moreover, with $\mathcal{E}$ (i.e., the values of the quantitative social attributes) increasing, the trust value never decreases. We can thus observe that a higher trust value will be obtained by the friend whose social attributes lead to more trustworthy. We therefore demonstrate that the input independent fuzzy model can correctly and effectively convert both quantitative and qualitative social attributes into trust values.

In our experiment, we randomize the values of the quantitative and qualitative social attributes between each two friends. After calculating the trust values by using the input independent fuzzy model, we adopt the trust propagation algorithm, which is proposed in Section \ref{subsec:tpm}, to generate the trust scores of each entity's friends or friends of friends.

%

\subsection{Experiment to Assess Secure Anonymity}
\label{sec:exsecure}
To evaluate the secure anonymity of Tor and STor, two different implementations of Tor and STor are considered and outlined in Table \ref{table:diffim}.

\begin{table}[ht!]
  \centering
  \caption{\footnotesize Different Implementations of Tor and STor.}
  \renewcommand{\arraystretch}{1.2}
  \label{table:diffim}
  \scriptsize
  \begin{tabular}{c|c}
  \toprule
  Implementation & Description \\
  \midrule
  Original Tor & Tor routers with higher self-reported bandwidth are malicious \\
  Opportunistic Tor & Random Tor routers are considered to be malicious \\
  Practical STor & Friends with smaller trust score are more likely to set up malicious routers \\
  Theoretical STor & Friendship circle excludes malicious routers \\
  \bottomrule
\end{tabular}
\end{table}

Originally, Tor routers self-report their available bandwidth to directory servers for circuit establishment and attackers can thus simply announce high bandwidth in order to launch attacks. We regard this implementation as \emph{Original Tor}, in which routers with higher bandwidth are considered to be malicious. This implementation illustrates the minimum secure anonymity obtained by Tor. To mitigate the affects caused by the falsely high bandwidth announcement, an opportunistic bandwidth measurement algorithm has been proposed \cite{tune-up,tuneupA}. Instead of the self-reporting approach, the opportunistic method allows directory servers to measure the authentic bandwidth provided by Tor routers. Therefore, random routers are considered to be malicious in this implementation, called \emph{Opportunistic Tor}.
%

In STor, users utilize social networks to select trustworthy routers from their friendship circles. Although friendship circles cannot guarantee to exclude all the malicious routers, friends with smaller trust score are more likely to set up malicious routers or possess vulnerable routers. Since STor allows users to establish circuits by taking the trust into consideration, we use \emph{Practical STor}, where routers with smaller trust score are more likely to be malicious, to demonstrate the expected implementation of STor. In contrast, the \emph{Theoretical STor}, in which malicious routers are excluded by users' friendship circle, is used to show STor with the theoretically maximum secure anonymity.
%

\subsubsection{Secure Anonymity in Simulation}
\label{sec:sisecure}
In the simulation, we conduct the experiments in four different malicious routers occupancies, representing $5\%$, $10\%$, $15\%$ and $20\%$ candidate routers are malicious respectively. We regard a round of simulation as an user selecting a router in $1000$ times and use the ratio of malicious routers to selected routers, denoted as $\mathcal{R}_\mathfrak{MR}$, in each round to measure secure anonymity. Therefore, smaller $\mathcal{R}_\mathfrak{MR}$ indicates better secure anonymity. Particularly, $\mathcal{R}_\mathfrak{MR}=0$ leads to the maximum secure anonymity.

\begin{figure*}[ht!]
     \centering
     \subfigure[\scriptsize $5\%$ Candidate Routers are Malicious.]{
           \label{fig:sai5}
          \includegraphics[width=.238\textwidth]{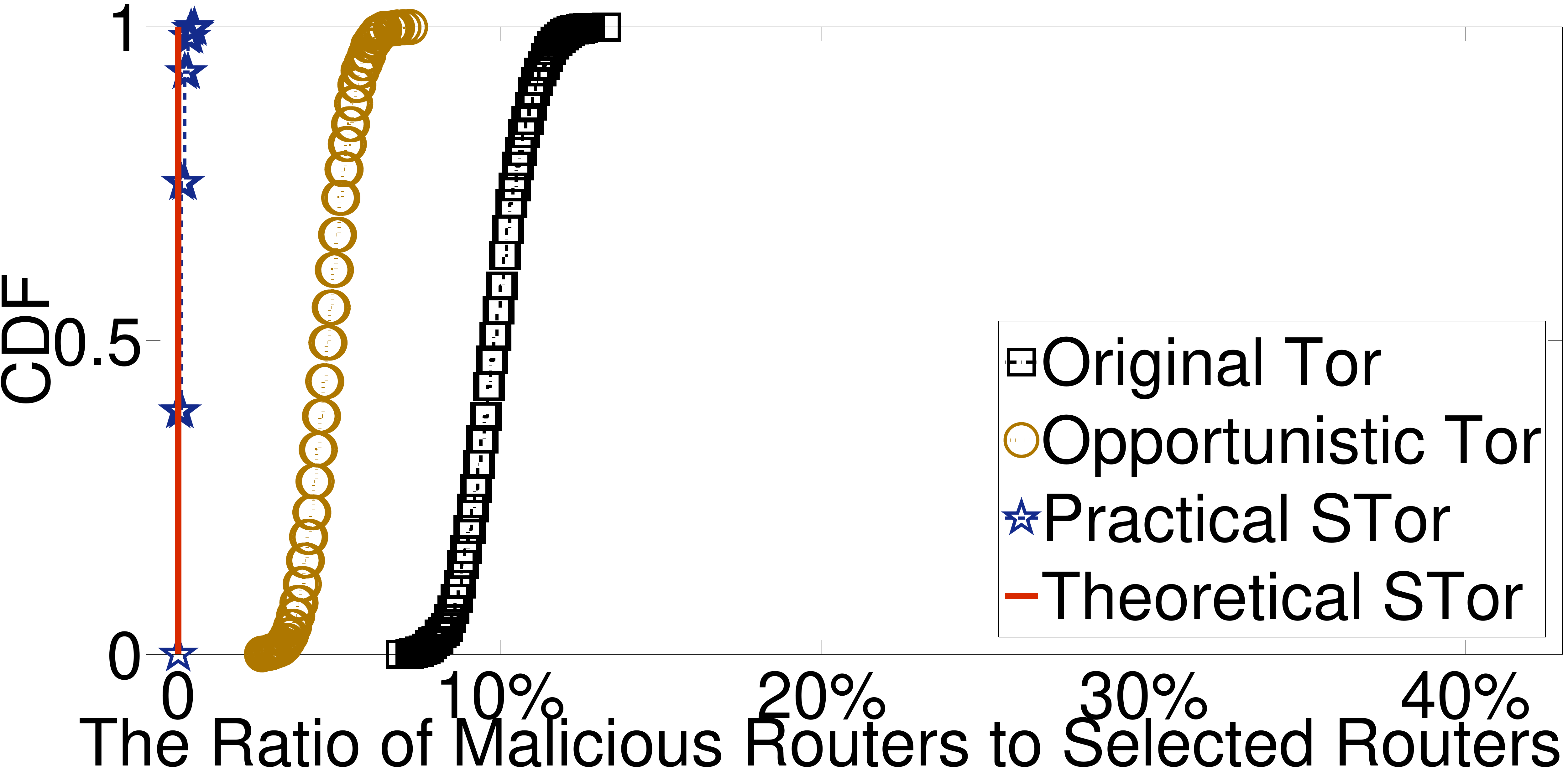}}
     \subfigure[\scriptsize $10\%$ Candidate Routers are Malicious.]{
           \label{fig:sai10}
          \includegraphics[width=.238\textwidth]{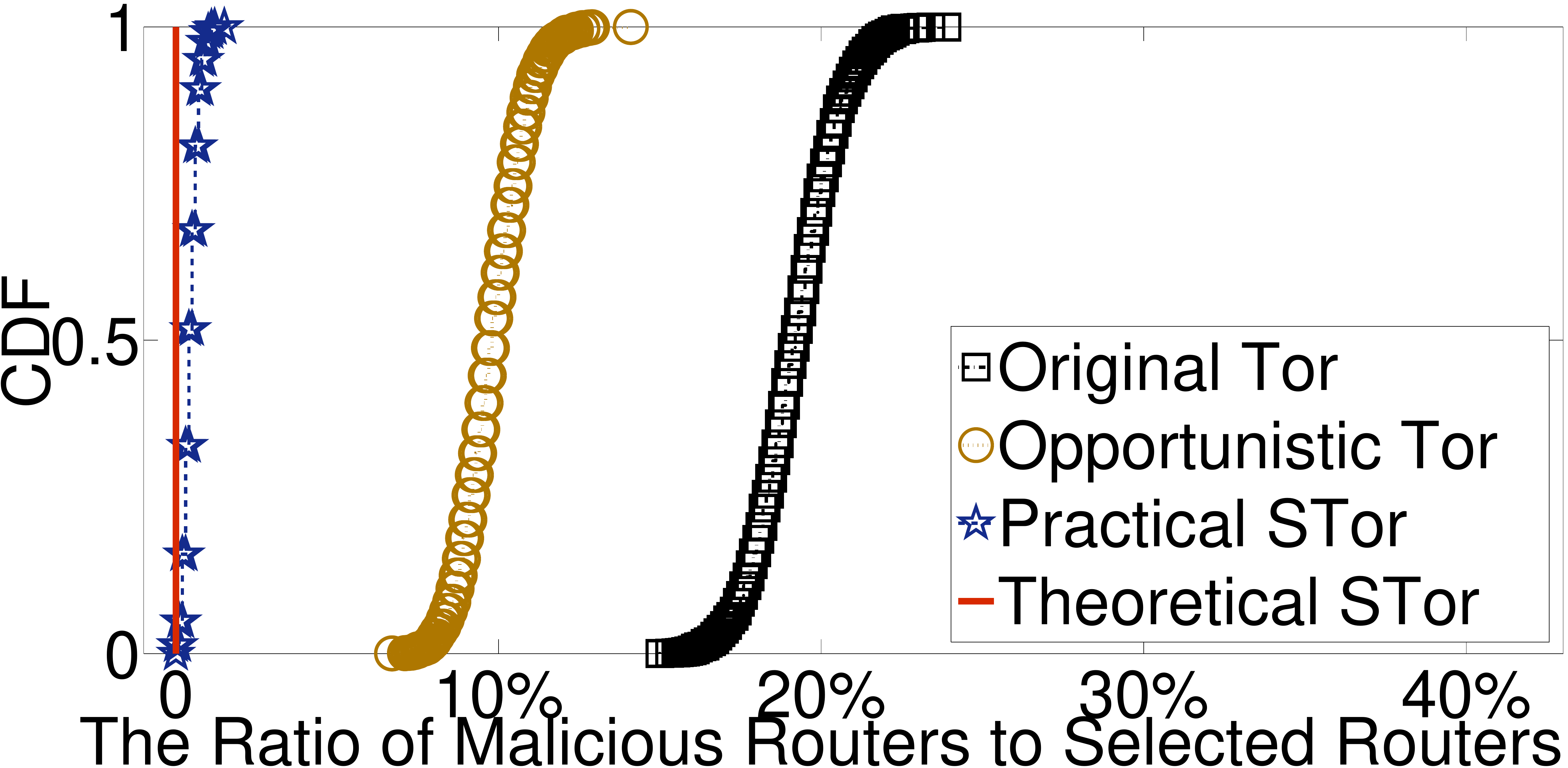}}
     \subfigure[\scriptsize $15\%$ Candidate Routers are Malicious.]{
           \label{fig:sai15}
          \includegraphics[width=.238\textwidth]{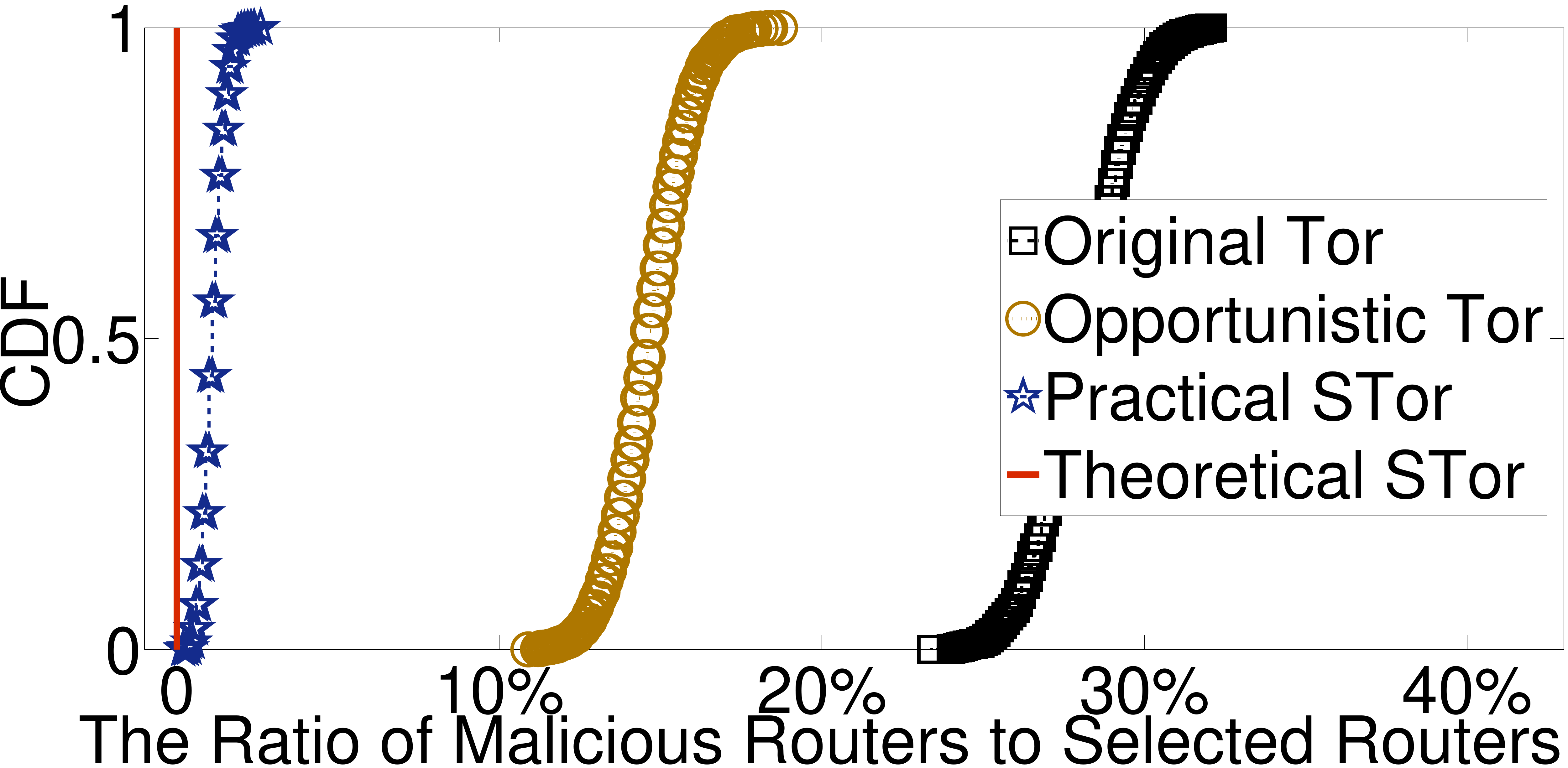}}
     \subfigure[\scriptsize $20\%$ Candidate Routers are Malicious.]{
           \label{fig:sai20}
          \includegraphics[width=.238\textwidth]{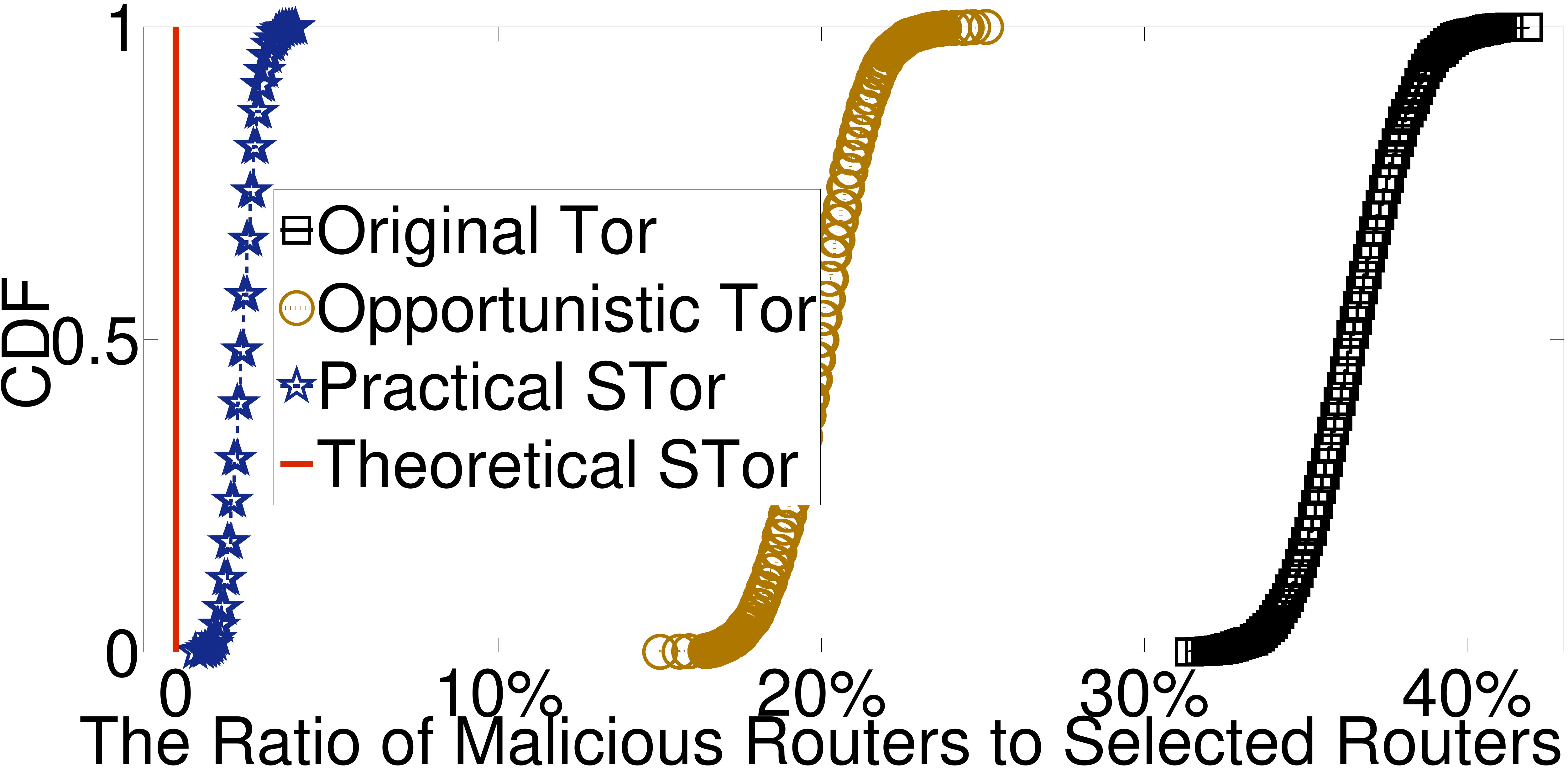}}
     \caption{The Ratio of Malicious Routers to Selected Routers (i.e., $\mathcal{R}_\mathfrak{MR}$) of Tor and STor in Different Implementations.}
     \label{fig:sidiffm}
\end{figure*}

To evaluate the secure anonymity, we conduct $1000$ rounds of simulation in different implementations of Tor and STor over different malicious routers occupancies. Note that Practical STor fixes both $\omega$, a parameter defined in Eqn.(\ref{equ:pj}), and $ts_h$, a threshold defined in section \ref{subsubsec:anoimpact}, to $0$ in this evaluation. Figs. \ref{fig:sai5}-\ref{fig:sai20} show the CDF of secure anonymity, measured by $\mathcal{R}_\mathfrak{MR}$, for different implementations of Tor and STor. Among Original Tor, Opportunistic Tor and Practical STor, although their secure anonymity decreases when the malicious routers occupancy climbs from $5\%$ to $20\%$, Practical STor always shows much better secure anonymity than the other two (e.g., smaller than one tenth of $\mathcal{R}_\mathfrak{MR}$ compared with that of Original Tor). Particularly, Practical STor gives out approximately $\mathcal{R}_\mathfrak{MR}=0$ distribution when $5\%$ candidate routers are malicious, thus demonstrating Practical STor can achieve around the maximum secure anonymity when the malicious routers occupancy is small. Beside that, Theoretical STor is constant to show the maximum secure anonymity (i.e., $\mathcal{R}_\mathfrak{MR}=0$) regardless the malicious routers occupancies. That is because malicious routers are excluded by friendship circles.
\begin{figure*}[ht!]
     \centering
     \subfigure[\scriptsize The Best Case when $5\%$ Candidate Routers are Malicious.]{
           \label{fig:sabi5}
          \includegraphics[width=.238\textwidth]{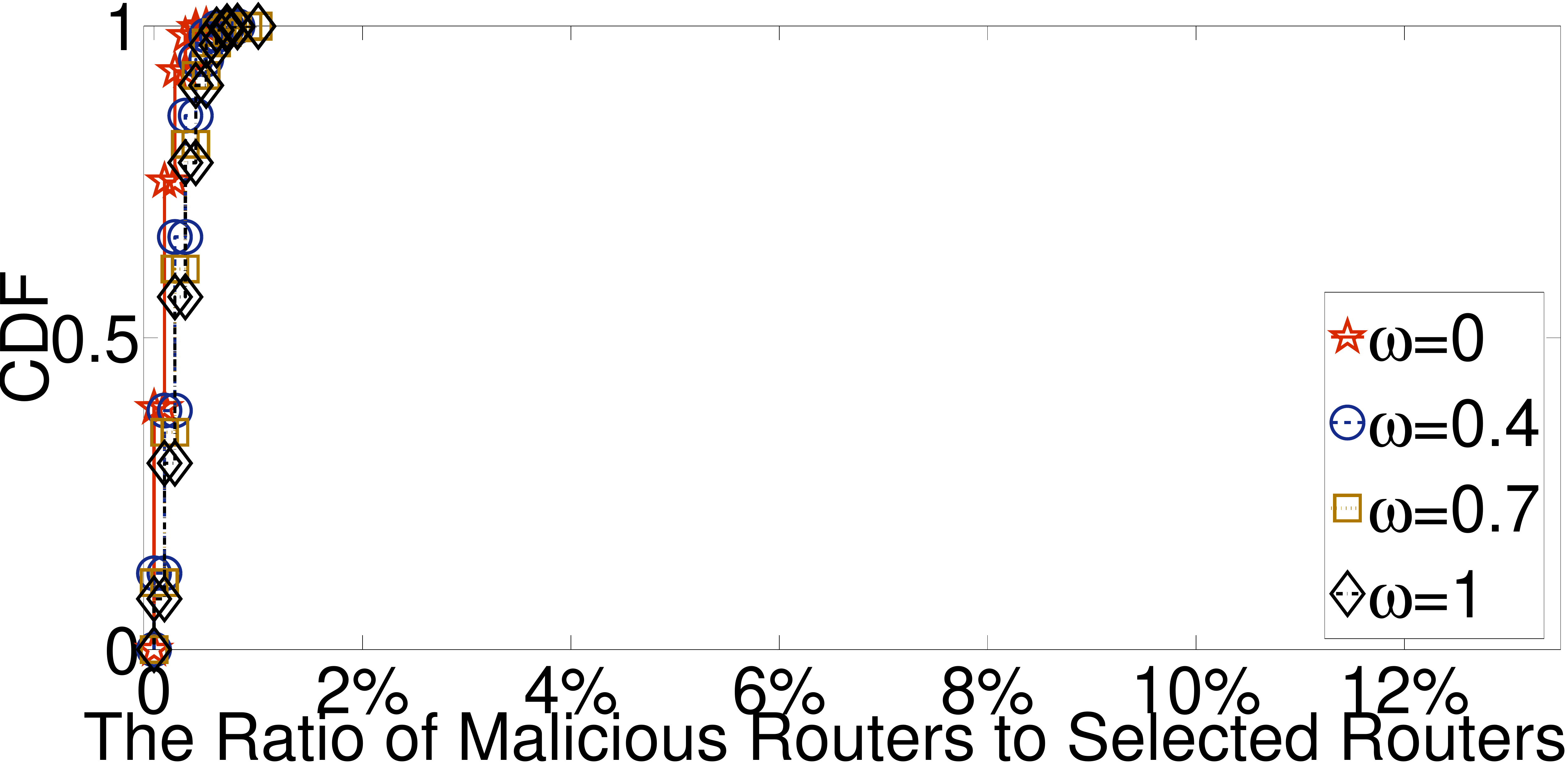}}
     \subfigure[\scriptsize The Worst Case when $5\%$ Candidate Routers are Malicious.]{
           \label{fig:sawi5}
          \includegraphics[width=.238\textwidth]{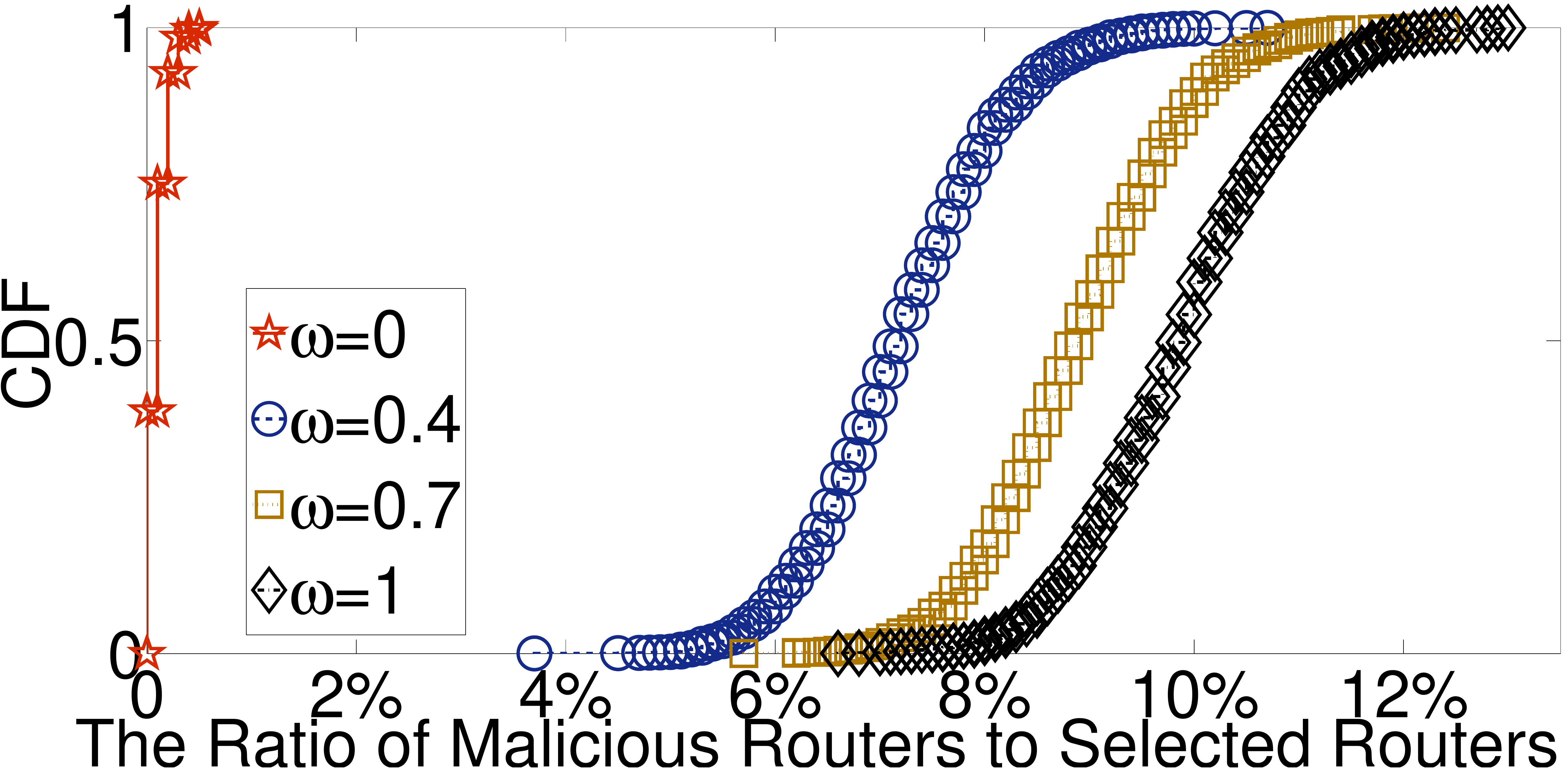}}
     \subfigure[\scriptsize The Best Case when $20\%$ Candidate Routers are Malicious.]{
           \label{fig:sabi20}
          \includegraphics[width=.238\textwidth]{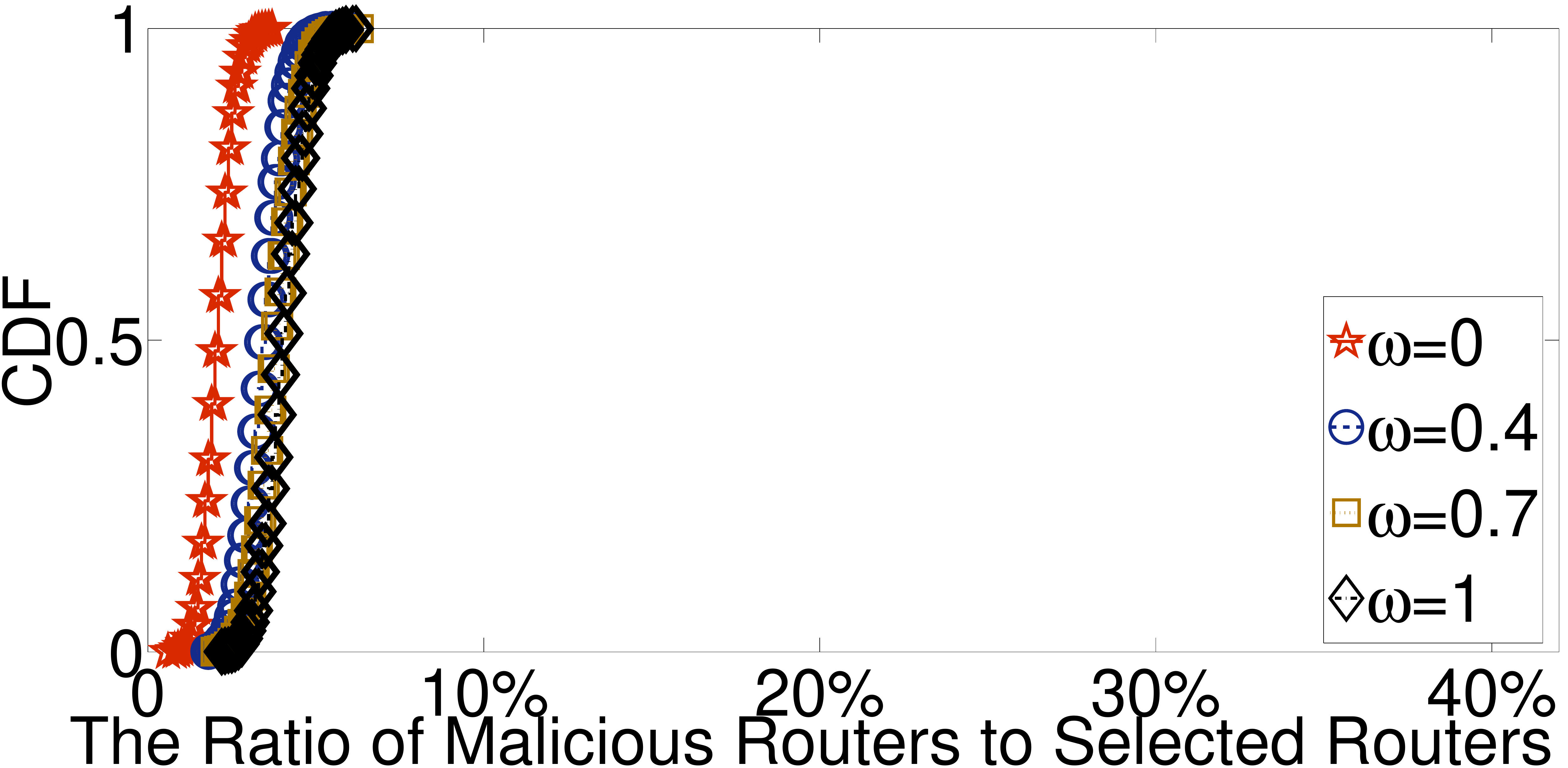}}
     \subfigure[\scriptsize The Worst Case when $20\%$ Candidate Routers are Malicious.]{
           \label{fig:sawi20}
          \includegraphics[width=.238\textwidth]{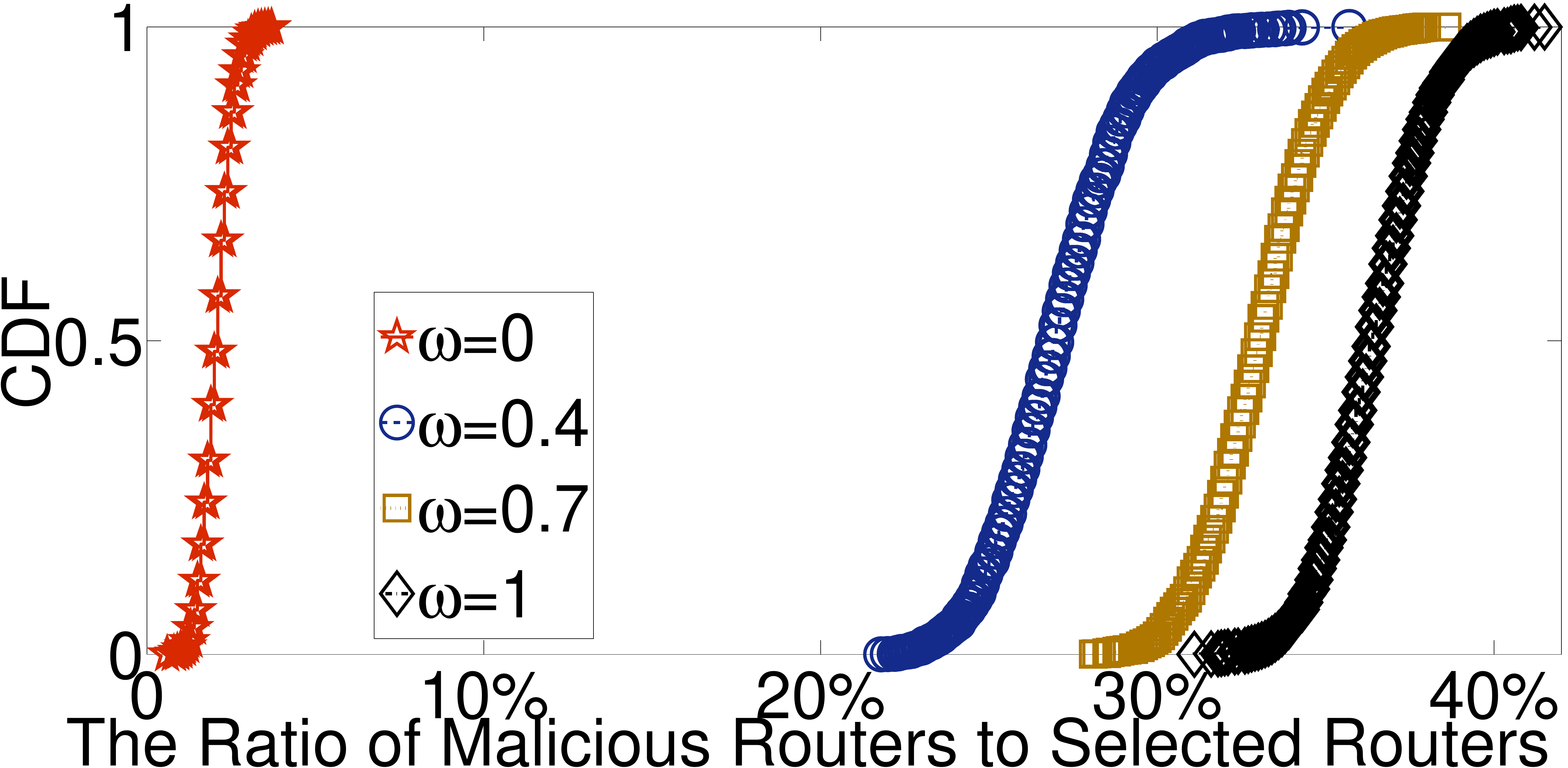}}
     \caption{The Ratio of Malicious Routers to Selected Routers of Practical STor in Different $\omega$.}
     \label{fig:sidiffw}
\end{figure*}

STor uses $\omega$, a parameter defined in Eqn.(\ref{equ:pj}), to balance the secure anonymity and performance for the trust-based routers selection. A large $\omega$ puts more weight to bandwidth, but a small $\omega$ has more weight to the trust score. According to entities with different trust score obtaining different bandwidth, we have two cases listed in Table \ref{table:diffcase}. The \emph{Best Case} is for entities outside of the friendship circle receive low bandwidth and friends with high trust score possess high bandwidth routers inside the friendship circle, while the \emph{Worst Case} is for entities outside of the friendship circle have high bandwidth and friends with high trust score obtain low bandwidth routers inside the friendship circle.
\begin{table*}[ht!]
  \centering
  \caption{\footnotesize Different Cases for Trust Score and Bandwidth in Each Router.}
  \renewcommand{\arraystretch}{1.2}
  \label{table:diffcase}
  \scriptsize
  \begin{tabular}{c|c}
  \toprule
  Case & Description \\
  \midrule
  the Best Case & Entities outside of the friendship circle possess \textbf{low} bandwidth routers and highly trusted friends with \textbf{high} bandwidth routers inside the friendship circle\\
  the Worst Case & Entities outside of the friendship circle possess \textbf{high} bandwidth routers and highly trusted friends with \textbf{low} bandwidth routers inside the friendship circle \\
  \bottomrule
\end{tabular}
\end{table*}


In the evaluation of $\omega$, $ts_h$ is set to $0$. Figs. \ref{fig:sabi5} and \ref{fig:sawi5} show the $\mathcal{R}_\mathfrak{MR}$ for the Best Case and the Worst Case in Practical STor with different $\omega$ when $5\%$ Candidate Routers are Malicious. With $\omega$ growing up from $0$ to $1$, the $\mathcal{R}_\mathfrak{MR}$ is sharply increasing from less than $0.5\%$ to around $10\%$ in the Worst Case, whereas the $\mathcal{R}_\mathfrak{MR}$ is slightly increased in the Best Case. Figs. \ref{fig:sabi20} and \ref{fig:sawi20} illustrate the similar result when $20\%$ Candidate Routers are Malicious. The effectiveness of $\omega$ in balancing the secure anonymity and performance for these two cases will be detailed in Section \ref{sec:eperformance}.

\begin{figure*}[ht!]
     \centering
     \subfigure[\scriptsize $5\%$ Candidate Routers are Malicious.]{
           \label{fig:sati5}
          \includegraphics[width=.238\textwidth]{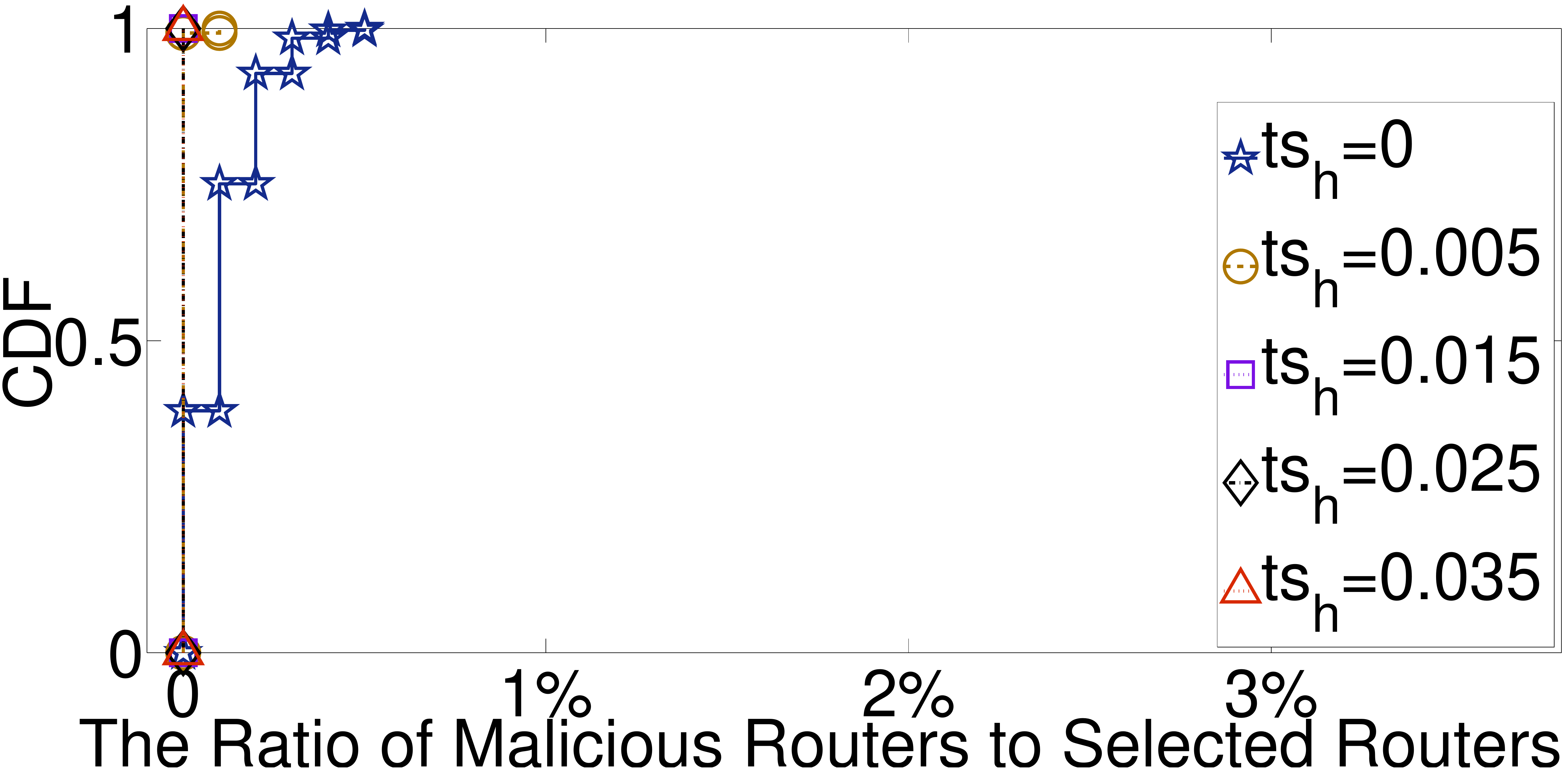}}
     \subfigure[\scriptsize $10\%$ Candidate Routers are Malicious.]{
           \label{fig:sati10}
          \includegraphics[width=.238\textwidth]{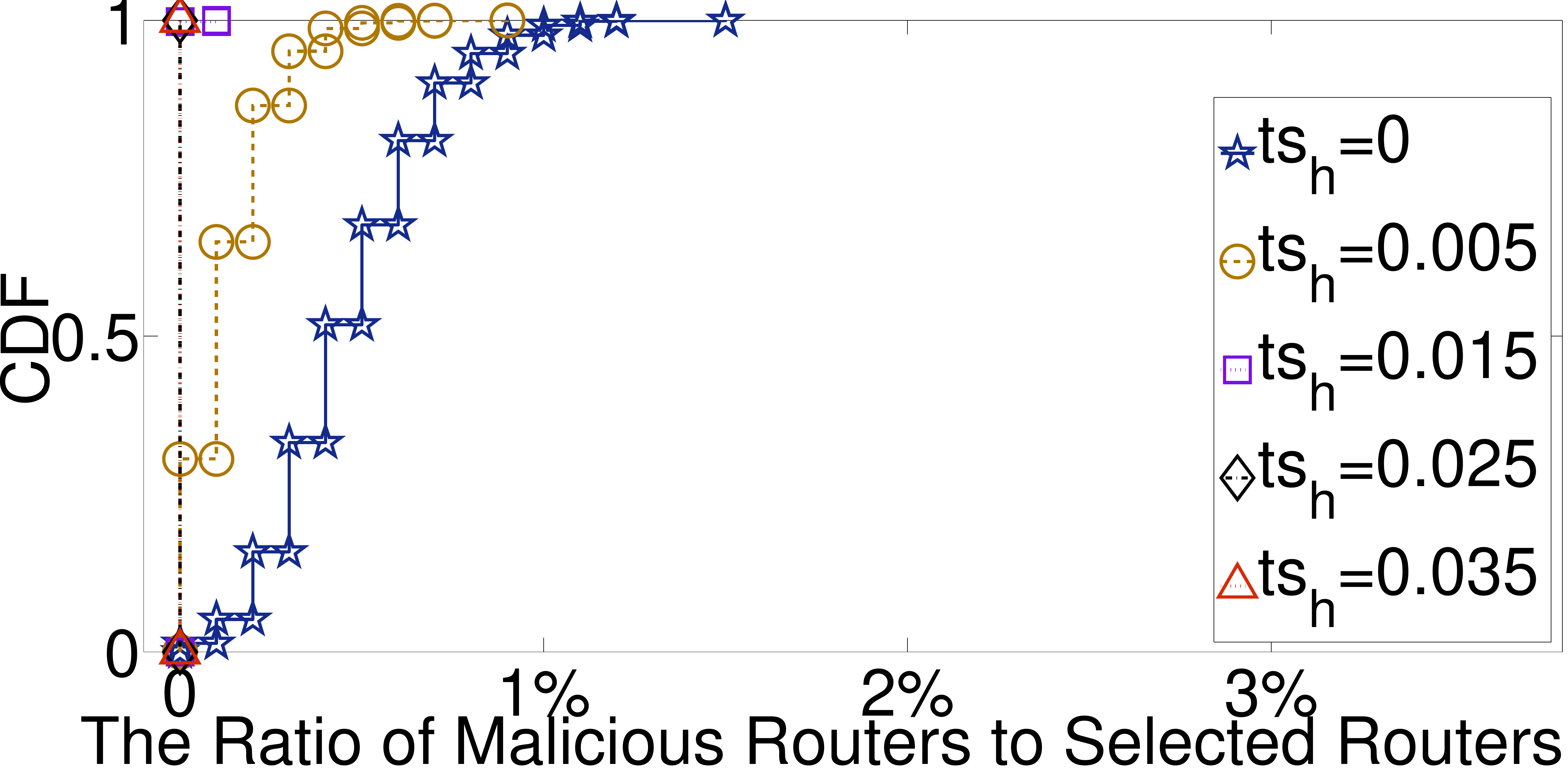}}
     \subfigure[\scriptsize $15\%$ Candidate Routers are Malicious.]{
           \label{fig:sati15}
          \includegraphics[width=.238\textwidth]{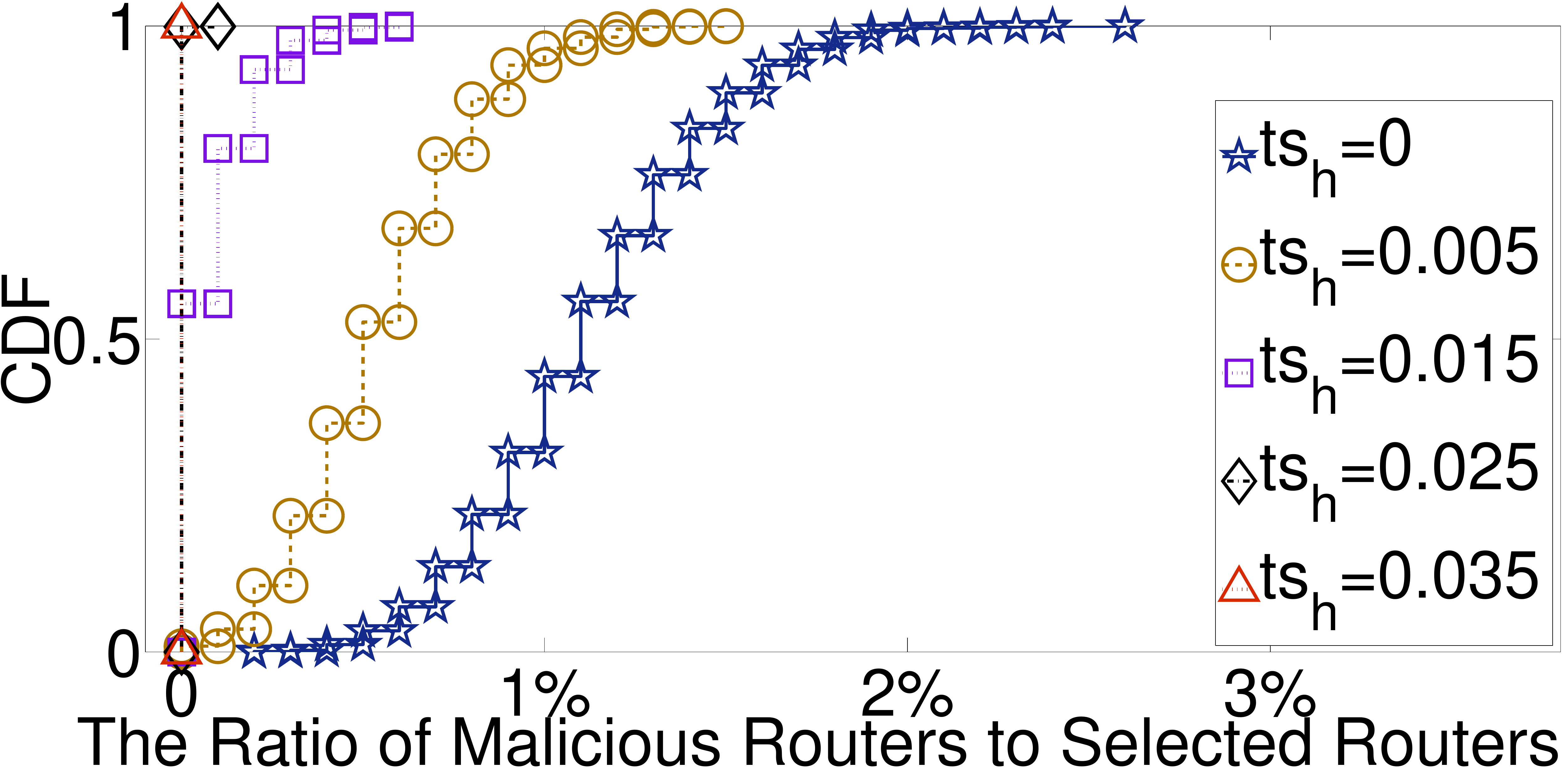}}
     \subfigure[\scriptsize $20\%$ Candidate Routers are Malicious.]{
           \label{fig:sati20}
          \includegraphics[width=.238\textwidth]{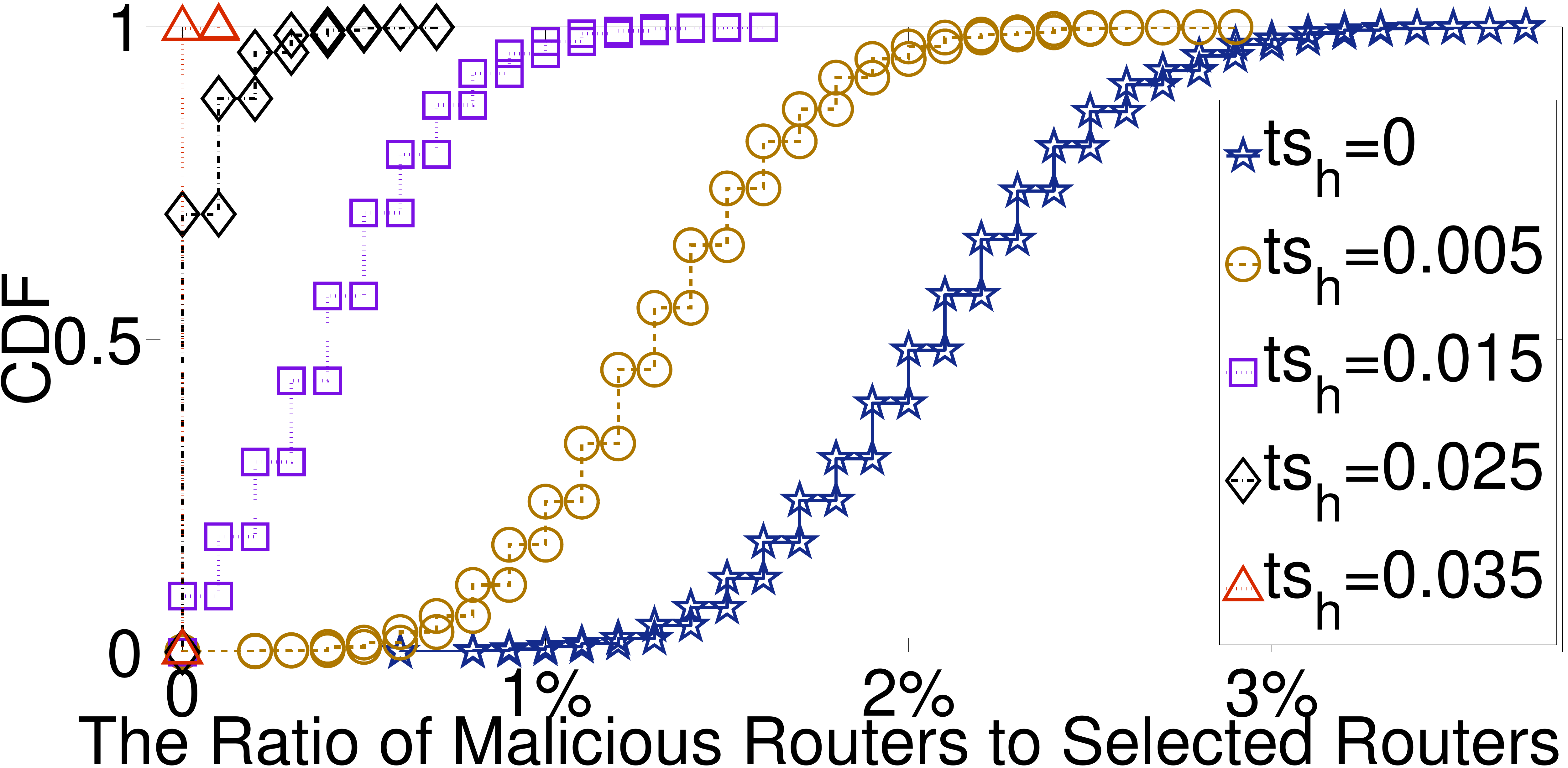}}
     \caption{The Ratio of Malicious Routers to Selected Routers (i.e., $\mathcal{R}_\mathfrak{MR}$) of Practical STor in Different $ts_h$.}
     \label{fig:sidifftsh}
\end{figure*}

To obtain better secure anonymity in Practical STor at the cost of baseline anonymity, the trustworthy friendship circle, defined in Section \ref{subsubsec:anoimpact}, can be used. More precisely, a trust score threshold, $th_s$, can filter out friends with smaller trust score, which are more likely to deploy malicious or vulnerable routers. In the evaluation of $ts_h$, $\omega$ is set to $0$. Figs. \ref{fig:sati5}-\ref{fig:sati20} illustrate the secure anonymity, measured by $\mathcal{R}_\mathfrak{MR}$, in Practical STor when malicious routers occupancy is $5\%$, $10\%$, $15\%$ and $20\%$. Coupled with $ts_h$ growing up from $0$ to $0.035$, the $\mathcal{R}_\mathfrak{MR}$ drops to around $0\%$, even in the case that $20\%$ candidate routers are malicious. We therefore observe that higher $ts_h$ leads to better secure anonymity. The corresponding impact caused by $ts_h$ to the baseline anonymity will be elaborated in Section \ref{sec:anonymity}.

\begin{figure*}[ht!]
     \centering
     \subfigure[\scriptsize Different Implementations in Tor and STor.]{
           \label{fig:plsai2}
          \includegraphics[width=.238\textwidth]{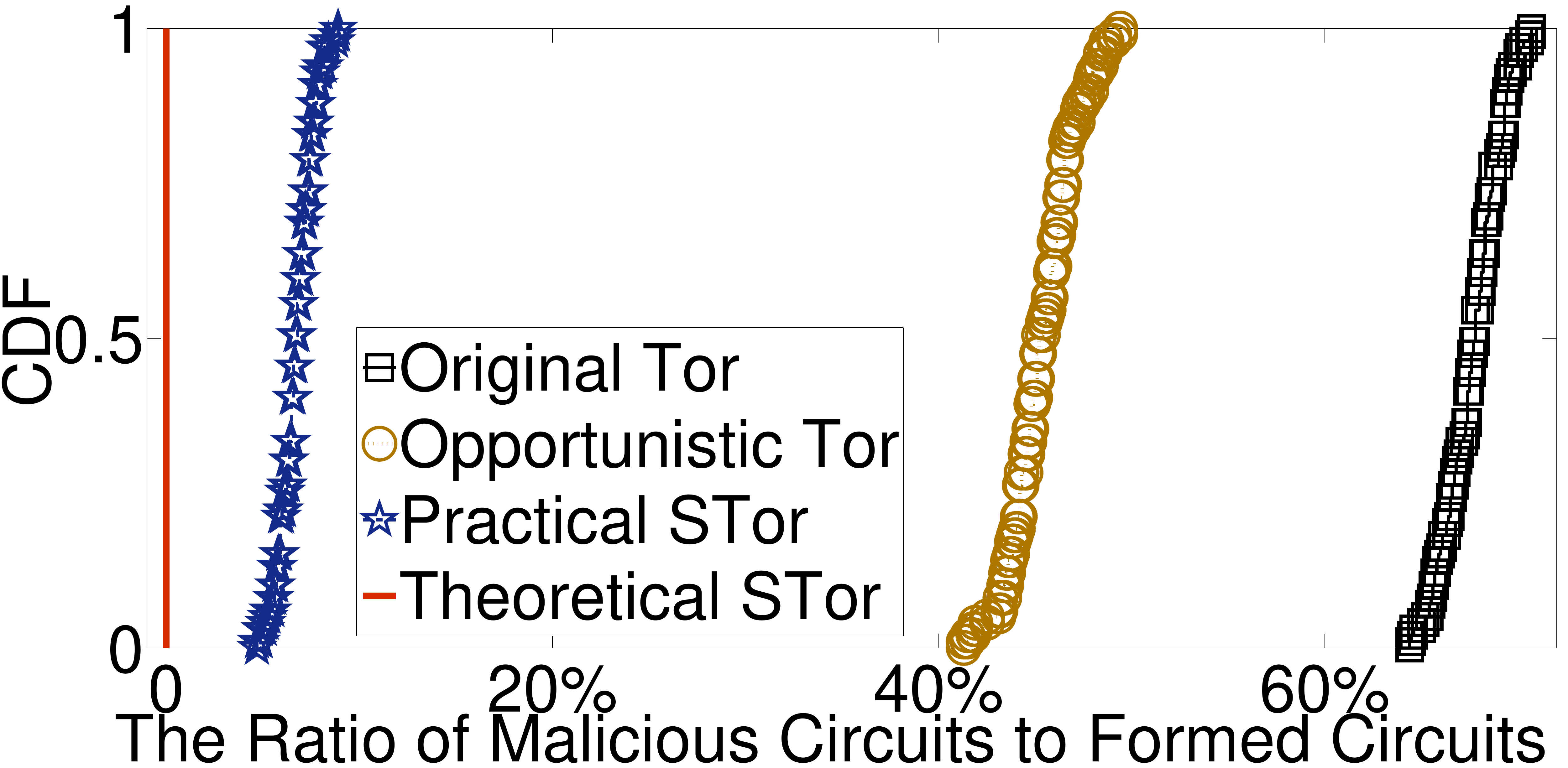}}
     \subfigure[\scriptsize Different $\omega$ in the Best Case.]{
           \label{fig:plsabi2}
          \includegraphics[width=.238\textwidth]{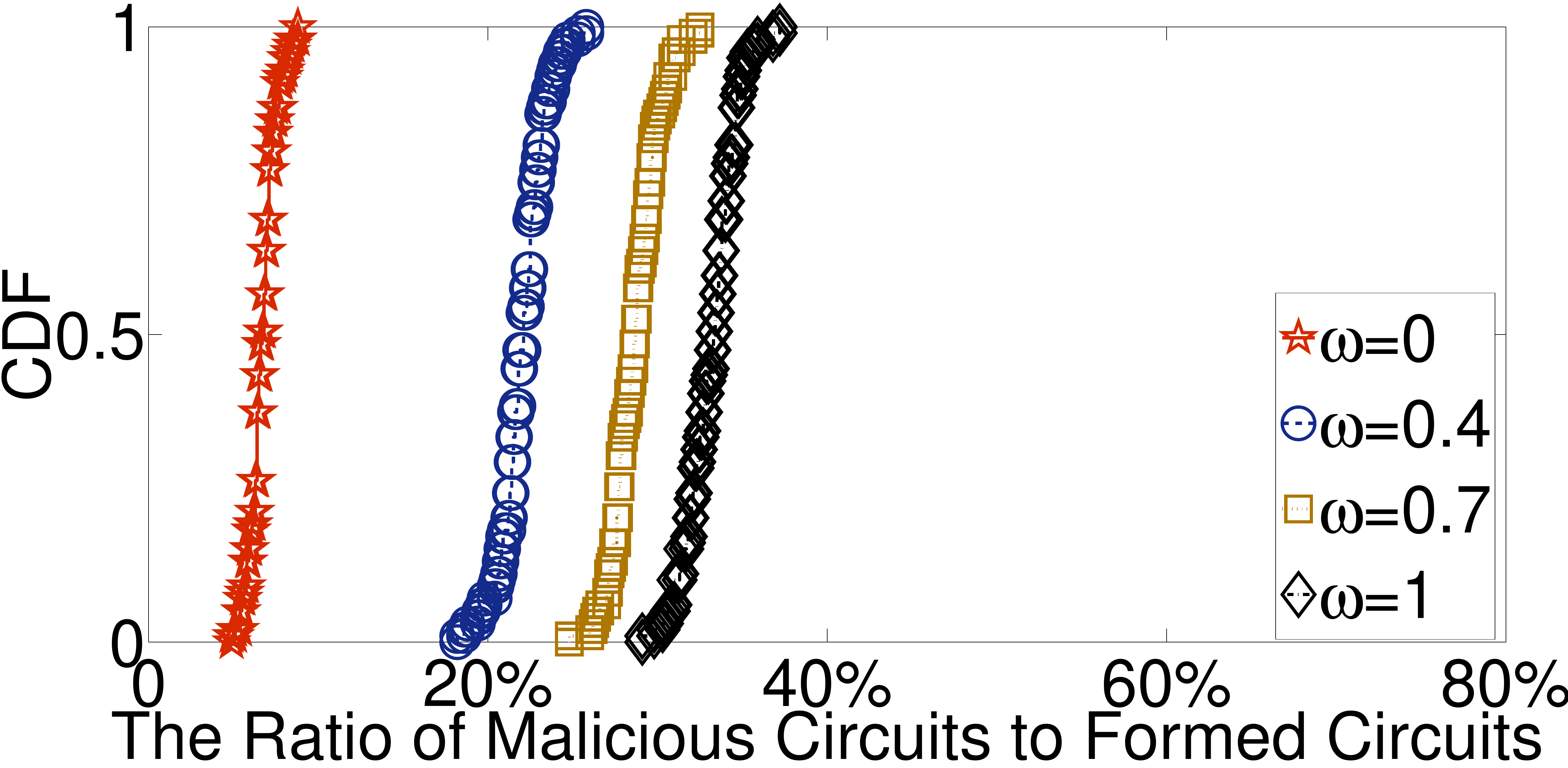}}
     \subfigure[\scriptsize Different $\omega$ in the Worst Case.]{
           \label{fig:plsawi2}
          \includegraphics[width=.238\textwidth]{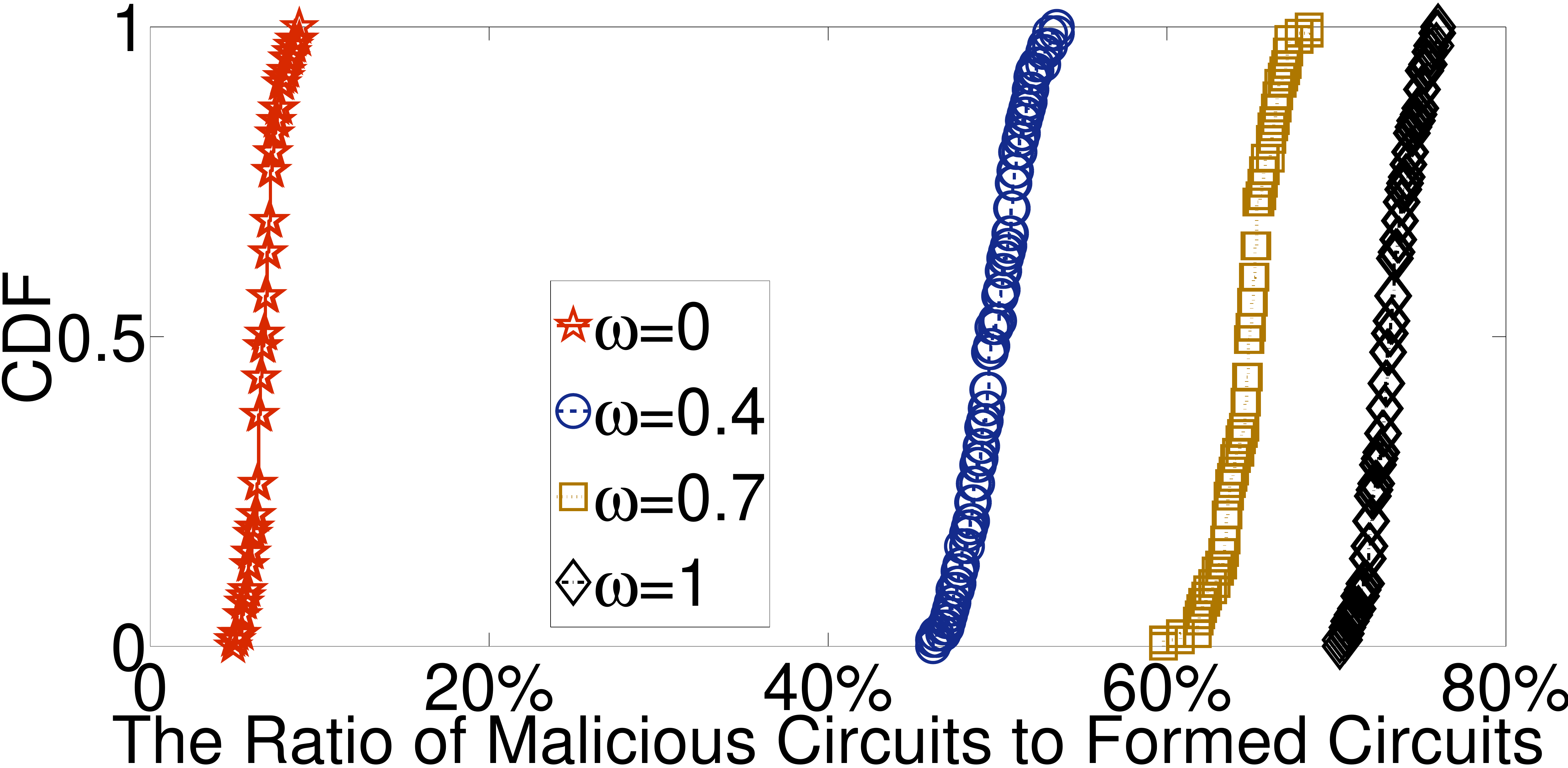}}
     \subfigure[\scriptsize Different $ts_h$ in STor.]{
           \label{fig:plsati2}
          \includegraphics[width=.238\textwidth]{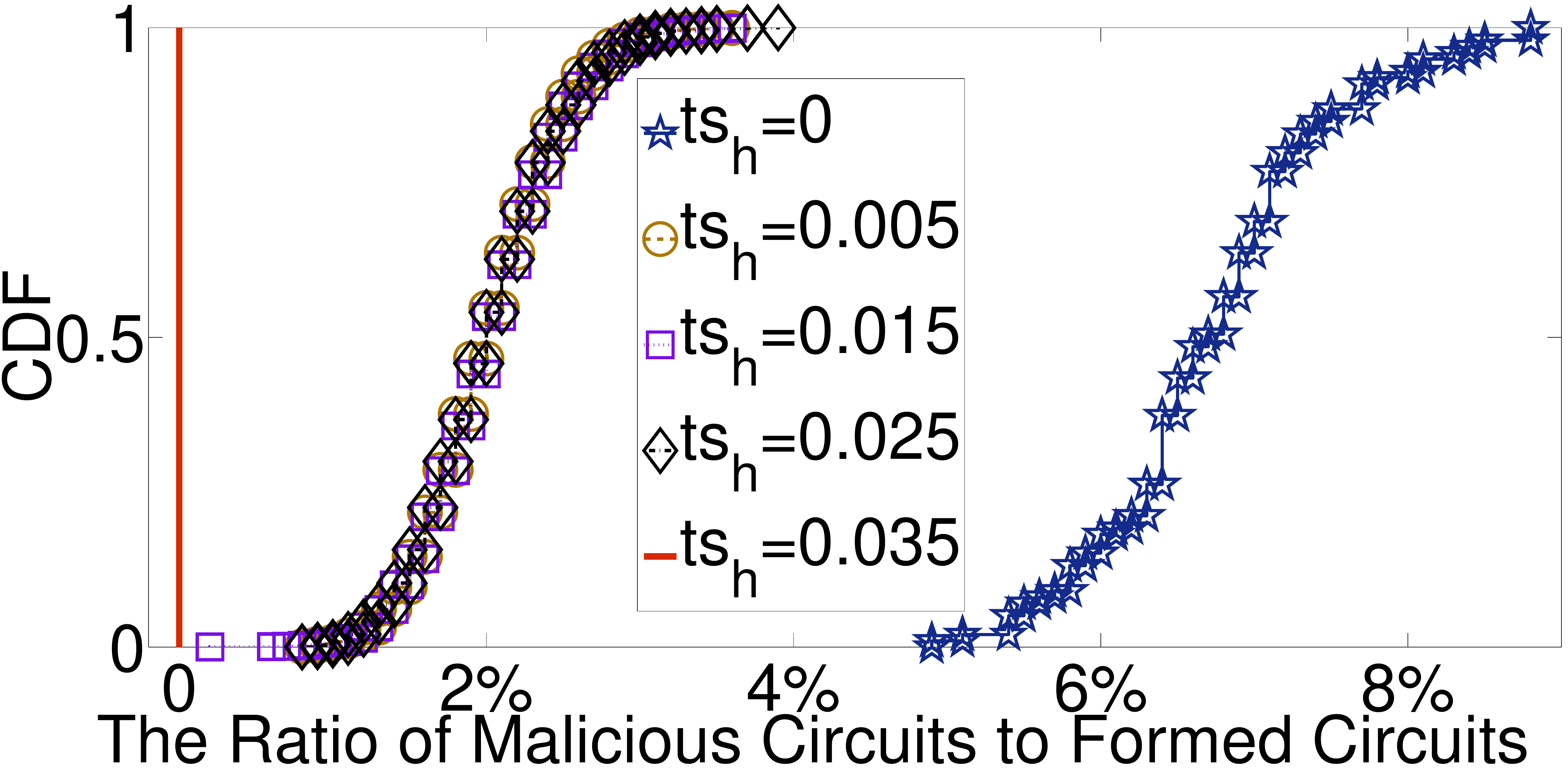}}
     \caption{The Ratio of Malicious Circuits to Formed Circuits (i.e., $\mathcal{R}_\mathfrak{MC}$) Obtained from PlanetLab Nodes when $20\%$ Candidate Routers are Malicious.}
     \label{fig:planetsecure}
\end{figure*}

\begin{figure*}[ht!]
     \centering
     \subfigure[\scriptsize The Best Case in Simulation.]{
           \label{fig:sibeper}
          \includegraphics[width=.238\textwidth]{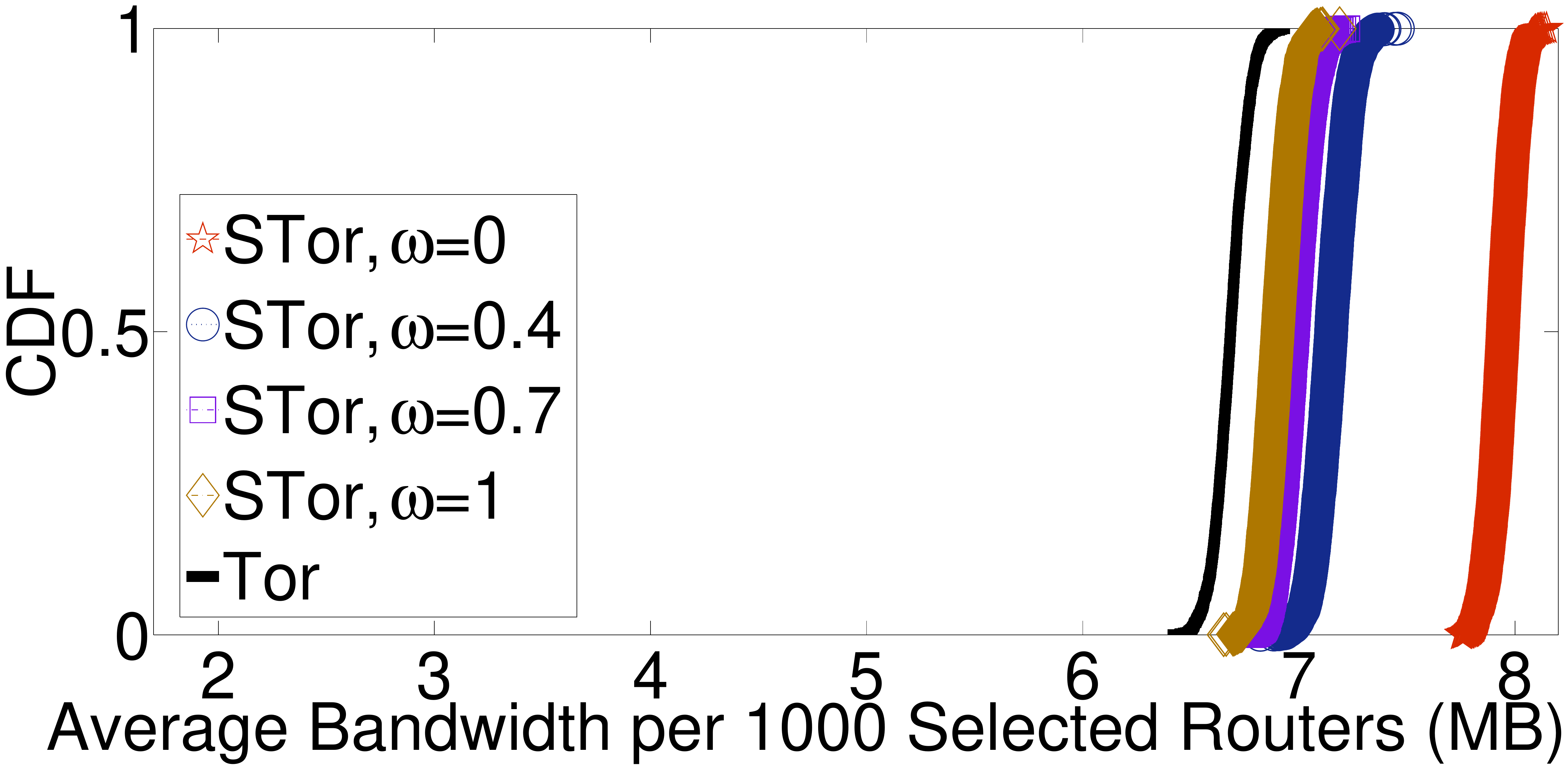}}
     \subfigure[\scriptsize The Worst Case in Simulation.]{
           \label{fig:siwoper}
          \includegraphics[width=.238\textwidth]{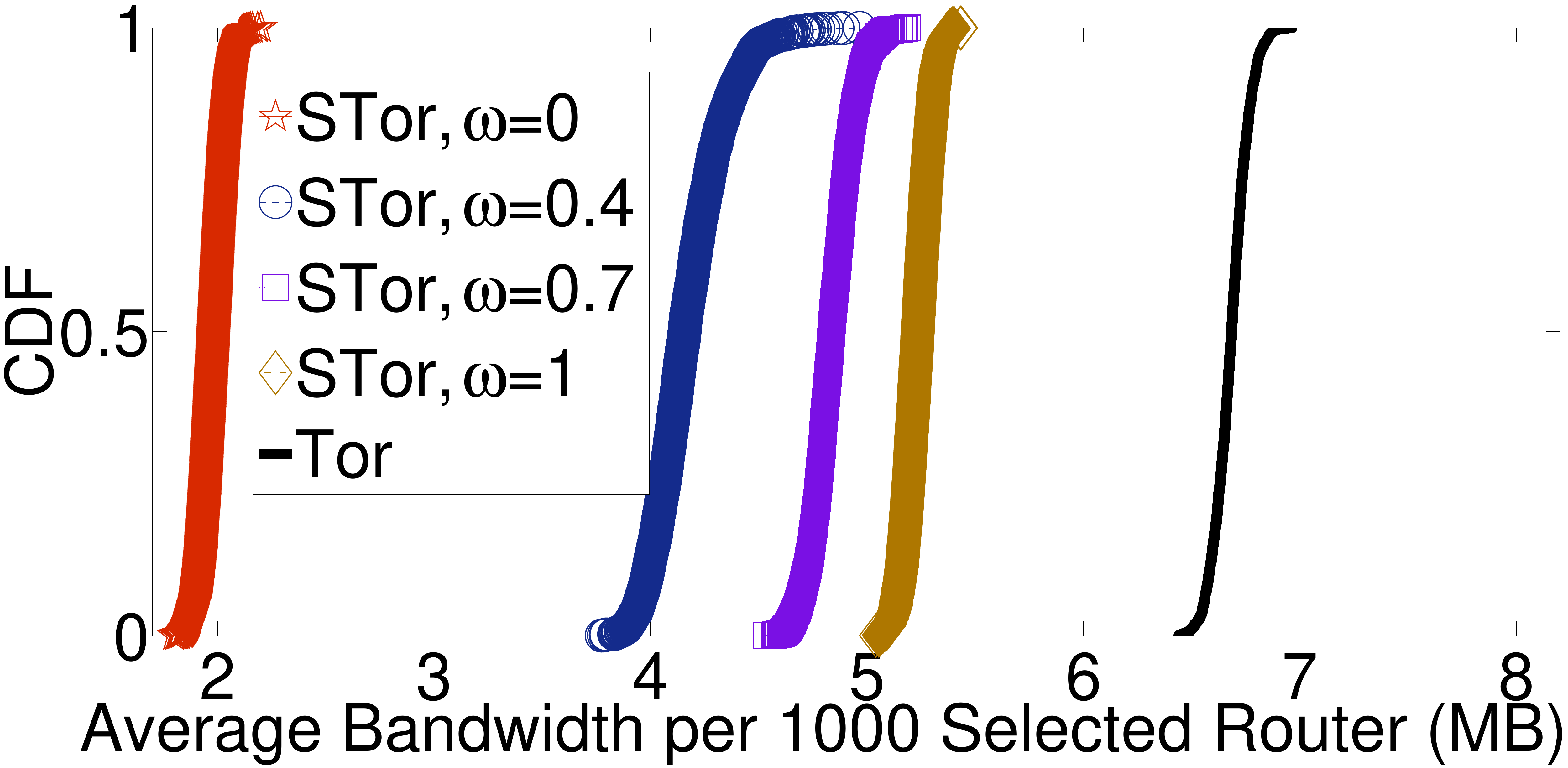}}
     \subfigure[\scriptsize The Best Case in Experiments over PlanetLab Nodes.]{
           \label{fig:plbeper}
          \includegraphics[width=.238\textwidth]{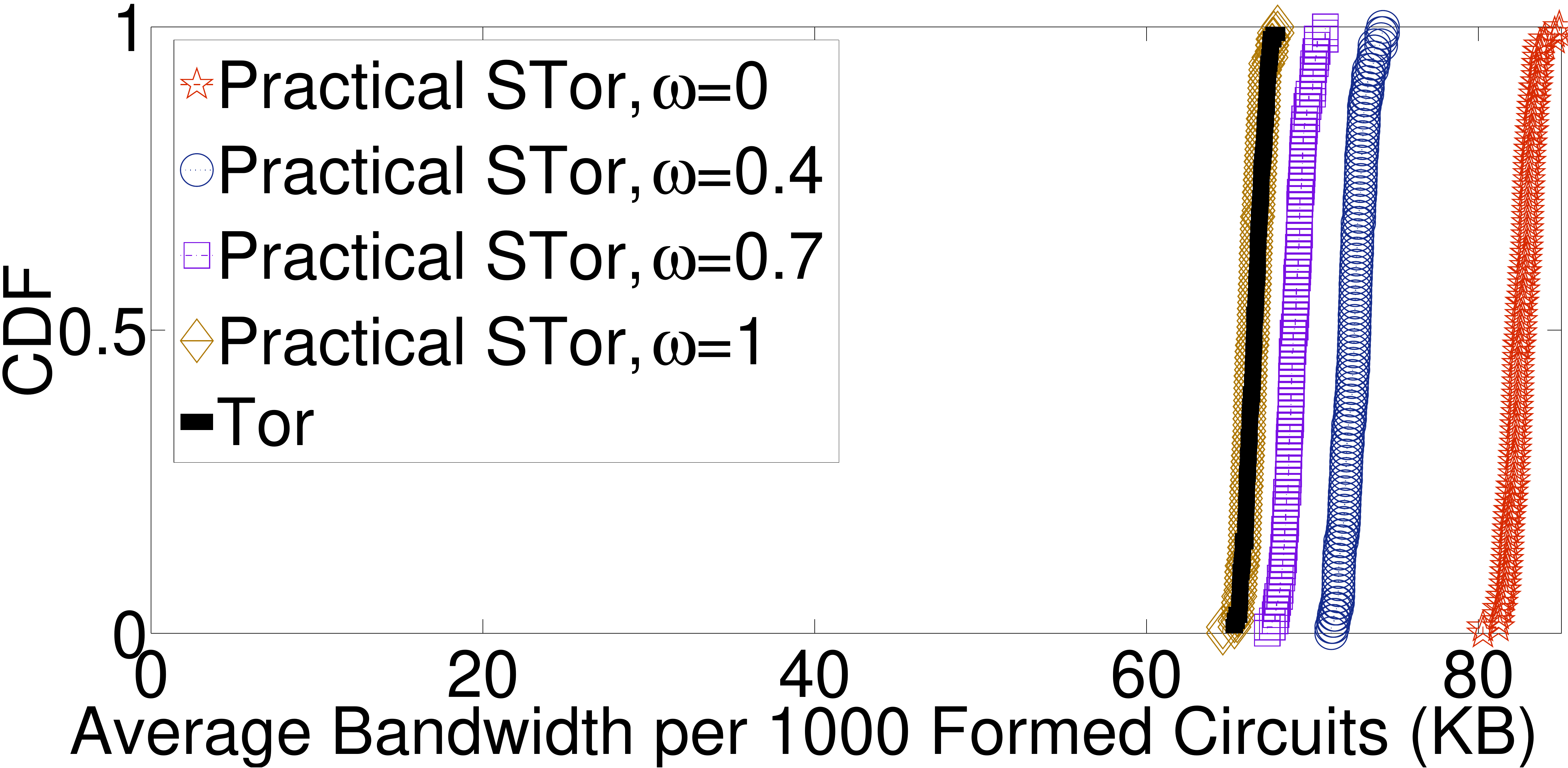}}
     \subfigure[\scriptsize The Worst Case Experiments over PlanetLab Nodes.]{
           \label{fig:plwoper}
          \includegraphics[width=.238\textwidth]{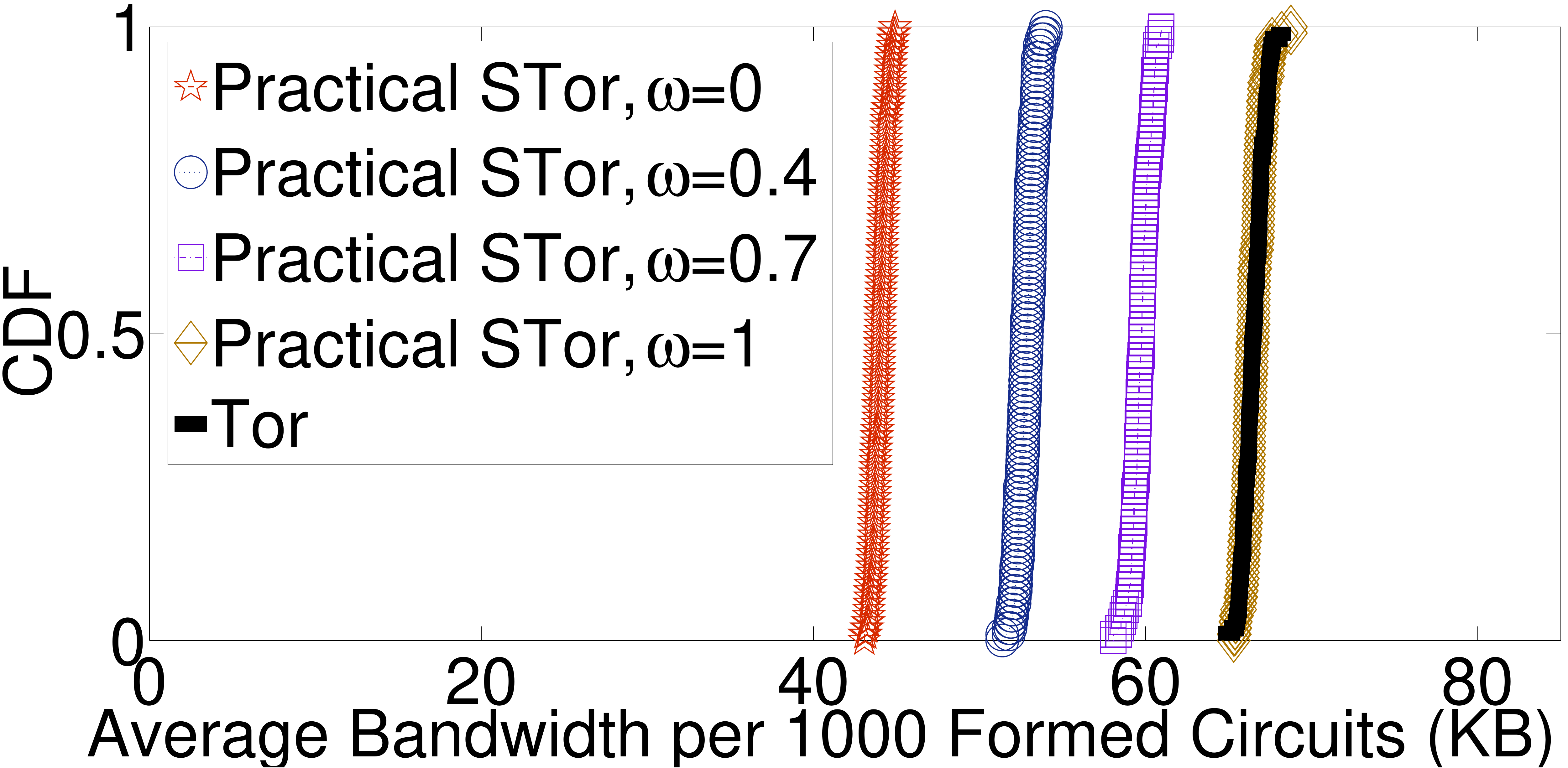}}
     \caption{The Average Bandwidth per $1000$ Selected Routers in Simulation (MBytes) and Experiments over PlanetLab Nodes (Kbytes).}
     \label{fig:siplper}
\end{figure*}

\begin{figure*}[ht!]
     \centering
     \subfigure[\scriptsize Size of Friendship Circle (i.e., $||F_{i}||$) in different Size of STor.]{
           \label{fig:bds}
          \includegraphics[width=.238\textwidth]{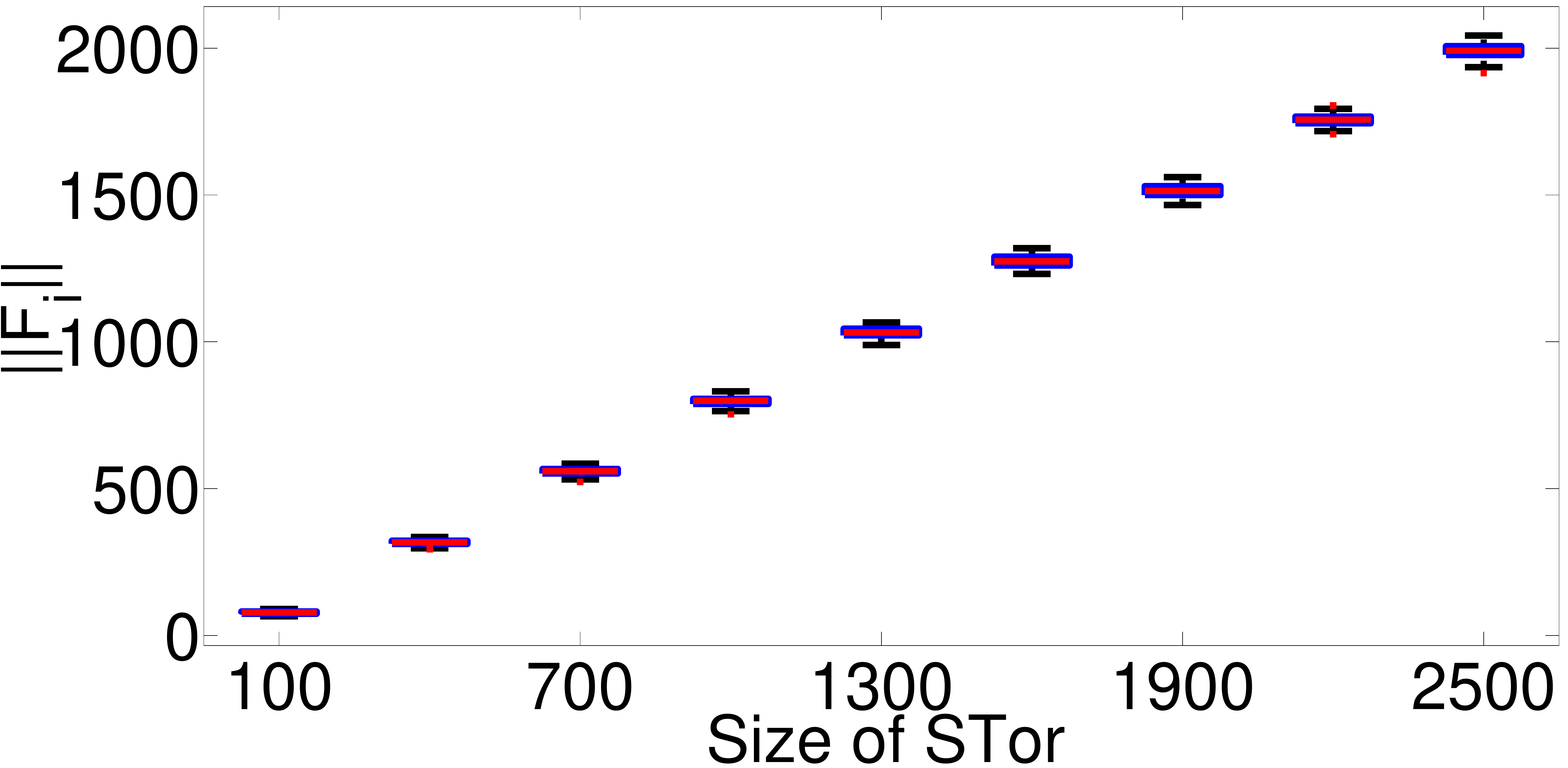}}
     \subfigure[\scriptsize Size of Trustworthy Friendship Circle (i.e., $||TF_{i}||$) in different $ts_{h}$ when the Size of STor is 100.]{
           \label{fig:bdst100}
          \includegraphics[width=.238\textwidth]{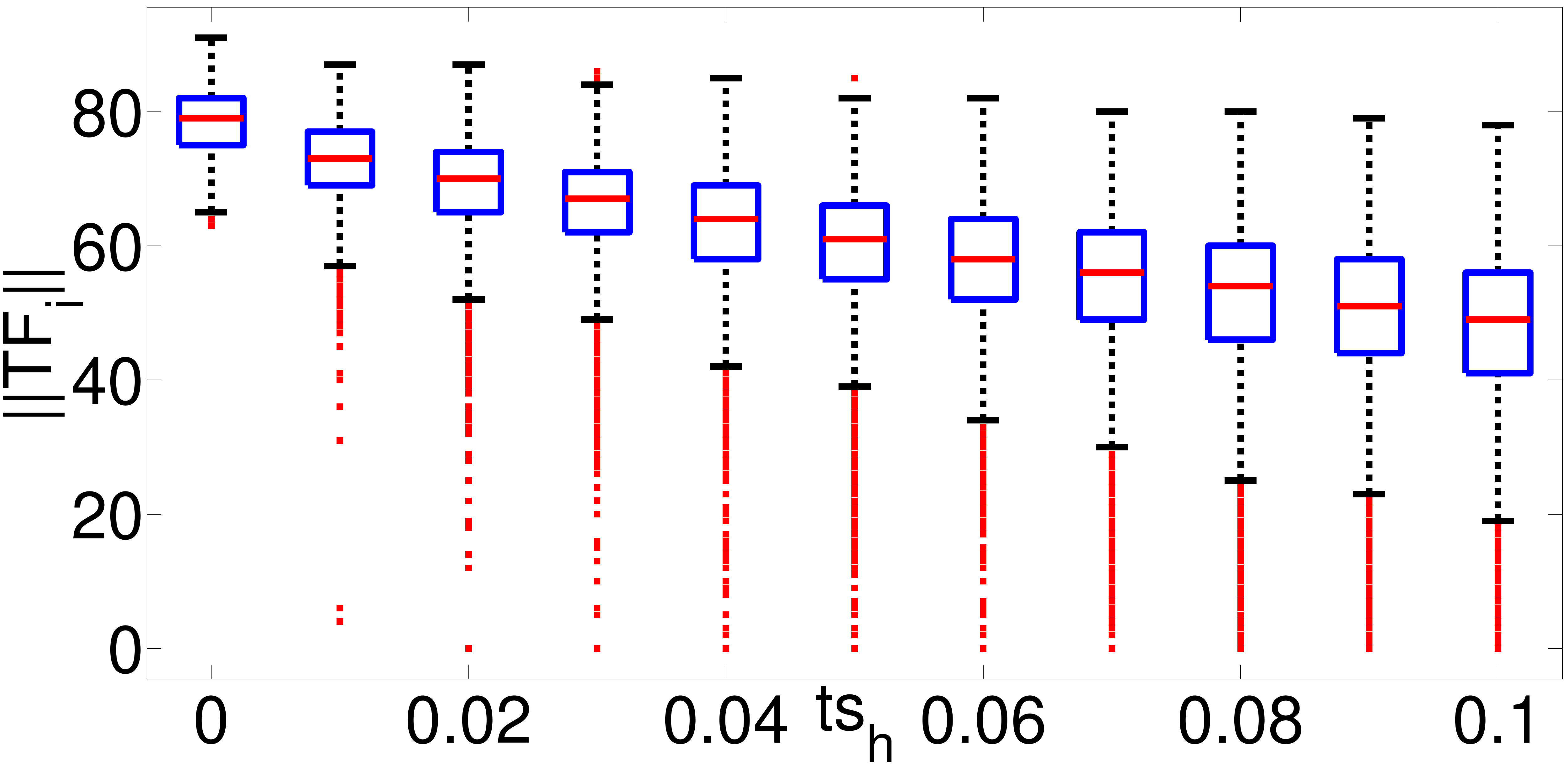}}
     \subfigure[\scriptsize Size of Trustworthy Friendship Circle (i.e., $||TF_{i}||$) in different $ts_{h}$ when the Size of STor is 1300.]{
           \label{fig:bdst1300}
          \includegraphics[width=.238\textwidth]{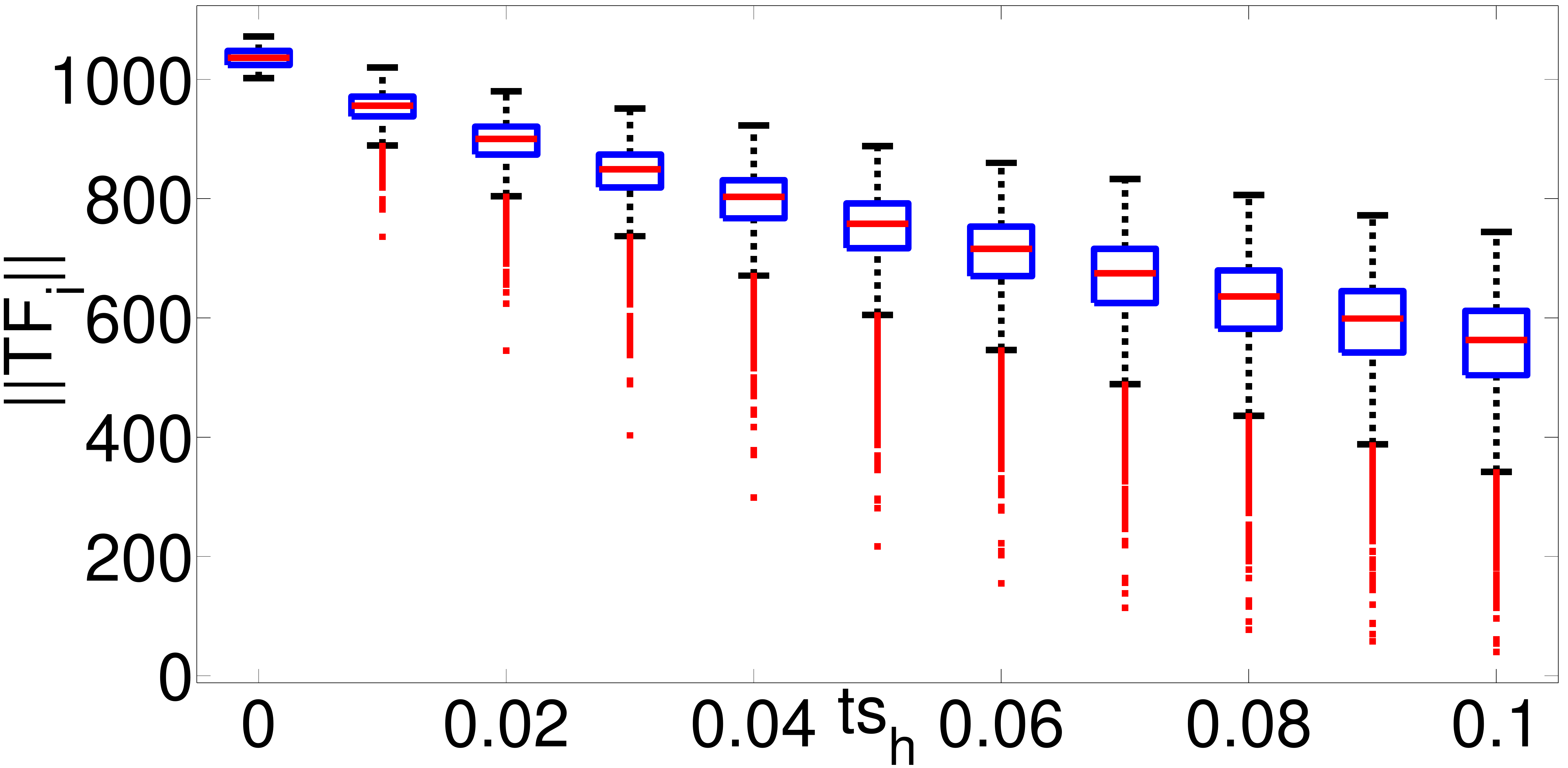}}
     \subfigure[\scriptsize Size of Trustworthy Friendship Circle (i.e., $||TF_{i}||$) in different $ts_{h}$ when the Size of STor is 2500.]{
           \label{fig:bdst2500}
          \includegraphics[width=.238\textwidth]{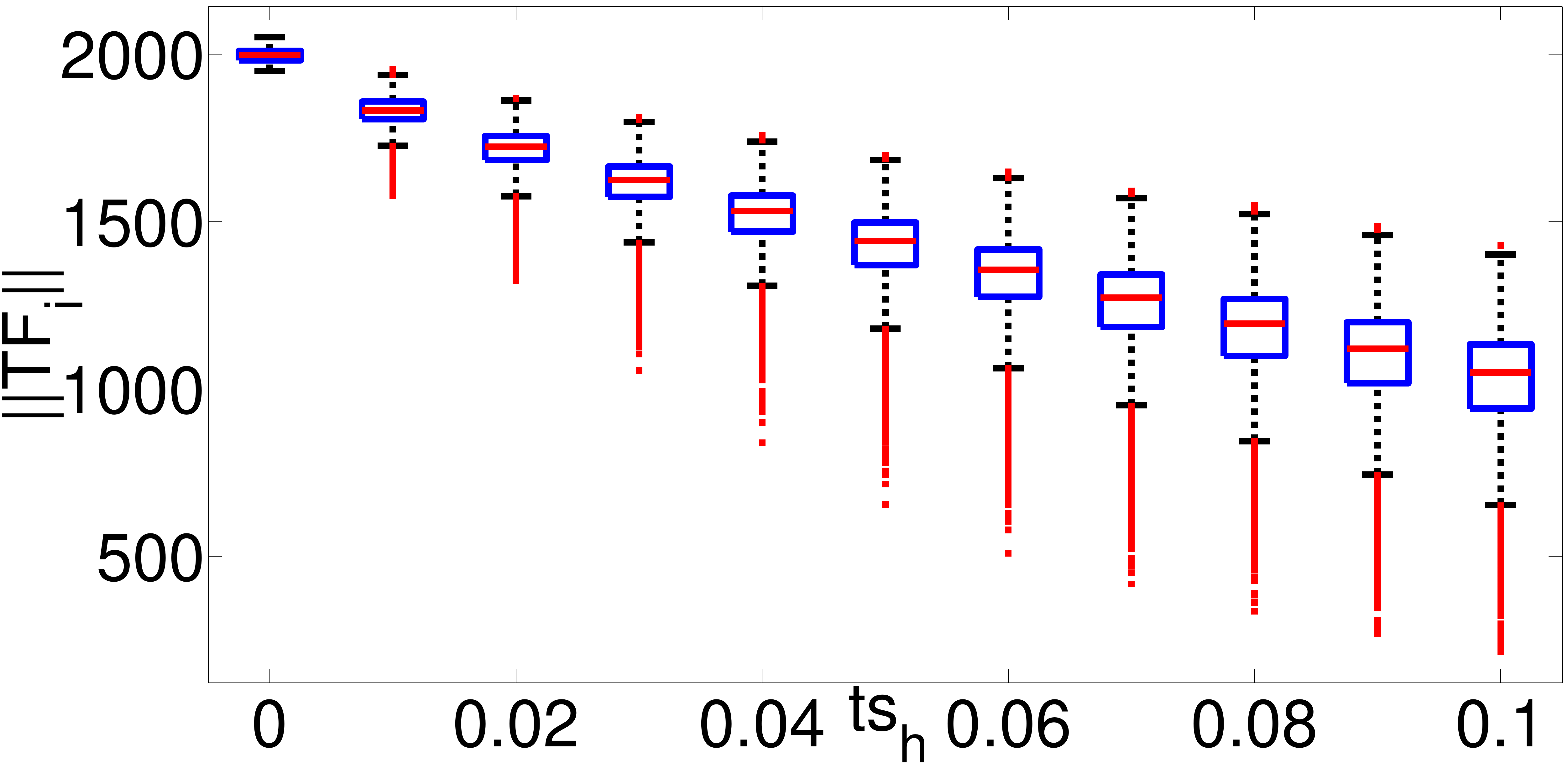}}
     \caption{Size of (Trustworthy) Friendship Circle (i.e., $||F_{i}||$ or $||TF_i||$) in different Size of STor and $ts_{h}$}
     \label{fig:sfcds}
\end{figure*}

\subsubsection{Secure Anonymity in Experiments over PlanetLab}
\label{sec:plsecure}
To conduct the experiments over PlanetLab platform, we modify the source code of Tor (V0.2.1.26) to let the onion proxy request to form circuits automatically and allow the directory server to record which routers are used to establish which circuits. The structure of Tor and STor have been generated in Section \ref{sec:plsetup}. Unlike the simulation, we regard a round of PlanetLab experiment as an user forming a circuit with $3$ routers in $1000$ times and use the ratio of malicious circuits to formed circuits, denoted as $\mathcal{R}_\mathfrak{MC}$, to measure secure anonymity. We consider a circuit as malicious if at least one router in this circuit is malicious.

Fig. \ref{fig:plsai2} shows the secure anonymity, measured by $\mathcal{R}_\mathfrak{MC}$, for Original Tor, Opportunistic Tor, Practical STor and Theoretical STor when $20\%$ candidate routers are malicious. Practical STor obtains less than one fifth of $\mathcal{R}_\mathfrak{MC}$ in the comparison with Opportunistic Tor and less than one eighth of $\mathcal{R}_\mathfrak{MC}$ compared with Original Tor, thus demonstrating much better secure anonymity. Theoretical STor still stays at $\mathcal{R}_\mathfrak{MC}=0$ as friendship circles exclude any malicious routers. Figs. \ref{fig:plsabi2} and \ref{fig:plsawi2} give out the secure anonymity for the Best and Worst cases in different $\omega$ when malicious routers occupancy is $20\%$. When $\omega$ increases from $0$ to $1$, the $\mathcal{R}_\mathfrak{MC}$ grows up from less than $10\%$ to around $30\%$ in the Best Case but to more than $70\%$ in the Worst Case. The balance between the secure anonymity and performance for these two cases over PlanetLab will be detailed in Section \ref{sec:eperformance}. Fig. \ref{fig:plsati2} illustrates that better secure anonymity is achieved with higher $ts_h$, which filters more friends with low trust score. We obtain the similar $\mathcal{R}_\mathfrak{MC}$ for $ts_h=0.005$, $0.015$ and $0.025$, because there are seldom friends with trust score between $0.005$ and $0.025$ in the friendship circle.

\subsection{Experiment to Assess Performance}
\label{sec:eperformance}

\subsubsection{Performance in Simulation}
\label{sec:siperformance}
In the simulation of performance, each router is randomly assigned with a value in ($0$,$10$MB] as its bandwidth. The average bandwidth for each round of simulation is calculated as the average value among $1000$ selected routers. Figs. \ref{fig:sibeper} and \ref{fig:siwoper} show the CDF of the average bandwidth for Tor and Practical STor over different $\omega$ in the Best and Worst Cases. Since the friendship circle excludes routers with large bandwidth in the Worst Case but eliminates small bandwidth ones in the Best Case, Practical STor with $\omega=1$ (i.e., selecting routers solely based on bandwidth in Practical STor) obtains smaller bandwidth distribution than Tor for the Worst Case, whereas larger bandwidth distribution for the Best Case. Therefore, it is necessary for users to encourage more friends with the ability to set up high bandwidth routers to participate in STor when their friendship circle meets the Worst Case.

With the $\omega$ increasing from $0$ to $1$, the average bandwidth of Practical STor decreases from about $8$MB to around $6.5$MB in the Best Case while grows up from less than $2$MB to around $5$MB in the Worst Case. In comparison with Figs. \ref{fig:sabi20} and \ref{fig:sawi20}, we can therefore observe that STor achieves the same trend of secure anonymity and performance for the Best Case, while $\omega$ can help users obtain better anonymity in the sacrifice of performance or vice versa. Particularly, lower $\omega$ leads to better performance (i.e., higher average bandwidth) and better secure anonymity (i.e., lower $\mathcal{R}_\mathfrak{MR}$) in Figs. \ref{fig:sibeper} and \ref{fig:sabi20}, respectively. That is because the rate of trust score's variance is a little bit larger than that of bandwidth's variance in our social network model and larger rate of the variance leads to both better performance and secure anonymity in the Best Case. Note that, if the rate of bandwidth's variance is larger than that of trust score's variance, larger $\omega$ obtains better performance and secure anonymity in the Best Case.

\subsubsection{Performance in Experiments over PlanetLab}
\label{sec:plperformance}
For the performance evaluation over PlanetLab platform, we regard the average bandwidth for each round as the average value among $1000$ formed circuits. We consider the lowest bandwidth among $3$ routers in a circuit as the bandwidth of this circuit. In Fig. \ref{fig:plbeper}, the average bandwidth of Practical STor is decreased from about $80$KB to less than $65$KB when the $\omega$ increases from $0$ to $1$ in the Best Case. By contrast, \ref{fig:plwoper} shows an escalating trend of the average bandwidth in Practical STor with the $\omega$ growing up in the Worst Case. By comparing Figs. \ref{fig:plsabi2} and \ref{fig:plsawi2}, we can see that STor achieves the same variation of secure anonymity and performance for the Best Case and larger rate of the variance leads to better results in this case. $\omega$, on the other hand, can be used to help users obtain better anonymity at the cost of performance and vice versa for the worst case. It is in accordance to the simulation results. Unlike that, experiments over PlanetLab nodes show that Tor obtains a similar bandwidth distribution as Practical STor with $\omega=1$. That is because, according to the setup that is elaborated in Section \ref{sec:plsetup}, the owners of Tor routers over PlanetLab platform are all belong to the friendship circle of STor in our experiments.

\subsection{Experiment to Assess Baseline Anonymity and Scalability}
\label{sec:anonymity}
Both baseline anonymity and scalability can be measured by the number of candidate routers. A large number of candidate routers provides a better baseline anonymity, but leads to a worse scalability. STor confine candidate routers to users' (trustworthy) friendship circle, thus $||F_i||$ and $||TF_i||$ are effective to reflect the baseline anonymity and scalability of STor. In this evaluation, the model used in Section \ref{sec:sisetup} is adopted to simulate STor with different number of user entities. The structure of STor in each size is generated for $100$ times.

Fig. \ref{fig:bds} demonstrates that, although $||F_i||$ proportionally grows up when the size of STor increases from $100$ to $2500$, the size of STor retains to be larger than $||F_i||$ (i.e., $||F_i||$ is around $80\%$ to the size of STor in our model). As a result, STor receives a reduction in its baseline anonymity but an improvement for its scalability. However, the baseline anonymity of STor can be enhanced with the inflation of STor. For instance, as STor with $2500$ entities obtains an average of $2000$ friends in $F_i$, it shows the similar baseline anonymity as Tor with $2000$ candidate routers.

As shown in Figs. \ref{fig:bdst100}-\ref{fig:bdst2500}, the $||TF_i||$ keeps a decreasing trend when $ts_h=0$ grows up to $ts_h=0.1$ in STor with $100$, $1300$ and $2500$ user entities. By comparing with Fig. \ref{fig:sidifftsh}, which shows large $ts_h$ helps STor obtain better secure anonymity, we can thus observe that $ts_h$ is able to help balance the secure anonymity (and the scalability) with the baseline anonymity. Particularly, even when $20\%$ candidate routers are malicious, $ts_h=0.035$ can help STor reach around the theoretically maximum secure anonymity (i.e., $\mathcal{R}_\mathfrak{MR}=0$). Meanwhile, this $ts_h=0.035$ facilitate STor with $2500$ entities obtain a similar baseline anonymity as Tor with $1600$ candidate routers.

\section{Related Work}
\label{sec:relatedwork}
Malicious router based attacks, such as correlation attacks \cite{low-cost,locating,low-resource,one-cell,cell-counter,ZFGBW09}, congestion attacks \cite{EDG09}, disclosure attacks \cite{AK03} and latency based client location attacks \cite{HVC10} etc., have demonstrated severe threats to the state-of-the-art anonymity networks. To launch these attacks, at least one malicious router is required to observe or manipulate the relayed traffic. Particularly, the anonymity networks suffers a fundamental limit in protecting the baseline anonymity when attackers possess sufficient malicious routers to observe the traffic \cite{KAPR2006}. To evade malicious routers, some pioneer mechanisms have been proposed \cite{path,tune-up,tuneupA,DFHM2001,reputation}. Tor uses guard routers at the entry point and selects exit routers according to the exit node policy to circumvent malicious routers \cite{path}. This method relies on a group of directory servers to check each routers' credibility only according to their uptime. Furthermore, an opportunistic bandwidth measurement algorithm has been proposed to replace the existing self-reporting method \cite{tune-up,tuneupA}, thus helping Tor users reduce the hurt caused by malicious routers that announce false bandwidth. Some reputation systems \cite{DFHM2001,reputation}, on the other hand, are designed to verify the correctness of each router in MIX Cascade networks, thus potentially detecting malicious routers and making the circuits more reliable. However, these mechanisms cannot help anonymity networks achieve trust-based anonymous communication, because they have not taken trust into consideration when selecting routers.

A pioneer security analysis for the Onion Routing has implicitly indicated the necessary of trust-based routing algorithm \cite{LRST00}. Furthermore, by understanding the importance of the trust, the adversary models and routing algorithms for the trust-based anonymous communication have been demonstrated \cite{Johnson_CSF_09,Syverson_CCS_11}. Unlike these studies that focus on why the trust is necessary for anonymous communication, STor is a practical solution on how to introduce the trust to anonymous communication. Beside that, many studies \cite{p2pTor1,p2pTor2,p2pTor3} have appeared to use peer-to-peer approaches for scalable anonymous communication. They mainly focus on the design of anonymous P2P lookup mechanisms in the scalable architecture. Unlike that, the social network based trust model introduces trust-based scalability to anonymity networks.

A number of fuzzy model based approaches have been proposed to calculate the trust according to quantitative social properties and propagate the trust over the semantic web social networks \cite{masys,sofuzzy1,sofuzzy2,sofuzzy3,sofuzzy4}. However, these studies calculate trust by using the traditional fuzzy model, thus loosing the functionality to convert the qualitative social attributes. Moreover, a basic model for the propagation of trust and distrust over a trust graph is proposed by \cite{trust}, as well as the Friend-to-Friend networks (e.g., \cite{turtle,oneswam}) have been designed to use the trust from real-world social networks for data sharing. STor, on the other hand, introduces trust and trust propagation to anonymity networks from the real-world social networks.


\section{Conclusions}
\label{sec:conclusion}
We have proposed a novel social network based trust model for trust-based anonymous communication. This model determine trust relationships from real-world social networks. More precisely, an input independent fuzzy model has been devised and used to convert both quantitative and qualitative social attributes into trust values in various social networks. We also propose a trust propagation algorithm to propagate trust over the anonymity networks. By applying this trust model to the Tor network, we have designed STor, a social network based Tor. Extensive experimental results have demonstrated that STor is effective in helping users circumvent malicious routers in a distributed style. With the help of trusted friends, STor users obtain secure anonymity without inducing severe performance and baseline anonymity degradation. This social network based trust model can be widely adopted to introduce trust to other anonymity networks.


\bibliographystyle{IEEEtran}
\bibliography{CCBib}

\section*{Appendix A}
\label{sec:calcu}
\subsection*{Rule $K.1.i$}
For a single social attribute, $\mathcal{A}_{k=K}$, with its matched rule, $K.1.i$, input fuzzy set $\mathcal{I}_{1}^{\mathcal{A}_{k=K}}$ is mapped to output fuzzy set $\mathcal{O}_{1}$. Based on Eq.~(\ref{eqn:inputm}) and Eq.~(\ref{eqn:outputm}), $\mu_{\mathcal{I}_{1}^{\mathcal{A}_{k=K}}}(\mathcal{E})=\mathcal{E},\ \mathcal{E}\in[0, 1]$ and $\mu_{\mathcal{O}_{1}}(tv)=4\times tv-3,\ tv\in[0.75, 1]$. Thus, $\mu_{K.1.i}(\mathcal{E},tv)$ can be computed according to Eq.~(\ref{eqn:single}):
\begin{eqnarray*}
\begin{array}{l}
\mu_{K.1.i}(\mathcal{E},tv)=\left\{
\begin{array}{ll}
\mathcal{E}, &\frac{\mathcal{E}+3}{4}\leq tv\leq 1\\
4\times tv-3, &\frac{3}{4}\leq tv\leq \frac{\mathcal{E}+3}{4}
\end{array}
\right.\\
\mathcal{E}\in [0,1],\ tv\in [0.75,1].
\end{array}
\end{eqnarray*}
As $\rho_{q=1}=2$ is defined in Eq.~(\ref{eqn:rho}), $\mathcal{MP}_{K.1.i}(\mathcal{E})$ can be calculated according to Eq.~(\ref{eqn:mptv}) as:
\begin{eqnarray*}
\begin{array}{l}
\mathcal{MP}_{K.1.i}(\mathcal{E})=\int_{\frac{3}{4}}^{\frac{\mathcal{E}+3}{4}}(tv\times\rho_{q=1}\times(4\times tv-3))d(tv)\\
+\int_{\frac{\mathcal{E}+3}{4}}^{1}(tv\times\rho_{q=1}\times \mathcal{E})d(tv)=-\frac{1}{48}(\mathcal{E}^3+9\mathcal{E}^2-21\mathcal{E}).
\end{array}
\end{eqnarray*}
Considering Eq.~(\ref{eqn:mtv}), $\mathcal{M}_{K.1.i}(\mathcal{E})$ can be determined to be:
\begin{eqnarray*}
\begin{array}{l}
\mathcal{M}_{K.1.i}(\mathcal{E})=\int_{\frac{\mathcal{E}+3}{4}}^{1}(\rho_{q=1}\times \mathcal{E})d(tv)\\
+\int_{\frac{3}{4}}^{\frac{\mathcal{E}+3}{4}}(\rho_{q=1}\times(4\times tv-3))d(tv)=-\frac{1}{4}(\mathcal{E}^2-2\mathcal{E}).
\end{array}
\end{eqnarray*}
Therefore, Eq.~(\ref{eqn:singletvstar}) can be used to calculate $tv^*_{K.1.i}(\mathcal{E})$ as:
\begin{eqnarray*}
\begin{array}{l}
tv^*_{K.1.i}(\mathcal{E})=\frac{\mathcal{MP}_{K.1.i}(\mathcal{E})}{\mathcal{M}_{K.1.i}(\mathcal{E})}=\frac{\mathcal{E}^2+9\mathcal{E}-21}{12(\mathcal{E}-2)}.
\end{array}
\end{eqnarray*}

\subsection*{Rule $K.1.ii$}
For a single social attribute, $\mathcal{A}_{k=K}$, with its matched rule, $K.1.ii$, input fuzzy set $\mathcal{I}_{1}^{\mathcal{A}_{k=K}}$ is mapped to output fuzzy set $\mathcal{O}_{2}$. Based on Eq.~(\ref{eqn:inputm}) and Eq.~(\ref{eqn:outputm}), $\mu_{\mathcal{I}_{1}^{\mathcal{A}_{k=K}}}(\mathcal{E})=\mathcal{E},\ \mathcal{E}\in[0, 1]$, $\mu_{\mathcal{O}_{2}}(tv)=4\times tv-2,\ tv\in[0.5, 0.75]$ and $\mu_{\mathcal{O}_{2}}(tv)=4-4\times tv,\ tv\in[0.75, 1]$. Thus, $\mu_{K.1.ii}(\mathcal{E},tv)$ can be computed according to Eq.~(\ref{eqn:single}) as:
\begin{eqnarray*}
\begin{array}{l}
\mu_{K.1.ii}(\mathcal{E},tv)=\left\{
\begin{array}{ll}
4\times tv-2, &\frac{1}{2}\leq tv\leq \frac{\mathcal{E}+2}{4}\\
\mathcal{E}, &\frac{\mathcal{E}+2}{4}\leq tv\leq \frac{4-\mathcal{E}}{4}\\
4-4\times tv, &\frac{4-\mathcal{E}}{4}\leq tv\leq 1
\end{array}
\right.\\
\mathcal{E}\in [0,1],\ tv\in [0.5,1].
\end{array}
\end{eqnarray*}
As $\rho_{q=2}=1$ is defined in Eq.~(\ref{eqn:rho}), $\mathcal{MP}_{K.1.ii}(\mathcal{E})$ can be calculated according to Eq.~(\ref{eqn:mptv}) as:
\begin{eqnarray*}
\begin{array}{l}
\mathcal{MP}_{K.1.ii}(\mathcal{E})=\int_{\frac{1}{2}}^{\frac{\mathcal{E}+2}{4}}(tv\times\rho_{q=2}\times(4\times tv-2))d(tv)\\
+\int_{\frac{\mathcal{E}+2}{4}}^{4-\frac{\mathcal{E}}{4}}(tv\times\rho_{q=2}\times \mathcal{E})d(tv)\\
+\int_{\frac{4-\mathcal{E}}{4}}^{1}(tv\times\rho_{q=2}\times(4-4\times tv))d(tv)=-\frac{3}{16}(\mathcal{E}^2-2\mathcal{E}).
\end{array}
\end{eqnarray*}
Considering Eq.~(\ref{eqn:mtv}), $\mathcal{M}_{K.1.ii}(\mathcal{E})$ can be determined to be:
\begin{eqnarray*}
\begin{array}{l}
\mathcal{M}_{K.1.ii}(\mathcal{E})=\int_{\frac{1}{2}}^{\frac{\mathcal{E}+2}{4}}(\rho_{q=2}\times(4\times tv-2))d(tv)\\
+\int_{\frac{\mathcal{E}+2}{4}}^{4-\frac{\mathcal{E}}{4}}(\rho_{q=2}\times \mathcal{E})d(tv)\\
+\int_{\frac{4-\mathcal{E}}{4}}^{1}(\rho_{q=2}\times(4-4\times tv))d(tv)=-\frac{1}{4}(\mathcal{E}^2-2\mathcal{E}).
\end{array}
\end{eqnarray*}
Therefore, Eq.~(\ref{eqn:singletvstar}) can be used to calculate $tv^*_{K.1.ii}(\mathcal{E})$ as:
\begin{eqnarray*}
\begin{array}{l}
tv^*_{K.1.ii}(\mathcal{E})=\frac{\mathcal{MP}_{K.1.ii}(\mathcal{E})}{\mathcal{M}_{K.1.ii}(\mathcal{E})}=\frac{3}{4}.
\end{array}
\end{eqnarray*}

\subsection*{Rule $K.2$}
For a single social attribute, $\mathcal{A}_{k=K}$, with its matched rule, $K.2$, input fuzzy set $\mathcal{I}_{2}^{\mathcal{A}_{k=K}}$ is mapped to output fuzzy set $\mathcal{O}_{3}$. Based on Eq.~(\ref{eqn:inputm}) and Eq.~(\ref{eqn:outputm}), $\mu_{\mathcal{I}_{2}^{\mathcal{A}_{k=K}}}(\mathcal{E})=\mathcal{E},\ \mathcal{E}\in[0, 0.5]$, $\mu_{\mathcal{I}_{2}^{\mathcal{A}_{k=K}}}(\mathcal{E})=1-\mathcal{E},\ \mathcal{E}\in[0.5, 1]$, $\mu_{\mathcal{O}_{3}}(tv)=4\times tv-1,\ tv\in[0.25, 0.5]$ and $\mu_{\mathcal{O}_{2}}(tv)=3-4\times tv,\ tv\in[0.5, 0.75]$. Thus, $\mu_{K.2}(\mathcal{E},tv)$ can be computed according to Eq.~(\ref{eqn:single}) when $\mathcal{E}\in [0,0.5]$ as:
\begin{eqnarray*}
\begin{array}{l}
\mu_{K.2}(\mathcal{E},tv)=\left\{
\begin{array}{ll}
4\times tv-1, &\frac{1}{4}\leq tv\leq \frac{\mathcal{E}+1}{4}\\
\mathcal{E}, &\frac{\mathcal{E}+1}{4}\leq tv\leq \frac{3-\mathcal{E}}{4}\\
3-4\times tv, &\frac{3-\mathcal{E}}{4}\leq tv\leq \frac{3}{4}
\end{array}
\right.\\
\mathcal{E}\in [0,0.5],\ tv\in [0.25,0.75].
\end{array}
\end{eqnarray*}
And when $\mathcal{E}\in [0.5,1]$ as
\begin{eqnarray*}
\begin{array}{l}
\mu_{K.2}(\mathcal{E},tv)=\left\{
\begin{array}{ll}
4\times tv-1, &\frac{1}{4}\leq tv\leq \frac{2-\mathcal{E}}{4}\\
1-\mathcal{E}, &\frac{2-\mathcal{E}}{4}\leq tv\leq \frac{2+\mathcal{E}}{4}\\
3-4\times tv, &\frac{2+\mathcal{E}}{4}\leq tv\leq \frac{3}{4}
\end{array}
\right.\\
\mathcal{E}\in [0.5,1],\ tv\in [0.25,0.75].
\end{array}
\end{eqnarray*}
As $\rho_{q=3}=1$ is defined in Eq.~(\ref{eqn:rho}), $\mathcal{MP}_{K.2}(\mathcal{E})$ can be calculated according to Eq.~(\ref{eqn:mptv}) as follows. When $\mathcal{E}\in [0,0.5]$:
\begin{eqnarray*}
\begin{array}{l}
\mathcal{MP}_{K.2}(\mathcal{E})=\int_{\frac{1}{4}}^{\frac{\mathcal{E}+1}{4}}(tv\times\rho_{q=3}\times(4\times tv-1))d(tv)\\
+\int_{\frac{\mathcal{E}+1}{4}}^{3-\frac{\mathcal{E}}{4}}(tv\times\rho_{q=3}\times \mathcal{E})d(tv)\\
+\int_{\frac{3-\mathcal{E}}{4}}^{\frac{3}{4}}(tv\times\rho_{q=3}\times(3-4\times tv))d(tv)=-\frac{1}{8}(\mathcal{E}^2-2\mathcal{E}).
\end{array}
\end{eqnarray*}
And when $\mathcal{E}\in [0.5,1]$:
\begin{eqnarray*}
\begin{array}{l}
\mathcal{M}_{K.2}(\mathcal{E})=\int_{\frac{1}{4}}^{\frac{2-\mathcal{E}}{4}}(tv\times\rho_{q=3}\times(4\times tv-1))d(tv)\\
+\int_{\frac{2-\mathcal{E}}{4}}^{\frac{2+\mathcal{E}}{4}}(tv\times\rho_{q=3}\times (1-\mathcal{E}))d(tv)\\
+\int_{\frac{2+\mathcal{E}}{4}}^{\frac{3}{4}}(tv\times\rho_{q=3}\times(3-4\times tv))d(tv)=-\frac{1}{8}(\mathcal{E}^2-1).
\end{array}
\end{eqnarray*}
Considering Eq.~(\ref{eqn:mtv}), $\mathcal{M}_{K.2}(\mathcal{E})$ can be determined as below. When $\mathcal{E}\in [0,0.5]$:
\begin{eqnarray*}
\begin{array}{l}
\mathcal{M}_{K.2}(\mathcal{E})=\int_{\frac{1}{4}}^{\frac{\mathcal{E}+1}{4}}(\rho_{q=3}\times(4\times tv-1))d(tv)\\
+\int_{\frac{\mathcal{E}+1}{4}}^{3-\frac{\mathcal{E}}{4}}(\rho_{q=3}\times \mathcal{E})d(tv)\\
+\int_{\frac{3-\mathcal{E}}{4}}^{\frac{3}{4}}(\rho_{q=3}\times(3-4\times tv))d(tv)=-\frac{1}{4}(\mathcal{E}^2-2\mathcal{E}).
\end{array}
\end{eqnarray*}
And when $\mathcal{E}\in [0.5,1]$:
\begin{eqnarray*}
\begin{array}{l}
\mathcal{M}_{K.2}(\mathcal{E})=\int_{\frac{1}{4}}^{\frac{2-\mathcal{E}}{4}}(\rho_{q=3}\times(4\times tv-1))d(tv)\\
+\int_{\frac{2-\mathcal{E}}{4}}^{\frac{2+\mathcal{E}}{4}}(\rho_{q=3}\times (1-\mathcal{E}))d(tv)\\
+\int_{\frac{2+\mathcal{E}}{4}}^{\frac{3}{4}}(\rho_{q=3}\times(3-4\times tv))d(tv)=-\frac{1}{4}(\mathcal{E}^2-1).
\end{array}
\end{eqnarray*}
Therefore, Eq.~(\ref{eqn:singletvstar}) can be used to calculate $tv^*_{K.2}(\mathcal{E})$ as follows. When $\mathcal{E}\in [0,0.5]$:
\begin{eqnarray*}
\begin{array}{l}
tv^*_{K.2}(\mathcal{E})=\frac{\mathcal{MP}_{K.2}(\mathcal{E})}{\mathcal{M}_{K.2}(\mathcal{E})}=\frac{-\frac{1}{8}(\mathcal{E}^2-2\mathcal{E})}{-\frac{1}{4}(\mathcal{E}^2-2\mathcal{E})}=\frac{1}{2}.
\end{array}
\end{eqnarray*}
And when $\mathcal{E}\in [0.5,1]$:
\begin{eqnarray*}
\begin{array}{l}
tv^*_{K.2}(\mathcal{E})=\frac{\mathcal{MP}_{K.2}(\mathcal{E})}{\mathcal{M}_{K.2}(\mathcal{E})}=\frac{-\frac{1}{8}(\mathcal{E}^2-1)}{-\frac{1}{4}(\mathcal{E}^2-1)}=\frac{1}{2}.
\end{array}
\end{eqnarray*}
Thus $tv^*_{K.2}(\mathcal{E})=\frac{1}{2},\ \mathcal{E}\in [0,1]$.

\subsection*{Rule $K.3.i$}
For a single social attribute, $\mathcal{A}_{k=K}$, with its matched rule, $K.3.i$, input fuzzy set $\mathcal{I}_{3}^{\mathcal{A}_{k=K}}$ is mapped to output fuzzy set $\mathcal{O}_{4}$. Based on Eq.~(\ref{eqn:inputm}) and Eq.~(\ref{eqn:outputm}), $\mu_{\mathcal{I}_{3}^{\mathcal{A}_{k=K}}}(\mathcal{E})=1-\mathcal{E},\ \mathcal{E}\in[0, 1]$, $\mu_{\mathcal{O}_{4}}(tv)=4\times tv,\ tv\in[0, 0.25]$ and $\mu_{\mathcal{O}_{4}}(tv)=2-4\times tv,\ tv\in[0.25, 0.5]$. Thus, $\mu_{K.3.i}(\mathcal{E},tv)$ can be computed according to Eq.~(\ref{eqn:single}) as:
\begin{eqnarray*}
\begin{array}{l}
\mu_{K.3.i}(\mathcal{E},tv)=\left\{
\begin{array}{ll}
4\times tv, &0\leq tv\leq \frac{1-\mathcal{E}}{4}\\
1-\mathcal{E}, &\frac{1-\mathcal{E}}{4}\leq tv\leq \frac{1+\mathcal{E}}{4}\\
2-4\times tv, &\frac{1+\mathcal{E}}{4}\leq tv\leq \frac{1}{2}
\end{array}
\right.\\
\mathcal{E}\in [0,1],\ tv\in [0,0.5].
\end{array}
\end{eqnarray*}
As $\rho_{q=4}=1$ is defined in Eq.~(\ref{eqn:rho}), $\mathcal{MP}_{K.3.i}(\mathcal{E})$ can be calculated according to Eq.~(\ref{eqn:mptv}) as:
\begin{eqnarray*}
\begin{array}{l}
\mathcal{MP}_{K.3.i}(\mathcal{E})=\int_{0}^{\frac{1-\mathcal{E}}{4}}(tv\times\rho_{q=4}\times (4\times tv))d(tv)\\
+\int_{\frac{1+\mathcal{E}}{4}}^{\frac{1-\mathcal{E}}{4}}(tv\times\rho_{q=4}\times (1-\mathcal{E}))d(tv)\\
+\int_{\frac{1+\mathcal{E}}{4}}^{\frac{1}{2}}(tv\times\rho_{q=4}\times (2-4\times tv))d(tv)=-\frac{1}{16}(1-\mathcal{E}^2).
\end{array}
\end{eqnarray*}
Considering Eq.~(\ref{eqn:mtv}), $\mathcal{M}_{K.3.i}(\mathcal{E})$ can be determined to be:
\begin{eqnarray*}
\begin{array}{l}
\mathcal{M}_{K.3.i}(\mathcal{E})=\int_{0}^{\frac{1-\mathcal{E}}{4}}(\rho_{q=4}\times (4\times tv))d(tv)\\
+\int_{\frac{1+\mathcal{E}}{4}}^{\frac{1-\mathcal{E}}{4}}(\rho_{q=4}\times (1-\mathcal{E}))d(tv)\\
+\int_{\frac{1+\mathcal{E}}{4}}^{\frac{1}{2}}(\rho_{q=4}\times (2-4\times tv))d(tv)=-\frac{1}{4}(1-\mathcal{E}^2).
\end{array}
\end{eqnarray*}
Therefore, Eq.~(\ref{eqn:singletvstar}) can be used to calculate $tv^*_{K.3.i}(\mathcal{E})$ as:
\begin{eqnarray*}
\begin{array}{l}
tv^*_{K.3.i}(\mathcal{E})=\frac{\mathcal{MP}_{K.3.i}(\mathcal{E})}{\mathcal{M}_{K.3.i}(\mathcal{E})}=\frac{1}{4}.
\end{array}
\end{eqnarray*}

\subsection*{Rule $K.3.ii$}
For a single social attribute, $\mathcal{A}_{k=K}$, with its matched rule, $K.3.ii$, input fuzzy set $\mathcal{I}_{3}^{\mathcal{A}_{k=K}}$ is mapped to output fuzzy set $\mathcal{O}_{5}$. Based on Eq.~(\ref{eqn:inputm}) and Eq.~(\ref{eqn:outputm}), $\mu_{\mathcal{I}_{3}^{\mathcal{A}_{k=K}}}(\mathcal{E})=1-\mathcal{E},\ \mathcal{E}\in[0, 1]$ and $\mu_{\mathcal{O}_{5}}(tv)=1-4\times tv,\ tv\in[0, 0.25]$. Thus, $\mu_{K.3.ii}(\mathcal{E},tv)$ can be computed according to Eq.~(\ref{eqn:single}) as:
\begin{eqnarray*}
\begin{array}{l}
\mu_{K.3.ii}(\mathcal{E},tv)=\left\{
\begin{array}{ll}
1-\mathcal{E}, &0\leq tv\leq \frac{\mathcal{E}}{4}\\
1-4\times tv, &\frac{\mathcal{E}}{4}\leq tv\leq \frac{1}{4}
\end{array}
\right.\\
\mathcal{E}\in [0,1],\ tv\in [0,0.25].
\end{array}
\end{eqnarray*}
As $\rho_{q=5}=2$ is defined in Eq.~(\ref{eqn:rho}), $\mathcal{MP}_{K.3.ii}(\mathcal{E})$ can be calculated according to Eq.~(\ref{eqn:mptv}) as:
\begin{eqnarray*}
\begin{array}{l}
\mathcal{MP}_{K.3.ii}(\mathcal{E})=\int_{0}^{\frac{\mathcal{E}}{4}}(tv\times\rho_{q=5}\times (1-\mathcal{E}))d(tv)\\
+\int_{\frac{\mathcal{E}}{4}}^{\frac{1}{4}}(tv\times\rho_{q=4}\times (1-4\times tv))d(tv)=-\frac{1}{48}(1-\mathcal{E}^3).
\end{array}
\end{eqnarray*}
Considering Eq.~(\ref{eqn:mtv}), $\mathcal{M}_{K.3.ii}(\mathcal{E})$ can be determined to be:
\begin{eqnarray*}
\begin{array}{l}
\mathcal{M}_{K.3.ii}(\mathcal{E})=\int_{0}^{\frac{\mathcal{E}}{4}}(\rho_{q=5}\times (1-\mathcal{E}))d(tv)\\
+\int_{\frac{\mathcal{E}}{4}}^{\frac{1}{4}}(\rho_{q=4}\times (1-4\times tv))d(tv)=-\frac{1}{4}(1-\mathcal{E}^2).
\end{array}
\end{eqnarray*}
Therefore, Eq.~(\ref{eqn:singletvstar}) can be used to calculate $tv^*_{K.3.ii}(\mathcal{E})$ as:
\begin{eqnarray*}
\begin{array}{l}
tv^*_{K.3.ii}(\mathcal{E})=\frac{\mathcal{MP}_{K.3.ii}(\mathcal{E})}{\mathcal{M}_{K.3.ii}(\mathcal{E})}=\frac{\mathcal{E}^2+\mathcal{E}+1}{12(\mathcal{E}+1)}.
\end{array}
\end{eqnarray*}

\end{document}